\documentclass[trackchanges,twocolumn]{aastex}

\usepackage{natbib,epsfig,siunitx,url,graphicx,amsmath,footnote,float,xcolor,cases}

\begin{document}

\submitted{Accepted for publication in Icarus}

\title{Born eccentric: constraints on Jupiter and Saturn's pre-instability orbits}

\author{Matthew S. Clement\altaffilmark{1}, Sean N. Raymond\altaffilmark{2}, Nathan A. Kaib\altaffilmark{3}, Rogerio Deienno\altaffilmark{4}, John E. Chambers\altaffilmark{1} \& Andr\'{e} Izidoro\altaffilmark{5}}

\altaffiltext{1}{Earth and Planets Laboratory, Carnegie Institution for Science, 5241 Broad Branch Road, NW, Washington, DC 20015, USA}
\altaffiltext{2}{Laboratoire d'Astrophysique de Bordeaux, Univ. Bordeaux, CNRS, B18N, all{\'e} Geoffroy Saint-Hilaire, 33615 Pessac, France}
\altaffiltext{3}{HL Dodge Department of Physics Astronomy, University of Oklahoma, Norman, OK 73019, USA}
\altaffiltext{4}{Southwest Research Institute, 1050 Walnut St. Suite 300, Boulder, CO 80302, USA}
\altaffiltext{5}{Department of Earth, Environmental and Planetary Sciences, MS 126, Rice University, Houston, TX 77005, USA}
\altaffiltext{*}{corresponding author email: mclement@carnegiescience.edu}

\begin{abstract}
An episode of dynamical instability is thought to have sculpted the orbital structure of the outer solar system.  When modeling this instability, a key constraint comes from Jupiter's fifth eccentric mode (quantified by its amplitude $M_{55}$), which is an important driver of the solar system's secular evolution.  Starting from commonly-assumed near-circular orbits, the present-day giant planets' architecture lies at the limit of numerically generated systems, and $M_{55}$ is rarely excited to its true value.  Here we perform a dynamical analysis of a large batch of artificially triggered instabilities, and test a variety of configurations for the giant planets' primordial orbits. In addition to more standard setups, and motivated by the results of modern hydrodynamical simulations of the giant planets' evolution within the primordial gaseous disk, we consider the possibility that Jupiter and Saturn emerged from the nebular gas locked in 2:1 resonance with non-zero eccentricities.  We show that, in such a scenario, the modern Jupiter-Saturn system represents a typical simulation outcome, and $M_{55}$ is commonly matched.  Furthermore, we show that Uranus and Neptune's final orbits are determined by a combination of the mass in the primordial Kuiper belt and that of an ejected ice giant.  

%\break
\end{abstract}

%\linenumbers
%\doublespacing
\section{Introduction}

The realization that the outer planets' orbits diverge over time \citep{fer84} as a consequence of small exchanges of angular momentum with nearby planetesimals has revolutionized our understanding of the solar system's global dynamical history over the past several decades.  Early work invoked migration to explain the orbital and resonant structures of the Kuiper belt \citep{malhotra93,malhotra95}.  A consequence of this migration is an epoch of dynamical instability that rapidly transforms the outer solar system from its primordially concentrated architecture into the radially diffuse system of orbits that exists today \citep{thommes99}.  These ideas were eventually built upon to form a comprehensive evolutionary framework for the solar system.  In this paradigm of a migration-shaped solar system, the Kuiper belt's modern structure is tied to the giant planets' early dynamics \citep{hahn05,levison08,batygin11,nesvorny15a,nesvorny15b,kaibsheppard16} which, in turn, are dependent on the properties of the primordial gas disk \citep{masset01,morby07,pierens08,zhang10} and first solid bodies \citep{hahn99,gomes04}.  These ideas led to the development of the so-called \textit{Nice Model}; a robustly formulated hypothesis for the solar system's dynamical evolution \citep{Tsi05,mor05,gomes05}.  In the current framework of the Nice Model, the giant planets emerge from the primordial gaseous nebula in a compact, resonant configuration \citep{morby07}.  After some period of time, the collection of synchronized orbits is cataclysmically destroyed when one or more of the planets are perturbed out of resonance \citep{levison11,nesvorny12,deienno17,quarles19}.  This departure from resonance triggers an epoch of dynamical instability that expediently reshapes the outer solar system into its modern form \citep[for a recent review, see][]{nesvorny18_rev}.

While researchers continue to debate specific aspects of the Nice Model (e.g.: its initial conditions, timing and strength; addressed in greater detail in section \ref{sect:motivation}), its wide acceptance within the field is clearly a result of the scenario's consistent ability to reproduce many peculiar aspects of the solar system in numerical simulations.  Among these are the orbital structure of the Kuiper \citep{levison08,nesvorny15a,nesvorny15b} and asteroid belts \citep{roig15,deienno16,deienno18,clement18_ab}, the capture and evolution of trojan asteroids in the outer solar system \citep{mor05,nesvorny13,nesvorny18}, and certain properties of the giant planets' moons \citep{barr10,deienno14,nesvorny14a,nesvorny14b}.  In particular, a sequence of events where the giant planets acquire their modern orbits through a series of planetary encounters, rather than via smooth migration, is the only known model capable of explaining the capture of irregular moons in the outer solar system \citep{nesvory07}, and preventing sweeping secular resonances from destroying the asteroid belt \citep{walshmorb11}.

The highly chaotic nature of the Nice Model instability makes it difficult to investigate with N-body simulations.  While the exact initial configuration of the outer solar system prior to the instability is unknown, the giant planets' initial orbits can be somewhat constrained by hydrodynamical simulations that study their migration in the gas disk phase \citep[e.g.:][]{pierens11,pierens14}, and by comparing large suites of instability outcomes with important aspects of the modern planets' architecture \citep[e.g.:][]{nesvorny11}. However, even the most successful sets of initial conditions tested in statistical studies of the instability reproduce the outer solar system in broad strokes just a few percent of the time \citep{nesvorny11,nesvorny12,batbro12,kaibcham16,deienno17,clement18}.  The stochastic nature of the giant planets' early evolution thus presents a significant complication for authors attempting to break degeneracies between the different possible primordial giant planet configurations and global disk properties.  

In this paper we present a robust dynamical analysis of the solar system's instability similar to previous works by \citet{nesvorny12} and \citet{deienno17}.  In particular, we summarize the problems with the common assumption that Jupiter and Saturn were captured in to a 3:2 mean motion resonance (MMR), and systematically test whether the 2:1 is a viable alternative as proposed by \citet{pierens14}.  We also consider the possibility that the giant planets already possessed moderate eccentricities ($\sim$0.025-0.15) before the instability \citep{pierens14}.  The structure of our manuscript is as follows.  In section \ref{sect:motivation} we review the Nice Model instability, analyze the problems with the current consensus scenario (the 3:2 Jupiter-Saturn resonance), establish an improved means of constraining our simulations, and discuss the application of these new success criteria.  We describe our numerical simulations and in section \ref{sect:methods}, analyze their results in section \ref{sect:results}, and discuss the consequences of our proposed scenario in section \ref{sect:discuss}.  Additional supplementary background information is provided in appendices \ref{sect:timing}-\ref{sect:alternatives}.

\section{Motivation}
\label{sect:motivation}

\subsection{Background}

\subsubsection{The 3:2 MMR}
\label{sect:3_2}

The original Nice Model simulations presented in \citet{Tsi05} configured the primordial outer solar system in a somewhat ad hoc manner.  Specifically, the gas giants' initial orbits were assigned such that Jupiter and Saturn would migrate through their mutual 2:1 resonance and trigger the instability.  However, such initial conditions are at odds with results from investigations of giant planet growth and evolution in the gas disk phase \citep[particularly foundational were the works of][]{masset01,morby07,morbidelli07}.  Hydrodynamical simulations of the giant planets migrating in gaseous disks find that the giant planets are most likely to emerge from the primordial nebula in a mutual resonant\footnote{Note that it may not be a strict requirement that the giant planets be in resonance prior to the instability \citep[as was found in the pebble accretion model of][]{levison15_gp}.  Indeed, such a scenario has been shown to be more successful at generating the giant planets' modern obliquities \citep{brasser_lee_15} than the standard resonant version of the Nice Model \citep{vokrouhicky15}.  We speculate further about this alternative scenario in section \ref{sect:discuss}.} configuration \citep{zhang10,pierens11,dangelo12,izidoro15_ig}.  Moreover, these initial conditions are largely consistent with observed resonant chains of giant exoplanets \citep[perhaps most famously by GJ 876:][]{rivera10} and gaps in proto-planetary disks presumably induced by growing planets \citep[e.g.:][]{bae19}.  Thus, authors investigating the Nice Model must first determine which particular chain of resonances the giant planets were born in.  

Early hydrodynamical simulations indicated rather convincingly that capture within the 3:2 MMR is the only possibility for a Jupiter-Saturn-like mass configuration \citep{morby07,pierens08}.  As a result, an overwhelming number of dynamical investigations dedicated to the study of the Nice Model over the past decade or so have almost exclusively considered Jupiter and Saturn in a primordial 3:2 MMR \citep[e.g.:][]{nesvorny11,nesvorny13,roig15,roig16,kaibcham16,deienno16,deienno18,clement18,clement18_ab}.  In general, instabilities originating from the 3:2 Jupiter-Saturn resonance are inherently more violent that those that begin with the planets in a 2:1 MMR \citep{nesvorny12}.  This is an obvious consequence of the fact that placing the solar system's two most massive planets in closer proximity to one another leads to stronger gravitational encounters within the instability.  As Uranus or Neptune are often ejected in dynamical simulations of such a violent instability \citep{morby09}, versions of the departure-from-resonance Nice Model adaptation \citep[in contrast to the resonance-crossing scenario originally envisioned in][]{Tsi05} tend to invoke either a fifth or sixth primordial giant planet.  In successful models, the additional ice giants are ejected from the system during the instability, thus leading to a higher fraction of simulations finishing with four giant planets \citep{nesvorny11}.  

As each individual numerically generated instability from a given resonant chain is unique, the planets' particular evolution \textit{within} the instability (in addition to the initial conditions) is also important to constrain.  The Kuiper belt's final structure \citep{nesvorny15a,nesvorny15b,nesvorny16} is somewhat sensitive to the instability's timing \citep[mainly Neptune's migration history.  However][showed that Neptune's eccentricity is important as well]{volk19}.  In contrast, the asteroid belt \citep{roig15,izidoro16,deienno16,deienno18} and terrestrial planets' \citep{bras09,agnorlin12,kaibcham16} survival is particularly tied to the violent and chaotic evolution within the instability.  Powerful secular resonances chaotically traversing regions of the inner solar system during the instability can lead to planet ejections, collisions, and levels of dynamical excitation inconsistent with that of the modern inner solar system.  To avoid this problem, \citet{bras09} proposed that Jupiter and Saturn's semi-major axes must strictly evolve in a step-wise manner.  While this so-called ``jumping Jupiter'' type of evolution can adequately reproduce the terrestrial planets' orbits \citep{roig16} and the asteroid belt's dynamical structure \citep{deienno16,deienno18}, the inner asteroid belt's high-inclination population can be over-populated as a result of the violent sweeping of the $\nu_{6}$ secular resonance \citep{morby10,walshmorb11,minton11} if the jump is not ideal.  However, many of these high-inclination asteroids can be removed when $\nu_{6}$ reverses direction as Jupiter and Saturn approach the 5:2 MMR \citep{clement20_mnras}, thereby making somewhat less-regular instabilities with less-ideal jumps potentially viable.  Moreover, such finely-tuned instabilities might also not be required to save the terrestrial planets from loss and over-excitation \citep[e.g.:][]{roig16} if the event occurs while they are still forming \citep{lykawaka13,clement18,ray18_rev}.  In such a scenario, Jupiter's enhanced eccentricity \citep{ray09a} and chaotically sweeping secular resonances \citep{clement18_ab} excite and remove material from the asteroid belt- and Mars-forming regions, while leaving Earth and Venus' growth largely undisturbed \citep{clement18_frag}.  As simulations in \citet{clement18} found satisfactory terrestrial planet outcomes in systems that experienced a variety of jumps, the range of possible viable instability evolutions might be broadened if the event occurs in situ with the process of terrestrial planet formation (we discuss the timing of the Nice Model in appendix \ref{sect:timing}).  While such a violent scenario (without an ideal jump) is capable of depleting the asteroid belt's total mass by 3-4 orders of magnitude \citep{clement18_ab}, it is still unclear whether such a powerful depletion event and a successful inner solar system are mutually-exclusive results.

In summary, for the reasons detailed above, a ``jumping Jupiter'' type instability originating from the 3:2 Jupiter-Saturn resonance remains the current consensus version of the Nice Model.  However, adequately exciting Jupiter's eccentricity without exceeding Jupiter and Saturn's modern orbital spacing is extremely challenging (see further discussion about the primordial 3:2 Jupiter-Saturn resonance's problem in section \ref{sect:e55_prob}).  Therefore, we argue that a detailed investigation of alternative primordial outer solar system configurations, specifically the 2:1 Jupiter-Saturn resonance, is warranted.

\subsubsection{The 2:1 MMR}
\label{sect:2_1}

As a result of hydrodynamical studies that largely considered isothermal disks of fixed viscosity \citep[e.g.:][]{morby07,pierens08} favoring the 3:2 Jupiter-Saturn resonance,  subsequent study of the primordial 2:1 MMR's dynamics and potential instability evolution was largely regarded as a purely academic endeavor \citep{nesvorny12}.  In general, instabilities beginning from the 2:1 are less violent than those born out of the 3:2 \citep{nesvorny12,deienno17}, and often yield evolutionary histories for the planets similar to those recorded in simulations of the original Nice Model \citep{Tsi05}.  Thus, the 2:1 is highly successful at generating a final system of planets with low eccentricities and inclinations, and less successful at exciting Jupiter's eccentricity (though the primordial 3:2 MMR struggles in this manner as well) and matching the modern Jupiter-Saturn period ratio.  As discussed in the previous section, exciting Jupiter's eccentricity requires a relatively strong encounter with the ejected ice giant \citep{morby09}.  When the planets emerge from the gas disk in the 2:1 MMR's broader radial spacing, the resulting instability encounters can often be weaker.  While this can lead to systematically smaller jumps, it is still relatively easy for a system to exceed the modern value of $P_{S}/P_{J}=$ 2.49 when Jupiter and Saturn migrate smoothly after the instability \citep{nesvorny12}. As the amount of residual migration is a function of the remnant planetesimal mass, and the magnitude of Saturn's jump is related to the mass of the ejected planet, it may be possible to improve the likelihood that simulations finish with $P_{S}/P_{J}$ near the modern value by adjusting these two parameters (we explore this in sections \ref{sect:kb_mass} and \ref{sect:ig_mass}).

Hydrodynamical simulations probing different disk thermal and viscosity profiles in \citet{pierens14} found that the solar system's gas giants' capture in the 2:1 MMR is a real possibility in relatively low-mass disks.  Additionally, the authors found that outward migration occurs in the 2:1 MMR when the disk viscosity parameter is low.  Of great interest to our present manuscript, \citet{pierens14} found several cases where Jupiter and Saturn attain relatively high eccentricities ($\sim$0.05-0.20) while migrating outward in the 2:1 MMR (see figures 2 and 4 in that paper).  This dynamical excitation occurs in the 2:1 MMR because the planets carve out a larger gap in the disk than when they are locked in the 3:2.  Because of this wider gap, the planets' orbits are damped less strongly by the gas disk.  Additionally, a less-massive, low-viscosity disk is a crucial prerequisite for capture in the 2:1 as the planets are less likely to migrate past the 2:1 and fall in to the 3:2.  However, it should be noted that the work of \citet{pierens14} was performed before the study of \citet{kanagawa18}; which demonstrated that type II migration is slowed and subdued \citep[e.g.:][]{mcnally19} during the process of radial gap opening as the surface density at the bottom of the gap decreases.  Thus, capture in the 2:1 might potentially be more favorable with the incorporation of an improved radial migration model.   \citet{pierens14} also presented a small set of instability simulations taking the Jupiter-Saturn 2:1 MMR as an initial condition.  Although the total number of runs performed was low, the authors reported reasonable success rates, and concluded that the 2:1 MMR is a viable evolutionary path for the solar system.  However, subsequent study of the 2:1 has been noticeably limited.  \citet{deienno17} reanalyzed the favored configurations of \citet{nesvorny12} against new constraints for the Kuiper belt's orbital structure \citep{nesvorny15a,nesvorny15b} and included two different resonant chains beginning with the 2:1 (from inside out: 2:1,3:2,3:2,3:2 and 2:1,4:3,3:2,3:2).  The configuration with Jupiter and Saturn in the 2:1 and the first ice giant in a 4:3 with Saturn was shown to be highly successful at yielding the proper migration history for Neptune ($\sim$40$\%$ of systems displaying the appropriate evolution).  However, \citet{deienno17} also found that wide resonant chains beginning with the gas giants and first ice giant in a 2:1,3:2 chain are highly stable, and typically do not decompose into an orbital instability.  Therefore, it is important to point out that tighter ice giant resonances (e.g. 4:3 or 5:4) are a possible pre-condition of the 2:1 Jupiter-Saturn configuration.

\subsection{Exciting $M_{55}$}
\label{sect:e55_prob}

An instability simulation where the final giant planets' mean eccentricities closely resemble those of the actual planets might still be a poor solar system analog if its secular architecture is incorrect.  Indeed, mean eccentricities are insufficient to fully characterize the secular structure of a planetary system because planets' eccentricities oscillate on timescales of order $P_{J}M_{\odot}/M_{J}$ \citep[as Jupiter is the most significant perturber in the N-body problem of the solar system, e.g.:][]{poincare1892,laskar90,morb09}.  For a system possessing $N$ massive planetary bodies there will be $2N$ fundamental frequencies describing the system's secular evolution.  These frequencies, denoted $g_{i}$ for the eccentricity vector precessions and $s_{i}$ for the nodal precession rates, and their magnitudes in each planet's orbit ($M_{ij}$; $i,j=$ 1-8) define the secular evolution of the solar system.  As Jupiter and Saturn are the most significant eccentric perturbing bodies in the solar system (see further discussion in appendix \ref{sect:secular}, it is reasonable to seek out an evolutionary model that consistently generates the correct orbital orientation and secular structure for the two planets.  For the reasons outlined in the introduction, the most successful hypothesis to date is the Nice Model \citep{Tsi05}.  Moreover, the arguments laid out in \citet{morby09} convincingly explain why the instability scenario is the only viable option to explain the solar system's unique secular architecture: particularly the large magnitude of $M_{55}$ and the fact that Jupiter's eccentricity lies in the $|M_{55}|>|M_{56}|$ regime.  \citet{morby09} also investigated the possibility of Jupiter's eccentricity being excited solely via planetary migration, MMR crossings, or three planet dynamics (e.g.: the migration of an eccentric Uranus or close encounters with Uranus).  The authors concluded that no alternative could provide strong enough excitation in as self-consistent of a manner as the Nice Model (we briefly explore the possibility of alternative excitation mechanisms in appendix \ref{sect:alternatives}).  

\begin{table*}
\centering
\begin{tabular}{c c c c c c c c}
\hline
Name  & $N_{Pln}$ &  $M_{disk}$ & $\delta$r & $r_{out}$ & $a_{nep}$ & Resonance Chain & $M_{ice}$\\
& & ($M_{\oplus}$) & (au) & (au) & (au) & & ($M_{\oplus}$)\\
\hline
$5GP_{control}$ & 5 & 35 & 1.5 & 30 & 17.4 & 3:2,3:2,3:2,3:2 & 16,16,16 \\
$6GP_{control}$ & 6 & 20 & 1.0 & 30 & 20.6 & 3:2,4:3,3:2,3:2,3:2 & 8,8,16,16 \\
\hline
\end{tabular}
\caption{Table of giant planet initial resonant configurations reproduced from \citet{clement18} and \citet{clement18_frag}.  The columns are: (1) the name of the simulation set, (2) the number of giant planets, (3) the mass of the planetesimal disk exterior to the giant planets, (4) the distance between the outermost ice giant and the planetesimal disk’s inner edge, (5) the semi-major axis of the outermost ice giant (commonly referred to as Neptune, however not necessarily the planet which completes the simulation at Neptune's present orbit), (6) the resonant configuration of the giant planets starting with the Jupiter/Saturn resonance, and (7) the masses of the ice giants from inside to outside.}
\label{table:gp}
\end{table*}

\begin{figure}
	\centering
	\includegraphics[width=0.49\textwidth]{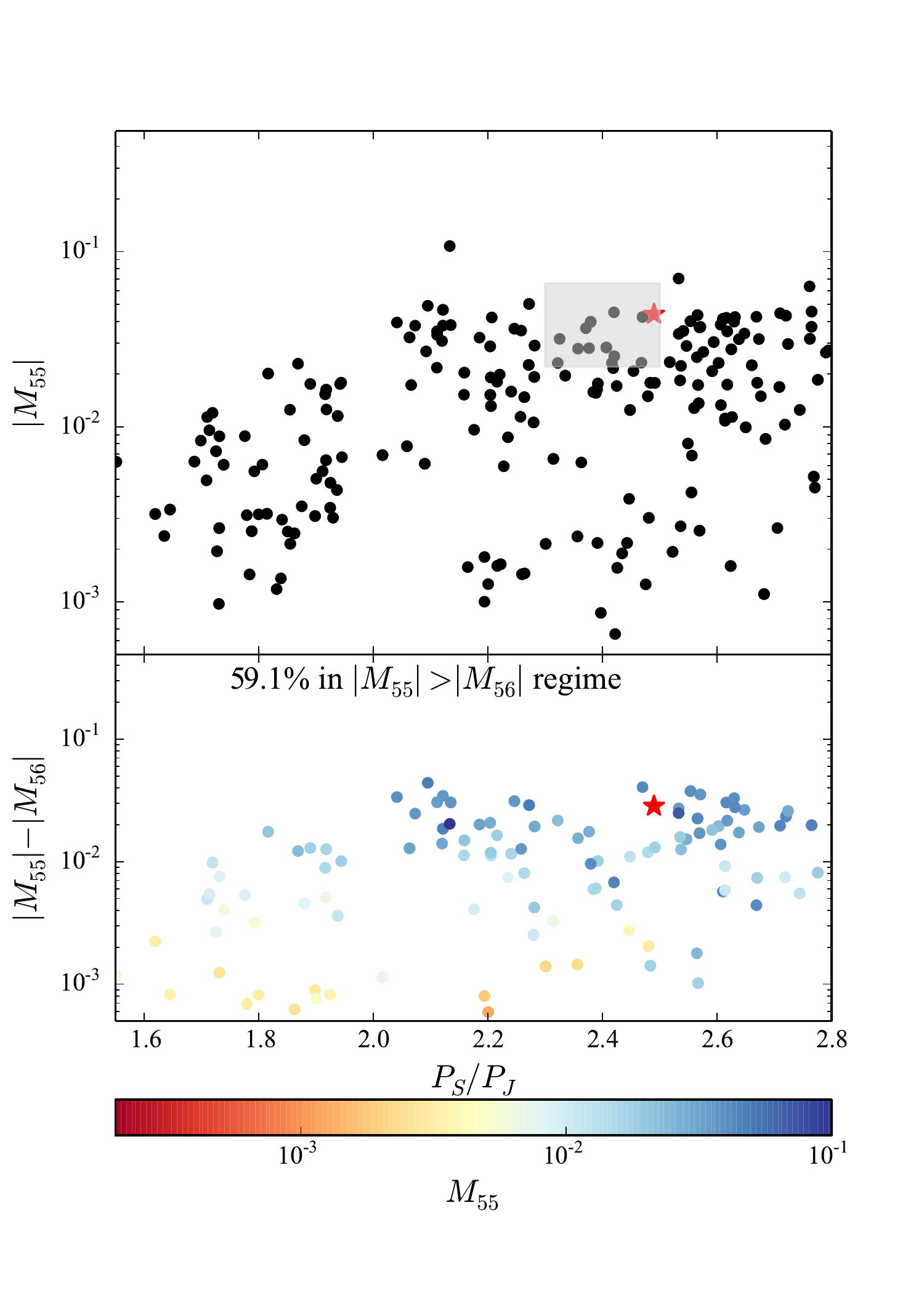}
	\caption{386 instability simulations from \citet{clement18} and \citet{clement18_frag}.  The top panel plots the final value of $M_{55}$ against the Jupiter-Saturn period ratio.  The bottom panel depicts $|M_{55}|-|M_{56}|$ vs $P_{S}/P_{J}<$ for 184 simulations that finished in the $|M_{55}|>|M_{56}|$ regime.  The shaded grey area delimits the region of 2.3 $<P_{S}/P_{J}<$ 2.5 and 0.022 $<M_{55}<$ 0.066 (related to our success criteria, see section \ref{sect:success}).}
	\label{fig:c18_19}
\end{figure}

To illustrate the types of final Jupiter-Saturn configurations that emerge from instabilities beginning with the gas giants locked in a 3:2 MMR, we analyze two batches of instability simulations from \citet{clement18} and \citet{clement18_frag}.  These initial conditions (table \ref{table:gp}) are based on the most successful five ($5GP_{control}$) and six giant planet ($6GP_{control}$) 3:2 configurations from \citet{nesvorny12}. In figure \ref{fig:c18_19}, we plot the 298 simulations (of 1,800 total) from both papers that finish with $P_{S}/P_{J}<$ 2.8 (success criteria \textbf{D}, see section \ref{sect:success}).  While higher values of $M_{55}$ are generated regularly in simulations where Saturn is scattered onto a distant orbit, it is clear from figure \ref{fig:c18_19} that the solar system result lies at the extreme of the distribution of possible outcomes in $M_{55}$-$P_{S}/P_{J}$ space for $P_{S}/P_{J}<$ 2.5.  Indeed, when we analyze the spectrum of values that these instabilities produce for $M_{55}$, $M_{56}$, $M_{65}$ and $M_{66}$ we find that it is extremely difficult to adequately excite $M_{55}$ compared to the other three modes.  In this manner, studies that take the standard success criteria of $P_{S}/P_{J}<$ 2.8 and $M_{55}>$ 0.022 \citep[half the modern value:][see further discussion in section \ref{sect:success}]{nesvorny12} inadvertently bias samples of successful simulations towards lower values of $M_{55}$ and systems with $P_{S}/P_{J}>$ 2.5.  This is not to say that successful outer solar systems cannot be produced from the primordial 3:2 Jupiter-Saturn resonance.  Indeed, a large sample of instability simulations need only generate one successful realization to represent a viable evolutionary pathway for the solar system.  

\begin{figure}
	\centering
	\includegraphics[width=.49\textwidth]{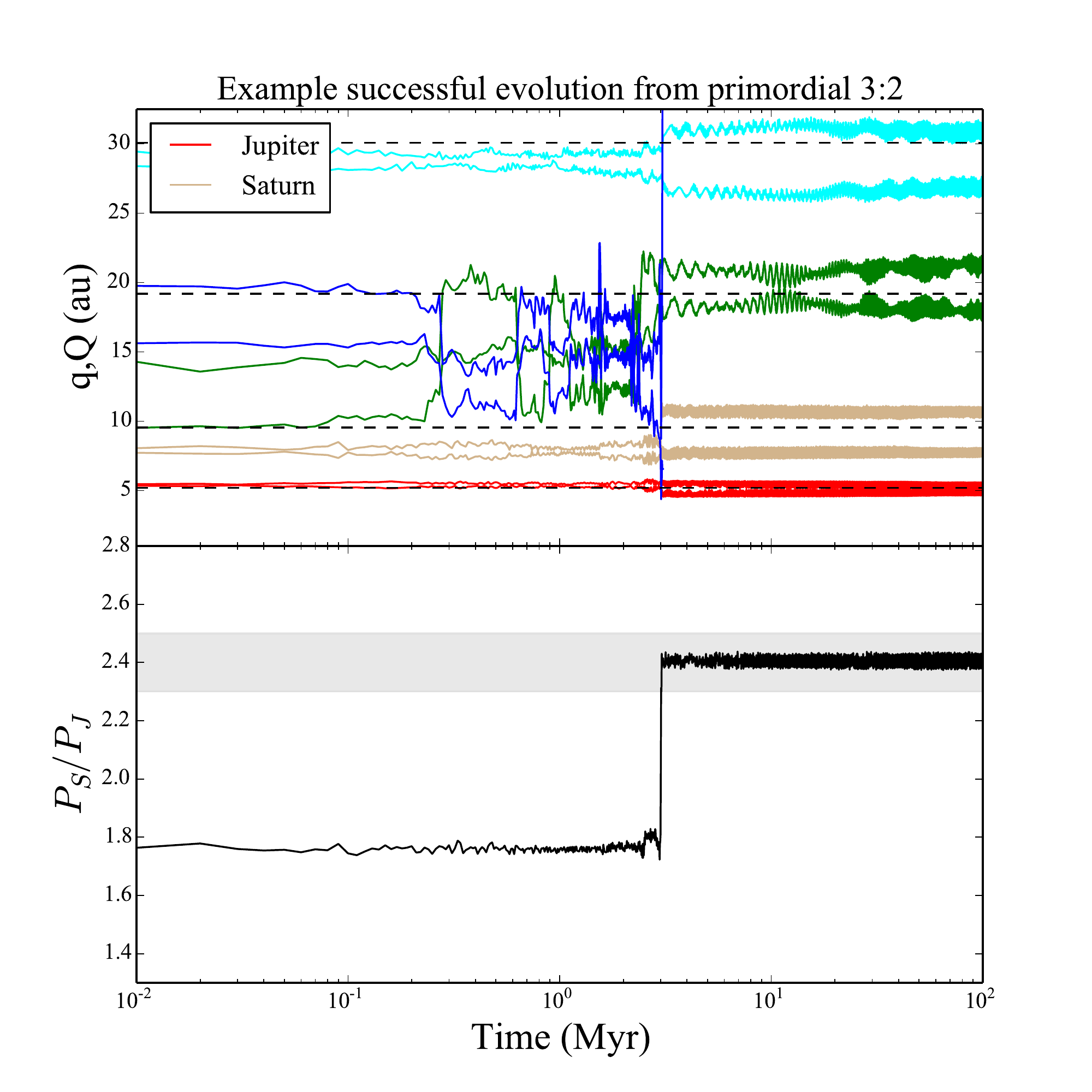}
	\caption{Example instability evolution from \citet{clement18_frag}.  Using the nomenclature of that work, this simulation was from set $5GP/1Myr$ (indicating it took the $5GP_{control}$ initial conditions and delayed the instability 1 Myr after the start of terrestrial planet formation).  The simulation finished with $P_{S}/P_{J}=$2.41 and $M_{55}=$.039.  \textit{It is important to note that this represents a very rare instability outcome of these initial conditions.}  More typical final systems eject too many planets and possess over-excited Saturn analogs on orbits that are too distant.  The top panel plots the perihelion and aphelion of each planet over the length of the simulation.  The bottom panel shows the Jupiter-Saturn period ratio.  The horizontal dashed lines in the upper panel indicate the locations of the giant planets' modern semi-major axes.  The shaded region in the lower panel delimits the range of 2.3 $<P_{S}/P_{J}<$ 2.5.}
	\label{fig:gp_c19}
\end{figure}

At present, initial conditions similar to those of our $5GP_{control}$ set are perhaps the best explanation for the solar system's dynamical state \citep{nesvorny12,batbro12,roig15,brasser_lee_15,vokrouhicky15,roig16,deienno16,deienno17}.  An example of such a successful, ``Jumping Jupiter'' \citep{bras09} type of evolution from \citet{clement18_frag} is plotted in figure \ref{fig:gp_c19}.  In this case, Jupiter's eccentricity is excited by a strong encounter with the ejected ice giant during the instability.  We also direct the reader to the three instabilities scrutinized by \citet{nesvorny13}, the one discussed in \citet{batbro12}, and the one in \citet{deienno18} for other examples of ideal evolutions from the primordial 3:2 Jupiter-Saturn resonance.  However it is important to note that these successful systems lie at the extreme of simulation generated outcomes in $M_{55}$-$P_{S}/P_{J}$ space, and thus at the extreme of the spectrum of systems satisfying the traditional success criteria ($P_{S}/P_{J}<$ 2.8 and $M_{55}>$ 0.022).  Therefore, our study aims to explore whether the solar system result might be brought closer to the heart of the distribution of outcomes by invoking alternative initial conditions for the giant planets' orbits \citep[namely exploring the viability of the 2:1 Jupiter-Saturn resonance and heightened eccentricities:][]{pierens14}.

\subsection{Updated success criteria}
\label{sect:success}

\begin{table*}
	\centering
	\begin{tabular}{c c c c}
	\hline
	Criteria & \citet{nesvorny12} & \citet{deienno17} & This Work \\
	\hline
	\textbf{A} & $N_{GP}=$ 4 & $N_{GP}=$ 4 & $N_{GP}=$ 4 \\
	\textbf{B} & $|\Delta\bar{a}/a_{ss}|<0.2$, $\bar{e}<0.11$, $\bar{i}<2.0^{\circ}$ & $|\Delta\bar{a}/a_{ss}|<0.2$, $\bar{e}<0.11$, $\bar{i}<2.0^{\circ}$ & $|\Delta\bar{a}/a_{ss}|<0.2$\\
	\textbf{C} & $M_{55}>$0.022 & $M_{55}>$0.022 & $|\Delta M_{ij}/M_{ij,ss}|<$0.50 ($i,j=$5,6), $M_{55}>M_{56}$\\
	\textbf{D} & $P_{S}/P_{J}: <$2.1 to 2.3-2.8 in $<$1 Myr & $P_{S}/P_{J}: <$2.1 to 2.3-2.8 in $<$1 Myr  & $P_{S}/P_{J}<$2.5 \\
	\textbf{E} & N/A & 27$<a_{N}<$29 au $@$ $t_{inst}$ & N/A \\
	\end{tabular}
	\caption{Summary of success criteria \textbf{A}-\textbf{D} for our present manuscript (right column), compared with previous work by \citet{nesvorny12} (left column) and \citet{deienno17} (center column).  The Subscripts $GP$, $SS$, $J$, $S$, and $N$ denote the giant planets, solar system modern values, Jupiter, Saturn, and Neptune, respectively.  Note that, unlike the other success criteria, criterion \textbf{B} is only evaluated when \textbf{A} is satisfied.}
	\label{table:success}
\end{table*}

Over the past decade, most numerical studies \citep[e.g.:][]{kaibcham16,deienno17,clement18,deienno18,quarles19} of the giant planet instability have invoked the success criteria established by \citet{nesvorny12} (column one of table \ref{table:success}).  Criterion \textbf{A} requires that a system finish with 4 giant planets, criterion \textbf{B} stipulates that planets' orbits (in terms of $a$, $e$ and $i$) be close to the real ones, and criteria \textbf{C} and \textbf{D} mandate $M_{55}>$ 0.022 and $P_{S}/P_{J}$ evolve from $<$2.1 to between 2.3 and 2.8 in $<$1 Myr, respectively.  Recently, \citet{deienno17} added criterion \textbf{E} for the migration of Neptune, based off the work of \citet{nesvorny15a,nesvorny15b}.  As  our present study is most interested in finding the initial giant planet configurations that best generate the modern Jupiter-Saturn system, we leave the analysis of Neptune's migration history to future work.  Moreover, factors such as Neptune's eccentricity evolution, the instability's strength, and the cold Kuiper Belt's formation location are also important for shaping the trans-Neptunian region \citep{gomes18,volk19}  Thus, we focus on the original four success criteria of \citet{nesvorny12}, with several minor modifications described below:

\subsubsection{The outer solar system's secular architecture}

\begin{figure}
	\centering
	\includegraphics[width=.49\textwidth]{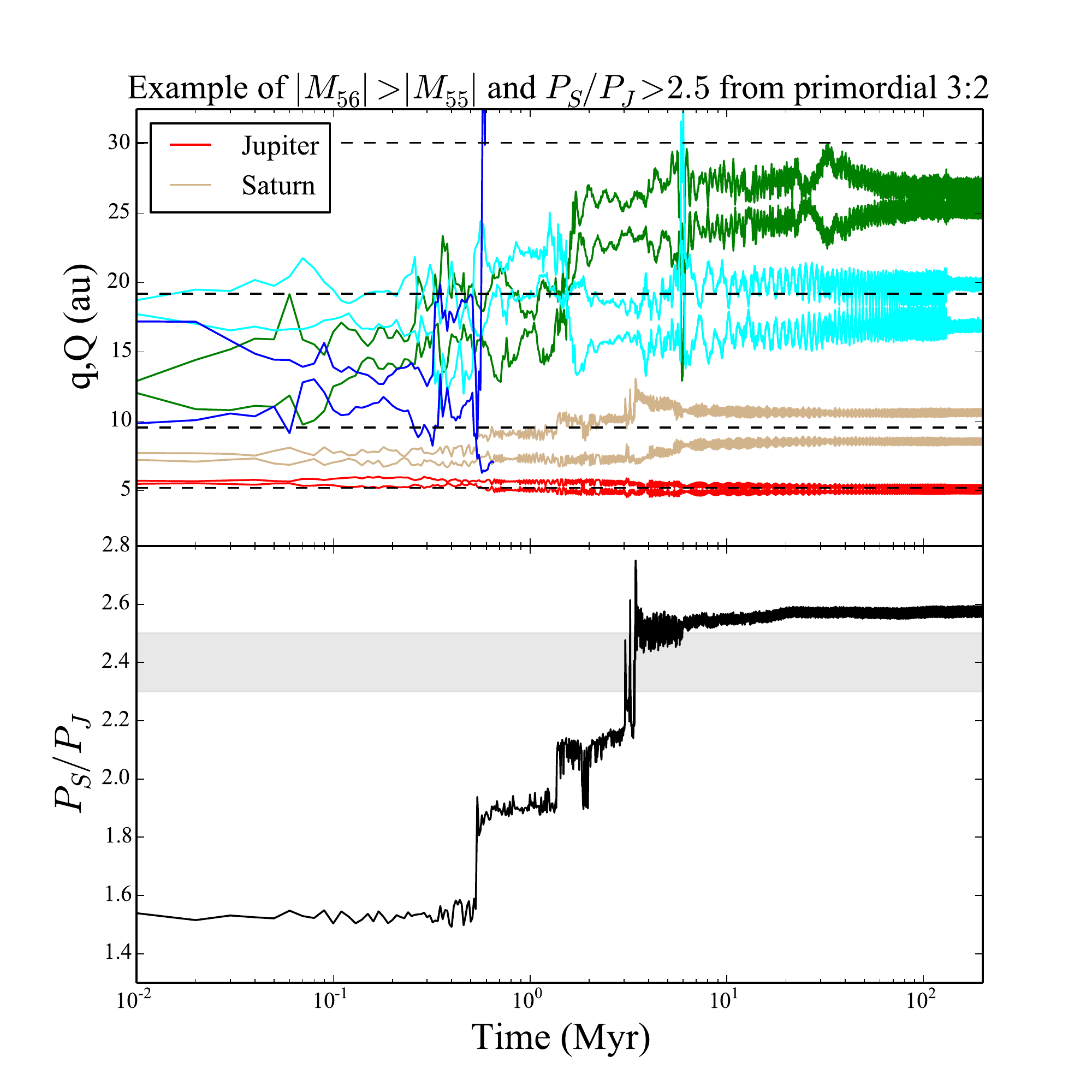}
	\caption{Example instability evolution from \citet{clement18_frag}.  Using the nomenclature of that work, this simulation was from set $5GP/1Myr$ (indicating it took the $5GP_{control}$ initial conditions and delayed the instability 1 Myr after the start of terrestrial planet formation).  The simulation finished with $P_{S}/P_{J}=$ 2.58 and $M_{55}=$.023.  The top panel plots the perihelion and aphelion of each planet over the length of the simulation.  The bottom panel shows the Jupiter-Saturn period ratio.  The horizontal dashed lines in the upper panel indicate the locations of the giant planets' modern semi-major axes.  The shaded region in the lower panel delimits the range of 2.3 $<P_{S}/P_{J}<$ 2.5.}
	\label{fig:gp_c19_bad}
\end{figure}

Figure \ref{fig:c18_19} illustrates the difficulty of matching $P_{S}/P_{J}$ and $M_{55}$ in a single numerical simulation.  Thus, criterion \textbf{C} satisfying simulations tend to have final $M_{55}$ values that are between 0.022 and the modern value of 0.044 (table \ref{table:Mij}).  This, coupled with criterion \textbf{B} allowing final average eccentricities up to 0.11, leads to ``successful'' systems that often inhabit the $M_{56}>M_{55}$ regime.  In this somewhat-typical scenario (40$\%$ of our $5GP_{control}$ and $6GP_{control}$ simulations that satisfy criteria the \textbf{C} and \textbf{D} of \citet{nesvorny12} simultaneously), Saturn's eccentricity is over-excited relative to Jupiter's; often yielding a semi-major axis jump that is too large ($P_{S}/P_{J}>$ 2.5; section \ref{sect:perrat}).  An example of this type of instability from \citet{clement18_frag} for the initial 3:2 Jupiter-Saturn resonance is plotted in figure \ref{fig:gp_c19_bad}.  In this system, Jupiter's final average eccentricity (0.035) is low relative to Saturn's (0.10) as the result of a strong encounter with the ejected ice giant (dark blue line) during the instability.  This powerful dynamical exchange leaves Jupiter's eccentricity with $M_{56}>M_{55}$ ($M_{56}=$ 0.037 and $M_{55}=$ 0.023), and the gas giant's final semi-major axes finish beyond of the 5:2 MMR ($P_{S}/P_{J}=$ 2.58).  However, because Saturn and the Uranus analog each have final eccentricities of $\simeq$0.10, and all four giant planets' final semi-major axes are within 20$\%$ of their modern values, this system satisfies all four success criteria of \citet{nesvorny12}.  It is important to point out that this problem does not imply that the primordial 3:2 Jupiter-Saturn resonance is nonviable.  Indeed, successful systems (e.g.: figure \ref{fig:gp_c19}) are produced from the 3:2 resonance in a small fraction of simulations.  However, the over-excitation of Saturn and under-excitation of Jupiter appear to be systematic traits of successful systems born out of these initial conditions.  Indeed, while $|M_{56}|$ is still less than $|M_{55}|$ in the successful simulation plotted in figure \ref{fig:gp_c19}, Saturn's final average eccentricity is still much higher ($\sim$0.09) than in the actual solar system.  Therefore, our work seeks to investigate whether the solar system is simply a statistical outlier in this manner, or if alternative instability scenarios might better generate the precise Jupiter-Saturn system.

We aim to establish a set of success criteria for our present work that prevent systems similar to the one depicted in figure \ref{fig:gp_c19_bad} from being categorized as ``successful,''  without creating an excessively strict classification scheme that no artificially-generated system can satisfy.  Therefore, we modify criterion \textbf{C} to require that $M_{55}$ be greater than $M_{56}$, and that each of the four magnitudes in the Jupiter-Saturn secular system (equation \ref{eqn:js}) be within 50$\%$ of their modern values.  As Saturn's eccentricity is relatively easy to excite, the additional requirements for $M_{65}$ and $M_{66}$ do not overly constrain our simulations.  Moreover, as Jupiter and Saturn's eccentricities are now fully evaluated by criterion \textbf{C}, and the ice giant's eccentricities and inclinations are somewhat easily damped (e.g.: figure \ref{fig:gp_c19_bad}) during the post-instability residual migration phase \citep{nesvorny12}, we relax criterion \textbf{B} to only scrutinize the giant planet's final semi-major axes.

\subsubsection{The Jupiter-Saturn period ratio}
\label{sect:perrat}

Recent work proposed that Jupiter and Saturn's precise migration towards the 5:2 MMR sculpted the inner asteroid belt's inclination distribution \citep[see full discussion in][]{clement20_mnras}.  Furthermore, the gas giants' ultimate approach to the modern value of $P_{S}/P_{J}=$ 2.49 is fossilized in the asteroid belt's orbital precession distribution as an absence objects with $g_{6}>g>g_{6}-$ 2 $''/yr$.  In fact, the only significant clustering of objects with precession rates in this range in the modern belt are members of the high-inclination collisional family (31) Euphrosyne near $\sim$3.1 au \citep{novakovic11}.  After the Euphrosyne formation event, the family members filled the gap in asteroidal precessions that had been cleared by the gas giants' residual migration via Yarkovsky drift \citep[eg:][]{bottke01}.  Therefore, while setting criterion \textbf{D} to $P_{S}/P_{J}<$ 2.8 is useful for determining which sets of initial conditions systematically yield small jumps, we limit our criterion \textbf{D} satisfying systems to those that finish with the planets interior to the 5:2 MMR.

\subsection{Definition of simulation success}

It is challenging to study a chaotic event like the Nice Model instability because it is impossible to know whether the actual outer solar system is a typical outcome of all possible evolutionary paths that could have been followed from the unknown authentic initial conditions, or an outlier.  Section 4 of \citet{nesvorny12} provides a good synopsis of this somewhat philosophical conundrum.  Following their example, we analyze our simulations' rates of satisfying our success criteria, \textbf{A}-\textbf{D}, both individually and collectively.  This is important because our success criteria jointly analyze nearly a dozen different parameters that may or may not be correlated with one another (table \ref{table:success}).  It seems reasonable to expect that any individual set of successful initial conditions might only meet all four criteria simultaneously a few times, if at all, in $\sim$100-200 realizations.  For example, if the rate of success for each success criteria were 1/3, and the respective rates were not mutually-exclusive, the fraction of systems satisfying all four criteria would only be 1/81 $\simeq$ 1$\%$.  Thus, while we do find many simulations that are quite triumphant in this manner (we scrutinize several in section \ref{sect:results}), additional information can still be gleaned from a careful analysis of the individual success criteria.  

A good example of this concept is examined by \citet{nesvorny12}.  If a batch of simulations originating from the same set of initial conditions finishes with roughly equal numbers of systems possessing 2, 3, 4 and 5 giant planets, it can still be considered successful when scrutinized against criterion \textbf{A} provided that the $N_{GP}=$ 4 systems are not systematically different than the others.  Conversely, if the $N_{GP}=$ 3 systems are the only ones within the batch that satisfy criterion \textbf{B}, \textbf{C} and \textbf{D}, the simulation set would be a failure in terms of criterion \textbf{A}.  In both scenarios, the total success rate for satisfying all 4 criteria simultaneously might be zero.  However, the former case would only possess a null population of totally successful simulations as a result of an over-multiplication of constraints.  Thus, if a large number of systems satisfy 2 or 3 of the criteria without preference to a particular combination, and only narrowly fail the others (for example $P_{S}/P_{J}=$ 2.51 or $M_{56}=$ 0.024), the set of simulations could still represent a viable evolutionary path for the solar system.  Because of this, we focus our analysis on both the success criteria themselves and how \textit{mutually exclusive} they are.

\section{Numerical Simulations}
\label{sect:methods}

\subsection{Generating eccentric resonant chains}

We perform nearly 6,000 instability simulations using the $Mercury6$ Hybrid integration package \citep{chambers99}.  We begin by assembling each giant planet resonant chain using fictitious forces designed to mimic the effects of the gas disk \citep{lee02}.  While this mechanism of generating initial conditions is obviously somewhat contrived, it is employed by numerous authors throughout the literature to consistently and quickly produce stable resonant chains \citep[e.g.:][]{matsu10,beauge12,nesvorny12,deienno17,clement18}.  We first place the giant planets on circular, co-planar orbits outside of the desired resonant chain.  In general, we find that originating Jupiter $\sim$0.5 au beyond its sought after pre-instability semi-major axis \citep[5.6 au, see:][]{deienno17} provides enough migration range to accomplish the required modifications to the planets' orbits.  Each successive planet is placed at a radial distance that is 1-3$\%$ outside of resonance with the interior planet.  Through trial and error, we find that assembling configurations containing six giant planets is simplified by starting the planets further out of resonance (i.e.: closer to 3$\%$).  This ensures that each consecutive planet is able to stabilize in resonance with the interior planet in the chain prior to the exterior one.  The planets are then migrated inward with an external force that modifies the equations of motion with forced migration ($\dot{a}$) and eccentricity damping ($\dot{e}$) terms \citep[see full derivation in][]{lee02}.  To initially place the planets in resonance, we establish a form of $\dot{a}=ka$ and $\dot{e}=ke/100$ \citep{bat10}, where $k$ is set to achieve a migration timescale of $\tau_{mig}\simeq$ 1.0 Myr.

For this phase of computation, we utilize the $Mercury6$ Bulirsch-Stoer integrator using a 6.0 day timestep \citep{chambers99,bs}.  The Bulirish-Stoer algorithm is required because the force on a particle is a function of both the momenta and positions \citep{bat10}.  Once all the planets are in the desired resonance, and $a_{J}<5.7$ au (for 2:1 Jupiter-Saturn configurations, 5.675 au when assembling 3:2 cases), we begin the process of exciting the eccentricities of Jupiter and Saturn.  Though hydrodynamical simulations in \citet{pierens14} only found high eccentricity outcomes for the primordial 2:1 Jupiter-Saturn resonance, for comparison and consistency (see section \ref{sect:3_2_why}) we also test some sets of eccentricity-pumped chains where Jupiter and Saturn inhabit the 3:2 MMR.  Depending on the desired primordial eccentricity, we accomplish this excitation by either reducing the value of $\dot{e}$ on Jupiter or Saturn, or reversing its sign.  We find that both planets' orbits are excited most efficiently in 2:1 MMR gas giant configurations by pumping Saturn's eccentricity and simultaneously reducing the degree of eccentricity damping on Jupiter.  3:2 MMR require the opposite prescription.  Furthermore, exciting Saturn's eccentricity while maintaining $e_{J}$ low (closer to $\simeq$ 0.025) within certain tighter resonant configurations (e.g.: 2:1,4:3,4:3,4:3) involves slightly relaxing the eccentricity damping on the innermost ice giant by around a factor of two.  In all other cases we find maintaining eccentricity damping on the ice giants to be crucial for preserving system integrity during the eccentricity pumping phase.  After Jupiter reaches 5.6 au, we remove all external forces, and integrate each system for an additional 5 Myr to ensure a degree of stability in the absence of artificial forcing.  The time-averaged eccentricities of Jupiter and Saturn from this final phase of integration are taken as $e_{J,o}$ and $e_{S,o}$.  We then verify that the planets are indeed locked in a mutual resonant chain by checking for libration about a series of resonant angles using the method described in \citet{clement18}.  For the precise values used to construct each individual resonant chain, we refer the reader to the online supplementary data files.

Of note, though beyond the scope of this paper, we find that certain configurations of primordial giant planet eccentricities are more difficult to generate than others.  In particular, some architectures are far less sensitive to the specific migration and eccentricity pumping sequence, and require substantially less fine tuning to produce.  In general, we find higher-$e_{J}$/lower-$e_{S}$ configurations to be less sensitive to changes in initial conditions within the 2:1 MMR, while the opposite combination (lower-$e_{J}$/higher-$e_{S}$) is more easily produced with Jupiter and Saturn in a 3:2 MMR.  Indeed Jupiter, being more massive, is more resilient to over-excitation during the eccentricity-pumping process.  As such, when Jupiter and Saturn are placed in the more isolated 2:1 MMR, a wider range of values for $\dot{e}_{J}$ can be used than can for $\dot{e}_{S}$.  For example, when generating 5 planet, 2:1 configurations we find that just a $\sim$1$\%$ change in $\dot{e}_{S}$ can be the difference between not exciting Saturn at all, adequate excitation, and pumping the planet's eccentricity to the point of ejection.  On the other hand, we find that $\dot{e}_{J}$ can be altered by several orders of magnitude and still yield the desired result. In the case of the 3:2 MMR, the strong eccentric forcing of Jupiter on Saturn due to the planets' close proximity makes it easy to over-excite Saturn while pumping Jupiter's eccentricity.

\subsection{Triggering the instability}
\label{sect:trigger}

Our goal is to study the relationship between the primordial eccentricities of the gas giants and their post-instability secular architecture. However, the orbits of our eccentricity-pumped systems of resonant giant planets tend to damp to near-zero eccentricity if they are allowed to interact with the external planetesimal disk for a significant period of time before fully destabilizing \citep[this occurred within about $\sim$10 Myr in our tests.  Note that our model does not include disk self-stirring:][]{levison11}.  In order to prevent this from happening in our simulations, we opt for an artificial instability trigger in order to originate each batch of instabilities from nearly the same values of $e_{J,o}$ and $e_{S,o}$.  As a consequence of this selection, we are unable to constrain our final systems by Neptune's pre-instability migration history \citep{nesvorny15a,nesvorny15b,deienno17}.  If Neptune's primordial migration was indeed significant, our instability simulations will yield a systematic under-estimate of the ice giants' final semi major axes.  However, since we are most interested in finding sets of initial conditions that best construct the modern Jupiter-Saturn system (section \ref{sect:success}), we relegate a detailed study of the ice giants' migration to future work.  Additionally, the growing body of work arguing for an early instability \citep[appendix \ref{sect:timing}:][]{morb18,clement18,deienno18,nesvorny18} is potentially consistent with our simulated scenario.

We utilize the same instability trigger as \citet{nesvorny11} and \citet{nesvorny12}.  Once each system of giant planets is placed in resonance, we add an external disk of 1,000 equal-mass planetesimals with a total mass of 20.0 $M_{\oplus}$ \citep[loosely based off][]{nesvorny12}\footnote{Note that our fiducial disk mass choice of 20.0 $M_{\oplus}$ might drive excessive residual migration in some of our 2:1 chains, thus causing Saturn to migrate past the 5:2 MMR after the instability \citep[e.g.:][]{nesvorny12,brasser_lee_15}.  For this reason, we experiment with different total disk masses in section \ref{sect:kb_mass}.}  It should be noted that these disk particles are unrealistically massive \citep{levison11,quarles19}, and thus our selection of initial masses is a compromise necessary to limit the computational cost of our study.  In future work, we intend to investigate the evolution of these systems using more realistic primordial planetesimal disks composed of $\gtrsim$10,000 objects.  In all cases, the inner edge of the planetesimal disk is set to 1.5 au beyond the final ice giant.  Because we artificially trigger instabilities, the precise radial offset between Neptune and the Kuiper belt is of less consequence to our work as we are more interested in its post-instability damping effects.  Planetesimal semi-major axes are selected to achieve a surface density profile that falls off as $r^{-1}$ \citep{williams11}.  Eccentricities and inclinations are chosen from Rayleigh distributions \citep[$\sigma_{e}=$ 0.01, $\sigma_{i}=$ 1$^{\circ}$; note that these might be unrealistically small if, for example, they were excited earlier by the ice giant's accretion:][]{ribeiro20}.  The entire system is then integrated for 100.0 Kyr with the $Mercury6$ hybrid integrator and a 50.0 day time-step \citep{chambers99}.  If the instability has not occurred after 100.0 Kyr, the mean anomaly of the innermost ice giant is shifted by $\sim$90$^{\circ}$ to trigger the instability \citep{nesvorny11}.  If an instability still does not ensue after 100 Myr, the simulation is discarded \citep[e.g.:][]{nesvorny18}.  Each system is evolved for an additional 20 Myr after the onset of the instability (as determined by either an ice giant ejection or a step change in the Jupiter-Saturn period ratio) in order to capture some of the remnant planetesimal disk's interactions with the post-instability outer solar system.  

It is important to recognize that our instability trigger is somewhat ad hoc, particularly in light of the fact that the pre-instability migration of the ice giants has been shown to be important for explaining the capture of D-type trojan asteroids by Jupiter \citep{nesvorny13} and the Kuiper Belt's modern orbital structure \citep{nesvorny15a,nesvorny15b}.  Moreover, Uranus and Neptune's survival probabilities (and the corresponding success rates for criterion \textbf{A} and \textbf{B}) are boosted when the planets undergo significant pre-instability migration.  It should also be noted that 20 Myr is likely insufficient to fully capture the giant planets' residual migration phase \citep[][for example, integrate for 100 Myr after the instability]{nesvorny12}.  Because we are most interested in studying the Jupiter-Saturn secular system that must be largely assembled through planetary encounters within the instability (discussed in section \ref{sect:e55_prob} and appendix \ref{sect:alternatives}), we make this choice of total integration time to limit the computational cost of our study.  However, to quantify the magnitude of our integration time's effect on our results, we extend one batch of 200 instability simulations (our 2:1,4:3,3:2,3:2; $e_{J,o}=e_{S,o}=$ 0.05 set: table \ref{table:ics}) to a total integration time of 100 Myr.  We find that the average change in $P_{S}/P_{J}$ and $M_{55}$ over this additional 80 Myr of integration is $\sim$0.06 and $\sim$0.0005, respectively, for systems with $P_{S}/P_{J}<$ 2.5 and $M_{55}>$ 0.022 at 20 Myr.  In contrast, the average additional migration for Neptune analogs in 4 planet systems is more significant ($\sim$1.5 au).  Thus, we find 20 Myr to be a reasonable total integration time for evaluating different sets of initial conditions' success at generating the modern Jupiter-Saturn system.  However, a caveat of our study is that our simulations are inadequate to fully resolve the ice giants' (particularly Neptune's) residual migration, and consequently that of Jupiter and Saturn \citep[necessary for their approach to the 5:2 MMR and the additional depletion of the objects above the $\nu_{6}$ resonance in the asteroid belt;][]{clement20_mnras}. We further discuss how this affects our results, in terms of our criteria for simulation success, in section \ref{sect:success}.

Finally, we remove all Kuiper belt objects from each fully evolved system and integrate only the remaining giant planets for an additional 100 Myr to ensure that the final system is stable.  We utilize the outputs of this final simulation phase to calculate the secular frequencies and amplitudes via frequency-modulated Fourier Transform \citep[][note that this step is only completed for systems that retain both Jupiter and Saturn]{nesvorny96}.  Thus, the results presented in section \ref{sect:results} are exclusively from systems that destabilize within 100 Myr, and do not eject Saturn.  For most of the sets of initial conditions we test, $\gtrsim$85$\%$ of configurations destabilize within 100 Myr, and $\lesssim$1$\%$ eject Saturn.

\subsection{Parameter space tested}

Our various batches of numerical simulations are summarized in table \ref{table:ics}.  As discussed by previous authors \citep[e.g.:][]{bat10,nesvorny11,nesvorny12,deienno17,clement18}, the parameter space of possible resonant chains, planetesimal disk properties, initial number of giant planets and primordial eccentricities is extensive.  Fortuitously, much of this phase space for low primordial eccentricities \citep{masset01} has already been probed by the comprehensive study of \citet{nesvorny12}.  This permits us to simplify our present manuscript by not repeating an analysis of previously investigated parameter space.  Instead, we perform three sets of control simulations utilizing the most successful configurations from \citet{nesvorny12}, in tandem with the identical disk conditions and calculation mechanisms described in section \ref{sect:methods}.  Thus, we are able to present a self-consistent comparison of our present results with those of \citet{nesvorny12}, and the $5GP_{control}/6GP_{control}$ sets from \citet{clement18,clement18_frag} discussed in section \ref{sect:e55_prob}.

The parameter space we seek to study is as follows: (1) a range of moderate primordial eccentricities for the gas giants ($\sim$0.025-0.05; we find higher eccentricities typically lead to overly-violent instabilities and poor solar system analogs), (2) the primordial 2:1 and 3:2 Jupiter-Saturn resonances, (3) initial configurations with 4, 5 and 6 giant planets, and (4), the various primordial ice giant resonances (e.g.: 2:1, 3:2, 4:3, etc.).  Though a comprehensive study of the various permutations of these variables would be very computationally expensive, we note several large swaths of this parameter space that systematically lead to poor solar system analogs.  

Resonant chains where this is the case are summarized below:

\subsubsection{3:2 with eccentricity pumping}
\label{sect:3_2_why}

We find that, in general, exciting the eccentricities of Jupiter and Saturn with the planets in a primordial 3:2 MMR (note that this case is not necessarily physically motivated) leads to instability evolutions with excessive jumps, and exceedingly high final values of $e_{S}$.  While primordial eccentricity pumping within the 3:2 does boost success rates for criterion \textbf{C} (i.e.: the instability more efficiently excites $M_{55}$, see section \ref{sect:results}), the corresponding success rates for the other 3 criteria are systematically lower than in the zero-eccentricity control case.  This is not surprising, as low-eccentricity instabilities beginning from the 3:2 are intrinsically more violent than those from the 2:1 (section \ref{sect:3_2}).  In our scenario, pumping Jupiter's primordial eccentricity only leads to systematically stronger encounters during the chaos of the instability.  In many cases, this leads to a strong scattering event between Jupiter and Saturn that essentially destroys the entire outer solar system.  We were also unable to develop a procedure for generating resonant chains with $e_{J,o}=e_{S,o}$ within the 3:2 MMR's tighter orbital spacing.  Indeed, the pre-instability eccentricity of Saturn for $e_{J,o} \approx$ 0.0 is $\sim$0.025 in classic simulations of the 3:2 version of the Nice Model \citep{nesvorny11,nesvorny12,batbro12}.  Thus, exciting $e_{J,o}$ to $\sim$0.05 or greater using our methodology (section \ref{sect:methods}) pumps Saturn's eccentricity to between 0.15-0.25.  We find that, when such a resonant chain is evolved through the instability, the percentage of system's that simultaneously satisfy criterion \textbf{C} and \textbf{D} (regardless of final $N_{GP}$) is near zero (thus, the distribution in figure \ref{fig:c18_19} moves to the right as a greater fraction of simulations experience excessively large jumps).  Therefore, we restrict our current study to three control cases \citep[table \ref{table:ics}, two of the most successful five planet configurations, and one six planet chain from][]{nesvorny12}, and two additional, eccentricity-pumped chains with $e_{J,o}\simeq$ 0.025 and $e_{S,o}\simeq$ 0.05.

\subsubsection{2:1 and four planets}

Broadly speaking, increasing the gas giants' primordial eccentricities leads to more violent evolutions for systems with Jupiter and Saturn beginning in a 2:1 MMR.  Thus, these systems now behave similarly to the circular, 3:2 cases (section \ref{sect:3_2}) in that they routinely eject one or more ice giants.  To illustrate this, we present one set of instabilities where four giant planets inhabit a 2:1-4:3-4:3 resonance chain ($e_{J,o} \approx e_{S,o} \approx 0.025$). Because we find that 100$\%$ of the systems in this batch that experience an orbital instability eject at least one ice giant, we do not study eccentricity pumped four planet cases further.

\subsubsection{Chains beginning with 2:1,3:2}
\label{sect:2_1_3_2}

On the opposite end of the spectrum, as we discuss in section \ref{sect:2_1}, we were unable to consistently generate orbital instabilities in resonant chains where Jupiter and Saturn inhabit a 2:1 MMR, and Saturn and the first ice giant are in a 3:2.  Even when the inner ice giant's mean anomaly is shifted, the planets in these widely-spaced configurations typically continue to migrate smoothly, and often fall back into resonance, as they interact with the external planetesimal disk.  Therefore, all the resonant chains we study that investigate the 2:1 Jupiter-Saturn resonance (table \ref{table:ics}) place the innermost ice giant in a 4:3 MMR with Saturn.

In summary, the majority of our simulations concentrate on the primordial 2:1 Jupiter-Saturn resonance with moderate eccentricities ($e_{J,o} \approx e_{S,o} \approx$ 0.025-0.05).  In our initial suite of simulations, the additional ice giants are assigned masses of $M_{IG}=$ 16.0 $M_{\oplus}$ for five planet configurations and $M_{IG}=$ 8.0 $M_{\oplus}$ for six planet cases.  This choice is based off the study of \citet{nesvorny12}, who find six planet instabilities with more massive additional planets to be too violent.  We also investigate the effect of varying $M_{IG}$ with additional simulations in section \ref{sect:ig_mass}.  As discussed in section \ref{sect:success}, a major limitation of our work is that it lacks a detailed analysis of the potential ice giant evolutionary paths.  Specifically, we do not test tighter resonances (e.g.: 5:4, 6:5, etc.) or initial conditions derived from models of ice giant formation \citep{ribeiro20} and obliquity evolution \citep{izidoro15_ig}.  Moreover, it is noticeably difficult to ascertain whether our initial conditions are realistic and physically motivated in the absence of robust hydrodynamical studies that follow the growth and evolution of the entire outer solar system in the gas disk phase.  For instance, it is unclear whether it is possible for the ice giants to migrate past a mutual 3:2 MMR and become trapped in any of the tighter first order resonance (specifically the 4:3) often tested in N-body studies \citep[e.g.:][]{bat10,batygin11,nesvorny11,batbro12,nesvorny12,deienno17,gomes18,quarles19}.  Indeed, the population of detected exoplanets locked in the 4:3 resonance has been argued to be in excess of theoretical predictions \citep{rein12,matsumura17,brasser18}.  To guarantee the capture of two planets in a particular resonance via convergent migration in the circularly restricted three body approximation, the smaller object's eccentricity must be less than a critical value \citep[see derivations in ][]{henrad83,petrovich13}.  For the innermost ice giants' capture with Saturn, this maximum eccentricity is $\sim$0.083 for the 3:2 MMR.  For the outer ice giants' resonant entrapment with one another, the smaller body's eccentricity must be less than 0.045 to guarantee the planets become locked in the 3:2.  In our simulations, the inner 1-2 ice giants' typically possess initial eccentricities in close to $\sim$0.10 (table \ref{table:ics}), and the outermost planets attain eccentricities as high as 0.05.  Thus, it might be possible for these planets to have migrated past their respective 3:2 MMRs without becoming locked in a such a looser chain.  However, this analysis does not fully account for many important phenomena; including gas dynamics, perturbations from additional planets, gas accretion and formation location.  Thus, while our tested resonant chains (table \ref{table:ics}) are rather fictitious and contrived in an effort to probe as much parameter space as possible, future work must validate their feasibility through comprehensive hydrodynamical simulations.  Indeed, simulations in \citet{izidoro15_ig} designed to replicate the modern obliquities of Uranus and Neptune via embryo impacts yielded cases where the respective ice giants were captured in the 4:3, 5:4, and even 6:5 MMRs.

\begin{table*}
	\centering
	\begin{tabular}{c c c c c c c}
	\hline
	$N_{pln}$ & Resonant Chain & $e_{J,o}$ & $e_{S,o}$ & $e_{IG,o}$ & $a_{N,o}$ & $N_{sim}$  \\
	\hline
	4 & 2:1,4:3,4:3 & 0.025 & 0.025 & 0.045, 0.012 & 13.2 & 187 \\
	\hline
	5 & 3:2,3:2,2:1,3:2* & 0.0 & 0.0 & 0.025, 0.007, 0.008 & 19.7 & 166 \\ 
	& 3:2,3:2,3:2,3:2* & 0.0 & 0.0 & 0.049, 0.023, 0.019 & 16.9 & 166 \\
	& & 0.025 & 0.05 & 0.115, 0.076, 0.051 & 16.9 & 174\\
	& 2:1,4:3,3:2,3:2* & 0.0 & 0.0 & 0.020, 0.007, 0.0 & 18.6 & 182 \\
	& & 0.025 & 0.025 & 0.067, 0.038, 0.034 & 18.6 & 184 \\
	& & 0.025 & 0.05&  0.078, 0.075, 0.061 & 18.6 & 177 \\
	& & 0.05 & 0.025 & 0.068, 0.027, 0.020 & 18.6 & 185 \\	
	& & 0.05 & 0.05 & 0.075, 0.031,0.022 & 18.6 & 177 \\
	& 2:1,4:3,4:3,3:2 & 0.025 & 0.025 & 0.060, 0.033, 0.020 & 17.2 & 183 \\
	& & 0.025 & 0.05 & 0.154, 0.114, 0.052 & 17.2 & 185 \\
	& & 0.05 & 0.025 & 0.076, 0.034, 0.032 & 17.2 & 171 \\	
	& & 0.05 & 0.05 & 0.073, 0.024, 0.013 & 17.2 & 191 \\
	& 2:1,4:3,4:3,4:3 & 0.025 & 0.025 & 0.058, 0.030, 0.025 & 15.9 & 190 \\
	& & 0.05 & 0.025 & 0.063, 0.020, 0.015 & 15.9 & 183 \\	
	& & 0.05 & 0.05 & 0.067, 0.027, 0.015 & 15.9 & 188 \\
	\hline
	6 & 3:2,4:3,3:2,3:2,3:2* & 0.0 & 0.0 & 0.030, 0.050, 0.036, 0.029 & 20.9 & 173 \\
	& & 0.025 & 0.05 &  0.097, 0.075, 0.022, 0.019 & 20.9 & 184 \\
	& 2:1,4:3,3:2,3:2,3:2 & 0.025 & 0.025 & 0.053, 0.042, 0.033, 0.026 & 24.4 & 182 \\
	& & 0.05 & 0.025 & 0.068, 0.023, 0.016, 0.012 & 24.4 & 184 \\	
	& & 0.05 & 0.05 & 0.088, 0.043, 0.028, 0.042 & 24.4 & 186 \\
	& 2:1,4:3,4:3,3:2,3:2 & 0.025 & 0.025 & 0.066, 0.041, 0.016, 0.010 & 22.5 & 152 \\
	& & 0.025 & 0.05 & 0.085, 0.040, 0.028, 0.025 & 22.5 & 183 \\
	& & 0.05 & 0.025 & 0.080, 0.022, 0.020, 0.019 & 22.5 & 185 \\	
	& & 0.05 & 0.05 & 0.075, 0.026, 0.017, 0.017 & 22.5 & 183 \\
	& 2:1,4:3,4:3,4:3,3:2 & 0.025 & 0.025 & 0.075, 0.048, 0.027, 0.022 & 20.2 & 168 \\
	& & 0.025 & 0.05 & 0.109, 0.067, 0.022, 0.019 & 20.2 & 184 \\
	& & 0.05 & 0.025 & 0.060, 0.026, 0.018, 0.013 & 20.2 & 184 \\	
	& & 0.05 & 0.05 & 0.079, 0.038, 0.023, 0.015 & 20.9 & 171 \\	
	\hline
	\end{tabular}
	\caption{Summary of resonant chains tested.  The columns are as follows: (1) the initial number of giant planets, (2) the resonant chain, beginning with the Jupiter-Saturn resonance, (3-4) the initial eccentricities of Jupiter and Saturn, (5) the initial eccentricities of the ice giants in order of increasing semi-major axis, (6) the initial semi-major axis of the outermost ice giant, and (7) the total number of simulations that produce an instability within 100 Myr and retain both Jupiter and Saturn.  In all cases, the planetesimal disk mass is set to 20.0 $M_{\oplus}$, the radial offset between Neptune and the disk is to 1.5 au, and Jupiter's initial semi-major axis is 5.6 au.  All four and five giant planet configurations have ice giant masses of 16.0 $M_{\oplus}$.  Ice giants in simulations beginning with six giant planets have masses of 8.0, 8.0, 16.0 and 16.0 $M_{\oplus}$, with increasing semi-major axis.  Note that, for control configurations \citep[based off the most successful five and six planet systems from][and denoted here by the ``*'' symbol]{nesvorny12} with $e_{J,o}$ and $e_{S,o}$ labeled as 0.0, the actual eccentricities are non-zero, but small.}
	\label{table:ics}
\end{table*}

\section{Results}
\label{sect:results}

Table \ref{table:results} gives the percentage of systems in each of our various simulation batches (table \ref{table:ics}) that satisfy our success criteria, \textbf{A}-\textbf{D} (table \ref{table:success}).  Our analysis is structured as follows: the subsequent four sections (\ref{sect:con_results}-\ref{sect:6gp_tight_res}) focus on the various resonant chains we test, section \ref{sect:ecc_results} discusses the dependency of our results on $e_{J,o}$ and $e_{S,o}$, and sections \ref{sect:kb_mass}-\ref{sect:ig_mass} present an additional suite of simulations (based off our most successful sets of initial conditions) where we vary the initial masses of the innermost ice giants and the planetesimal disk.  Because the goal of our work is to find sets of initial conditions where the solar system is not an outlier in $M_{55}$-$P_{S}/P_{J}$ space, we provide a plot of this parameter space for each of our tested resonant chains in figures \ref{fig:control_e55}, \ref{fig:control_e55_2_1}, \ref{fig:5_1ig_e55}, \ref{fig:5_2ig_e55}, \ref{fig:5_3ig_e55}, \ref{fig:6_1ig_e55}, \ref{fig:6_2ig_e55}, and \ref{fig:6_3ig_e55}.

\begin{table*}
	\centering
	\begin{tabular}{c c c c c c c c c}
	\hline
	$N_{pln}$ & Resonant Chain & $e_{J,o}$ & $e_{S,o}$ & \textbf{A} & \textbf{B} & \textbf{C} & \textbf{D} & \textbf{ALL} \\
	\hline
	4 & 2:1,4:3,4:3 & 0.025 & 0.025 & 0 & 0 & 9 & 18 & 0 \\
	\hline
	5 & 3:2,3:2,2:1,3:2* & 0.0 & 0.0 & 17 & 13 & 5 & 17 & 1 \\ 
	& 3:2,3:2,3:2,3:2* & 0.0 & 0.0 & 14 & 5 & 8 & 17 & 0 \\
	& & 0.025 & 0.05 & 13 & 1 & 5 & 14 & 1 \\
	& 2:1,4:3,3:2,3:2 & 0.0 & 0.0 & 32 & 21 & 6 & 25 & 0 \\
	& & 0.025 & 0.025 & 30 & 21 & 9 & 16 & 1 \\
	& & 0.025 & 0.05 & 30 & 19 & 6 & 11 & 0 \\
	& & 0.05 & 0.025 & 30 & 25 & 9 & 18 & 1 \\	
	& & 0.05 & 0.05 & 31 & 20 & 9 & 17 & 2 \\
	& 2:1,4:3,4:3,3:2 & 0.025 & 0.025 & 17 & 6 & 7 & 14 & 0 \\
	& & 0.025 & 0.05 & 19 & 12 & 6 & 10 & 0 \\
	& & 0.05 & 0.025 & 14 & 4 & 11 & 11 & 1\\	
	& & 0.05 & 0.05 & 17 & 5 & 12 & 16 & 1 \\
	& 2:1,4:3,4:3,4:3 & 0.025 & 0.025 & 17 & 3 & 9 & 12 & 1 \\
	& & 0.05 & 0.025 & 7 & 2 & 9 & 8 & 0 \\	
	& & 0.05 & 0.05 & 9 & 1 & 9 & 10 & 1 \\
	\hline
	6 & 3:2,4:3,3:2,3:2,3:2* & 0.0 & 0.0 & 17 & 9 & 5 & 25 & 1 \\
	& & 0.025 & 0.05 & 7 & 5 & 5 & 5 & 1 \\
	& 2:1,4:3,3:2,3:2,3:2 & 0.025 & 0.025 & 56 & 7 & 14 & 38 & 0 \\
	& & 0.05 & 0.025 & 55 & 14 & 8 & 36 & 1 \\	
	& & 0.05 & 0.05 & 48 & 15 & 8 & 43 & 1 \\
	& 2:1,4:3,4:3,3:2,3:2 & 0.025 & 0.025 & 55 & 19 & 9 & 22 & 0 \\
	& & 0.025 & 0.05 & 51 & 25 & 8 & 31 & 1 \\
	& & 0.05 & 0.025 & 60 & 28 & 6 & 37 & 1 \\	
	& & 0.05 & 0.05 & 54 & 26 & 14 & 31 & 1 \\
	& 2:1,4:3,4:3,4:3,3:2 & 0.025 & 0.025 & 53 & 26 & 12 & 21 & 0 \\
	& & 0.025 & 0.05 & 45 & 25 & 11 & 17 & 1 \\
	& & 0.05 & 0.025 & 55 & 24 & 13 & 24 & 0 \\	
	& & 0.05 & 0.05 & 53 & 30 & 17 & 22 & 3 \\
	\hline
	\end{tabular}
	\caption{Summary of results for our different simulations (table \ref{table:ics}) measured against our various success criteria (table \ref{table:success}).  The columns are as follows: (1) the initial number of giant planets, (2) the resonant chain, beginning with the Jupiter-Saturn resonance, (3-4) the initial eccentricities of Jupiter and Saturn, (5) the percentage of systems satisfying criterion \textbf{A} ($N_{GP}=$4), (6) criterion \textbf{B} (the planets' final semi-major axes within 20$\%$ of the real ones), (7) criterion \textbf{C} ($|\Delta M_{ij}/M_{ij,ss}|<$0.50 ($i,j=$5,6), $M_{55}>M_{56}$), (8) criterion \textbf{D} ($P_{S}/P_{J}<$2.5), and (9) the percentage of systems satisfying all four success criteria simultaneously.}
	\label{table:results}
\end{table*}

\subsection{Control runs: the 3:2 Jupiter-Saturn resonance and the circular 2:1 case}
\label{sect:con_results}

\subsubsection{3:2; circular orbits}

\begin{figure*}
	\centering
	\includegraphics[width=.7\textwidth]{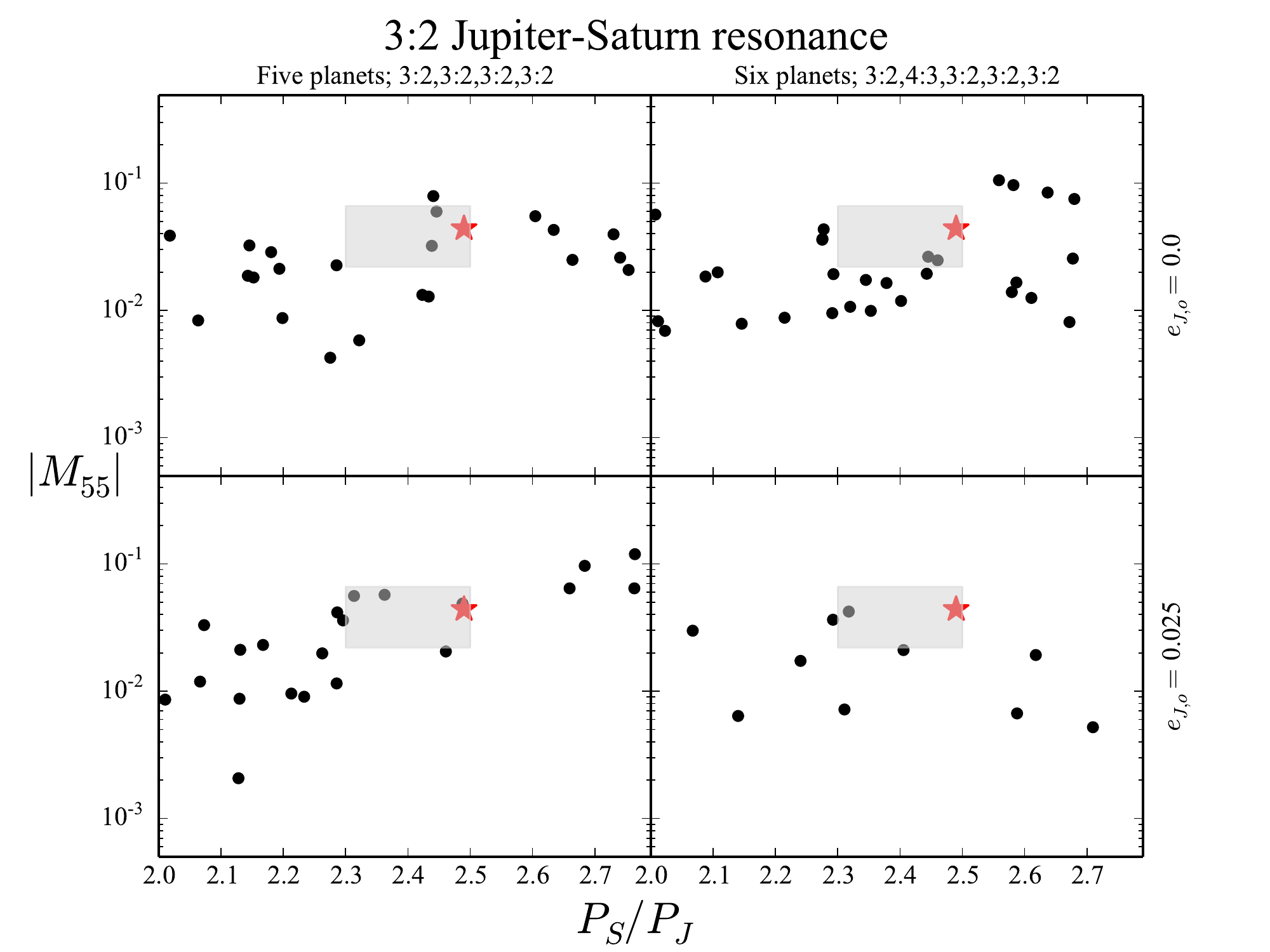}
	\caption{Similar to figure \ref{fig:c18_19}.  $M_{55}$-$P_{S}/P_{J}$ space for our simulations beginning from the 3:2 Jupiter Saturn resonance (table \ref{table:ics}).  The left two plots depict five giant planet chains (3:2,3:2,3:2,3:2) and the right plots show runs with six planets (3:2,4:3,3:2,3:2,3:2).  The plots on top are control runs without primordial eccentricity pumping.  The bottom two panels have minor eccentricity pumping ($e_{J}=$ 0.025, $e_{S}=$ 0.05).  The red stars indicate modern solar system values.  The shaded grey area delimits the region of 2.3 $<P_{S}/P_{J}<$ 2.5 (criterion \textbf{D}) and 0.022 $<M_{55}<$ 0.066 (criterion \textbf{C}).  While one simulation provides in the five planet, $e_{J}=$ 0.025, $e_{S}=$ 0.05 batch (bottom right panel) provides an excellent match to the Jupiter-Saturn system, this is not a typical outcome of these initial conditions.  Indeed, 86$\%$ of these simulations finish with $P_{S}/P_{J}>$2.5 (the majority of which are beyond the scale of this figure).}	
	\label{fig:control_e55}
\end{figure*}

\citet{nesvorny12} favored several different aspects of both the five planet, 3:2,3:2,3:2,3:2 and 3:2,3:2,2:1,3:2 chains' evolution.  For consistency, we present a set of simulations for each chain.  In general, the broader spacing of the 3:2,3:2,2:1,3:2 leads to higher success rates for criterion  \textbf{B}, and the two sets of initial conditions perform similarly when scrutinized against our other constraints.  However, we find that the 3:2,3:2,3:2,3:2 case yields the best results in terms of final distributions in  $M_{55}$-$P_{S}/P_{J}$ space (figure \ref{fig:control_e55}).  As our work seeks to find sets of initial configurations that best yield realistic magnitudes of $M_{55}$ and low jumps, we focus our analysis on this chain, and the six planet, 3:2,4:3,3:2,3:2,3:2 configuration.  

We find that the rates of success for our control simulations (no eccentricity pumping) are  similar to those reported by \citet{nesvorny12} and \citet{clement18} (our $5GP_{control}$ and $6GP_{control}$ sets presented in section \ref{sect:e55_prob}).  Compared to these previous studies, our control simulations possess slightly lower success rates for criterion \textbf{B}.  This is a consequence of the ice giants' residual migration being subdued in our simulations due to shorter integration times and a lower planetesimal disk mass for the five planet case (20 $M_{\oplus}$ as opposed to $\sim$35; we revisit the role of the disk's mass in section \ref{sect:kb_mass}).

All of our simulations beginning from the 3:2 Jupiter-Saturn resonance systematically struggle to satisfy criterion \textbf{C}.  As discussed in section \ref{sect:success}, establishing criterion \textbf{C} such that any simulation with $M_{55}>$ 0.022 is categorized as successful leads to many simulations with over-excited Saturn analogs, or those inhabiting the $|M_{56}|>|M_{55}|$ regime, satisfying the constraint.  Indeed, 75$\%$ of our five planet, 3:2,3:2,3:2,3:2 chains without primordial eccentricity pumping (79$\%$ when we include primordial excitation) excite $M_{55}$ to greater than 0.022 \citep[note that the majority of these are in system's that fail criterion \textbf{A},  the corresponding rate reported by][is lower because the authors only claim success for criterion \textbf{C} if \textbf{A} and \textbf{B} are met as well]{nesvorny12}.   However, only 8$\%$ of the runs in our same control five planet simulation batch satisfy our updated version of criterion \textbf{C} that scrutinizes all four eccentric amplitudes of the Jupiter-Saturn system.  In a similar manner, our simulations originating from six planet, zero-eccentricity, 3:2,4:3,3:2,3:2,3:2 chains finish with $M_{55}>$ 0.022 65$\%$ of the time (72$\%$ when we include mild primordial eccentricity pumping), but only satisfy our new criterion \textbf{C} at a rate of 5$\%$.  The most difficult amplitude for our 3:2 Jupiter-Saturn resonance simulations to match is $M_{56}$, with only 12$\%$ of systems finishing within the appropriate range.  Conversely, the number of simulations possessing proper values of $M_{55}$, $M_{65}$ and $M_{66}$ are 32$\%$ (note that this is not $\sim$70$\%$ as discussed above by virtue of our new constraint imposing a maximum limit on $M_{55}$ as well as a minimum), 54$\%$ and 27$\%$, respectively.  Furthermore, 67$\%$ of the systems that finish with adequate values of $M_{56}$ are in the $|M_{56}|>|M_{55}|$ regime.  Thus, while it is important to avoid over-constraining instability simulations when studying the Nice Model statistically, our results are indicative of a strong anti-correlation between the adequate excitation of $M_{55}$ and the broad replication of the complete Jupiter-Saturn system for the primordial 3:2 resonance.  While high $M_{55}$ magnitudes are common outcomes in our simulation sets studying these initial conditions, they occur preferentially in systems that experience large jumps (i.e.: those that fail criterion \textbf{D}) and possess overexcited values of $M_{56}$.

\subsubsection{3:2; primordially excited eccentricities}

Our simulation sets investigating the effects of mild primordial eccentricity pumping in the 3:2 Jupiter-Saturn resonance yield success rates for criteria \textbf{A}-\textbf{D} that are consistently similar or worse to those generated in the low-$e$ case.  In general, exciting the eccentricities of Jupiter and Saturn in the 3:2 MMR scenario tends to inject more violence into an already brutal event.  In our batch studying five planet configurations, this manifests as low success rates for criteria \textbf{A} and \textbf{B}.  Indeed, only 1$\%$ of these simulations satisfy both constraints simultaneously.  As discussed in section \ref{sect:methods}, stable 3:2 resonant chains with heightened values of $e_{J,o}$ and $e_{S,o}$ are more challenging to generate because, in the tighter configuration, Jupiter's dynamic excitation readily bleeds to the other giant planets via stochastic diffusion.  In spite of the fact that we maintain artificial damping forces on the ice giants throughout the process of generating these resonant chains, the inner two ice giants typically possess eccentricities around $\sim$0.10 once the chain is fully assembled (table \ref{table:ics}).  As a consequence of this resulting configuration, each consecutive pair of planets following Jupiter is already on nearly crossing orbits when the instability ensues.  When this is the case, the ice giants are more likely to experience stronger encounters and scattering events between one another during the chaos of the instability.  Thus, the subset of systems that do finish with $N_{GP}=$ 4 contain Uranus and Neptune analogs with orbits that are systematically more distant and excited than the real ones.  This effect is less severe in our six planet configurations because the inner two ice giants are less massive (8.0 $M_{\oplus}$ as opposed to 16.0 $M_{\oplus}$).  In this scenario, the more massive outer ice giants that typically go on to become the Uranus and Neptune analogs in criterion \textbf{A} satisfying systems are more dynamically detached from the excited gas giants, and possess lower eccentricities (e$\sim$0.05) at the beginning of our simulations.  As a result, when the inner two ice giants scatter off of Jupiter and Saturn, they tend to undergo weaker and fewer additional encounters with the Uranus and Neptune analogs.  Thus, our six planet, 3:2 configurations with primordial excitation boast slightly higher success rates for criterion \textbf{B}.

Our simulations indicate that the primordial 3:2 Jupiter-Saturn resonance still represents a viable post-formation evolutionary pathway for the solar system \citep{bat10,nesvorny11,nesvorny12,batbro12,deienno17,clement18}.  Indeed, our tested six planet configurations (with and without primordial excitation) each produce one system that simultaneously satisfies all 4 of our success criteria.  This is also the case for our five planet, 3:2,3:2,2:1,3:2 chain, and our five planet set that includes primordial eccentricity pumping.  As pointed out by previous authors, the challenge with the 3:2 version of the Nice Model (section \ref{sect:3_2}) is that it is extremely violent.  Thus, only $\sim$20$\%$ of simulations experience appropriately ``weak'' instabilities, and yield correspondingly realistic Jupiter-Saturn period ratio jumps (criterion \textbf{D}, this manifests as fewer points plotted in figure \ref{fig:control_e55} compared to the corresponding plots for our 2:1 simulations).  Our results indicate that there is very little difference in the performance of this subset of weaker instabilities when $e_{J,o}$ and $e_{S,o}$ are artificially elevated.  The main difference between the two sets is that the total size of the population of less-violent outcomes decreases with primordial eccentricity pumping.  This is evidenced by a 25$\%$ success rate for criterion \textbf{D} in our low-$e$ six planet batch, compared to just 5$\%$ in the moderate-$e$ case.  Therefore, we find that primordial eccentricity pumping in the 3:2 Jupiter-Saturn resonance does not bring the solar system result closer to the center of the distribution of possible outcomes in $M_{55}$-$P_{S}/P_{J}$ space (figure \ref{fig:control_e55}).  Instead, it pushes the actual Jupiter-Saturn period ratio farther towards the extreme of possible results.

\subsubsection{2:1; circular orbits}

\begin{figure*}
	\centering
	\includegraphics[width=.7\textwidth]{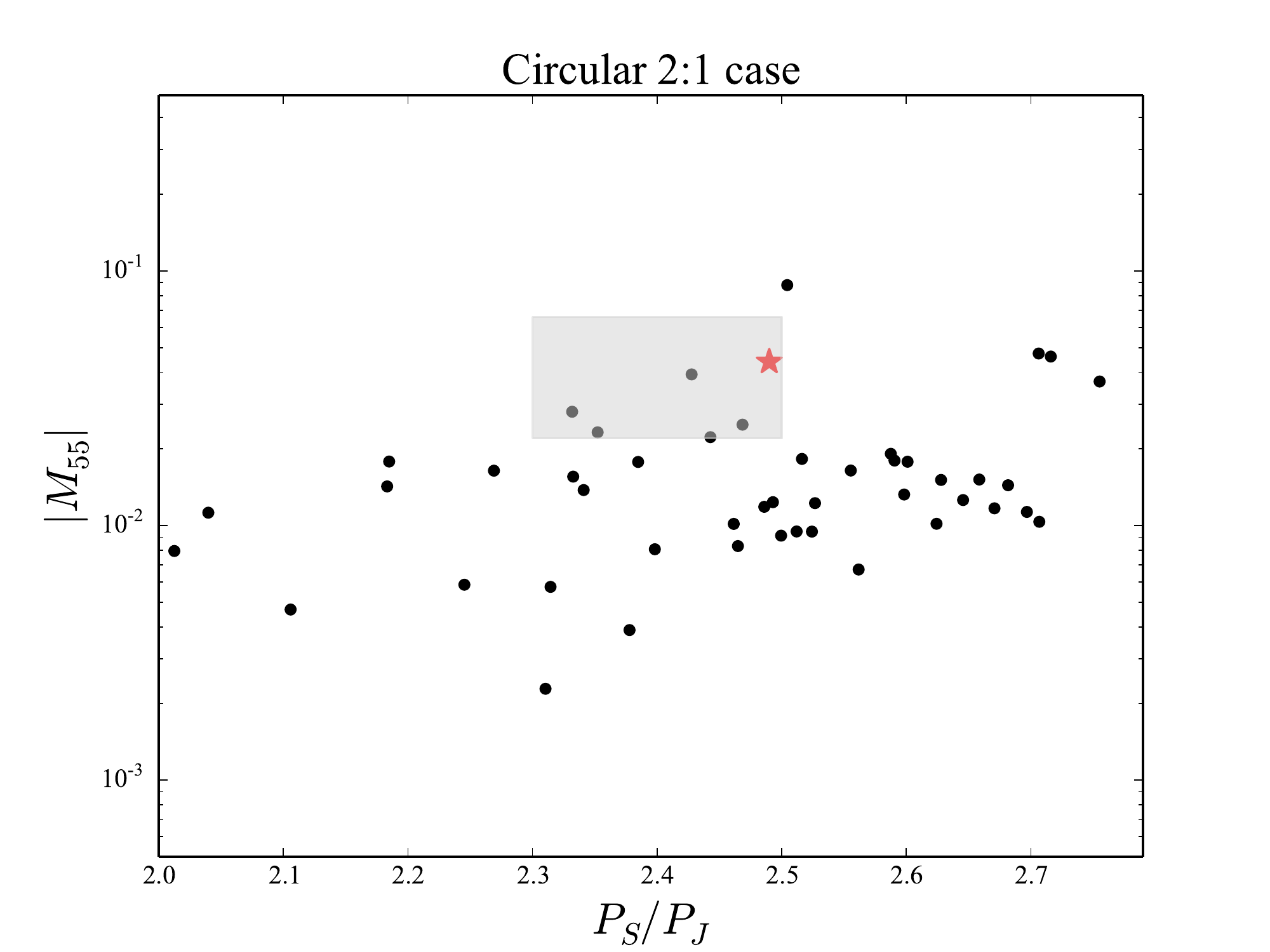}
	\caption{$M_{55}$-$P_{S}/P_{J}$ space for our simulations beginning from the circular 2:1 Jupiter Saturn resonance (table \ref{table:ics}).  The red stars indicate modern solar system values.  The shaded grey area delimits the region of 2.3 $<P_{S}/P_{J}<$ 2.5 (criterion \textbf{D}) and 0.022 $<M_{55}<$ 0.066 (criterion \textbf{C}).}	
	\label{fig:control_e55_2_1}
\end{figure*}

Figure \ref{fig:control_e55_2_1} plots the results of our 2:1 control runs in $M_{55}$-$P_{S}/P_{J}$ space.  In contrast to the 3:2 cases depicted in figure \ref{fig:control_e55}, no simulation in our circular 2:1 batch finishes with $M_{55}$ excited to at least the solar system value without exceeding $P_{S}/P_{J}=$ 2.5.  This is consistent with the results of \citet{nesvorny12} and \citet{deienno17}.  In both studies, the authors concluded that it is extremely difficult to adequately excite Jupiter's eccentricity out of the primordial 2:1 resonance.  Indeed, while the individual rates of success for each of our four success criteria are reasonable for our circular 2:1 instabilities (table \ref{table:results}), none are successful at simultaneously satisfying \textbf{C} and \textbf{D}.  Specifically, these systems systematically struggle to adequately excite the eccentricities of \textit{both} Jupiter and Saturn when the instability yields an appropriately small jump ($P_{S}/P_{J}<$ 2.5).  Indeed, only 23$\%$ of the systems that finish with Jupiter and Saturn inside of their mutual 5:2 MMR excite $M_{55}$ to greater than half its modern value, and only 10$\%$ possess $M_{65}$ magnitudes in excess of 0.016 (half the modern magnitude).  As none of our control, 2:1 simulations are successful at simultaneously satisfying criteria \textbf{C} and \textbf{D}, we conclude that some degree of primordial excitation is likely an important prerequisite to the viability of any 2:1 instability scenario.

\subsection{2:1, five planets, and loose}

In general we find that the primordial 2:1 Jupiter-Saturn resonance with heightened eccentricities is also a viable evolutionary pathway for the outer solar system.  In many ways, our 2:1 batches of simulations outperform the 3:2 sets discussed in the previous section.  However, we caution the reader that our results should be taken as motivation for follow-on study of the 2:1 resonance, and not as reason to abandon the 3:2.  As previously discussed, the 2:1 version of the Nice Model's (section \ref{sect:2_1}) effects on the solar system's global dynamics are not as well-studied as are those of the 3:2 \citep[e.g.:][]{nesvorny13,nesvorny15a,nesvorny15b,roig15,roig16}.  In particular, the asteroid belt is most sensitive to the instability's particular dynamics \citep{deienno16,deienno18} and the precise motions of the dominant secular resonances' locations in belt \citep{morby10,izidoro16,clement20_mnras}.  Thus, our simulations of the 2:1 instability are limited in that they only analyze the various sets of initial conditions' success at replicating the modern orbital configuration of the four giant planets (success criteria \textbf{A}-\textbf{D}).  Future work to fully validate the scenario must focus on the evolution of the orbital distributions in the asteroid \citep[e.g.:][]{roig15} and Kuiper \citep[e.g.:][]{nesvorny16} belts, and the obliquity evolution of the giant planets \citep{vokrouhicky15,brasser_lee_15}.

We also find it difficult to finely control the instability's timing, and therefore minimize the amount of damping in $e_{J,o}$ and $e_{S,o}$ that occurs prior to the instability.  Even with our artificial instability trigger (section \ref{sect:methods}), systems often take a few Myr (the median instability time for our simulation batches varies between $\sim$0.02-3.0 Myr) to fully evolve in to an orbital instability.  As the giant planets are still interacting with the exterior planetesimal disk during this time, the gas giants' eccentricities can damp out appreciably.  While we are unable to find any correlations between $t_{inst}$ and the final properties of our simulated systems that are statistically significant, we remind the reader that the gas giants in a subset (albeit, only a small set) of the systems analyzed in these sections have damped to near-zero eccentricity by the time the instability ensues.

\begin{figure*}
	\centering
	\includegraphics[width=.7\textwidth]{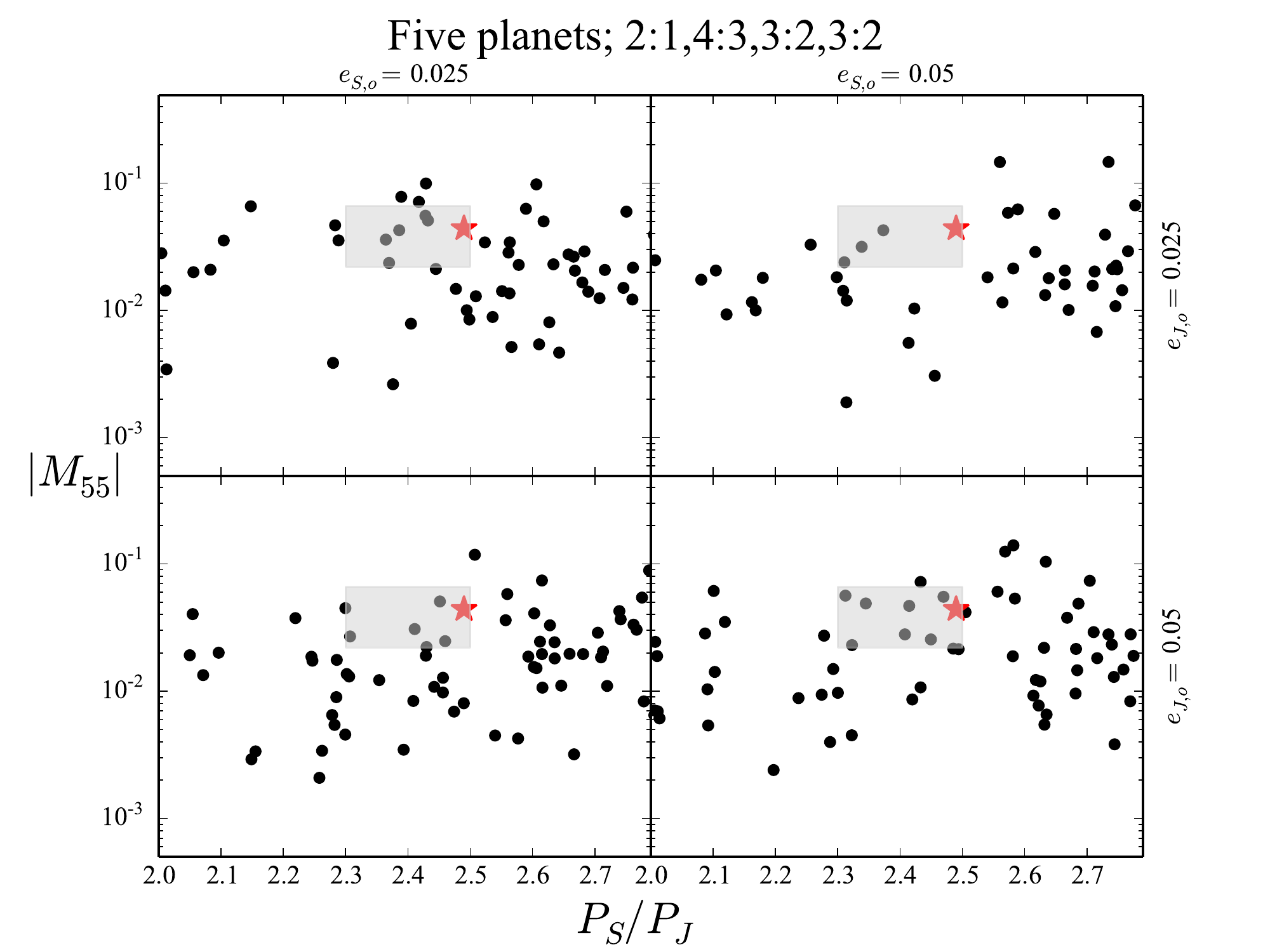}
	\caption{$M_{55}$-$P_{S}/P_{J}$ space for our simulations with five planets beginning from a 2:1,4:3,3:2,3:2 resonant chain.  The upper two panels plot simulations begin with $e_{J}=$ 0.025, and the bottom panels show systems where Jupiter's initial eccentricity is 0.05.  Similarly, the left two panels plot simulations begin with $e_{S}=$ 0.025, and the panels on the right show systems where Saturn's initial eccentricity is 0.05.  The shaded grey area delimits the region of 2.3 $<P_{S}/P_{J}<$ 2.5 (criterion \textbf{D}) and 0.022 $<M_{55}<$ 0.066 (criterion \textbf{C}).}	
	\label{fig:5_1ig_e55}
\end{figure*}

Figure \ref{fig:5_1ig_e55} plots the results of our 2:1,4:3,3:2,3:2 instabilities in $M_{55}$-$P_{S}/P_{J}$ space (we refer to this as our "loose," ``broad'' or ``wide'' configuration in the subsequent text).  We find these wider resonant chains to generally be more successful than our compact (or ``tight'') configurations of five planets (section \ref{sect:2_1_tight}).  As our preliminary work indicated that even broader chains beginning with 2:1 (e.g.: 2:1,3:2,3:2,3:2, see discussion in section \ref{sect:2_1_3_2}) seldom degenerate in to an instability \citep{deienno17}, the success of the 2:1,4:3,3:2,3:2 chain over tighter configurations places a fairly strict constraint on the range potentially viable five planet configurations.  Interestingly, these systems possess rates of success for criteria \textbf{C} ($\sim$5-10$\%$) and \textbf{D} ($\sim$15-20$\%$) that are similar to those of our 3:2 Jupiter-Saturn resonance five planet cases (section \ref{sect:con_results}).  However, the solar system seems to be less of an outlier in the overall distribution of $M_{55}$-$P_{S}/P_{J}$ outcomes for our 2:1 instabilities (red star in figure \ref{fig:5_1ig_e55}).  Thus, while high $|M_{55}|$ values occur preferentially in systems that experience a large jump out of the primordial 3:2 resonance, properly excited Jupiter analogs occur with similar frequencies in systems across the full spectrum of $P_{S}/P_{J}$ outcomes in simulations that begin from the 2:1 resonance.  This is largely a consequence of $M_{55}$ already being excited prior to the instability. \footnote{In general, the coefficients $M_{ij}$ ($i,j=$5,6) of our initial giant planet configurations are partitioned such that the magnitudes of $M_{55}$ and $M_{66}$ are approximately twice those of $M_{56}$ and $M_{65}$ prior to the instability.  However, the precise relative values also depend on the whether or not $e_{J,o}>e_{S,o}$.}

Additionally, our looser, 2:1,4:3,3:2,3:2 resonant chains yield systematically higher success rates for criteria \textbf{A} (30-40$\%$) and \textbf{B} (20-25$\%$) than our 3:2 control cases, and our tighter 2:1, five planet chains.  In a sense, these initial conditions place the planets in a more dynamically isolated configuration that already somewhat resembles that of the modern solar system.  Therefore, these systems typically undergo instabilities that are weaker and more regularly behaved than those experienced in more compact configurations (3:2,3:2,3:2,3:2, 3:2,3:2,2:1,3:2, 2:1,4:3,4:3,3:2, or 2:1,4:3,4:3,4:3).  The wider initial chain also tends to lead to weaker scattering events between Jupiter, Saturn, and the ejected ice giant that might excite the relevant secular eigenmodes.  Therefore, our most successful outcomes occur in systems with higher initial values of $e_{J,o}$ and $e_{S,o}$ ($\sim$0.05).  As depicted in the bottom right panel of figure \ref{fig:5_1ig_e55}, such a configuration is quite successful at placing the solar system outcome near the heart of the distribution of $M_{55}$-$P_{S}/P_{J}$ outcomes.  Moreover, both cases with $e_{J,o}=$ 0.05 are more successful at satisfying criterion \textbf{C} and fully replicating the complete Jupiter-Saturn secular system than those with  $e_{J,o}=$0.025.

\begin{figure}
	\centering
	\includegraphics[width=.49\textwidth]{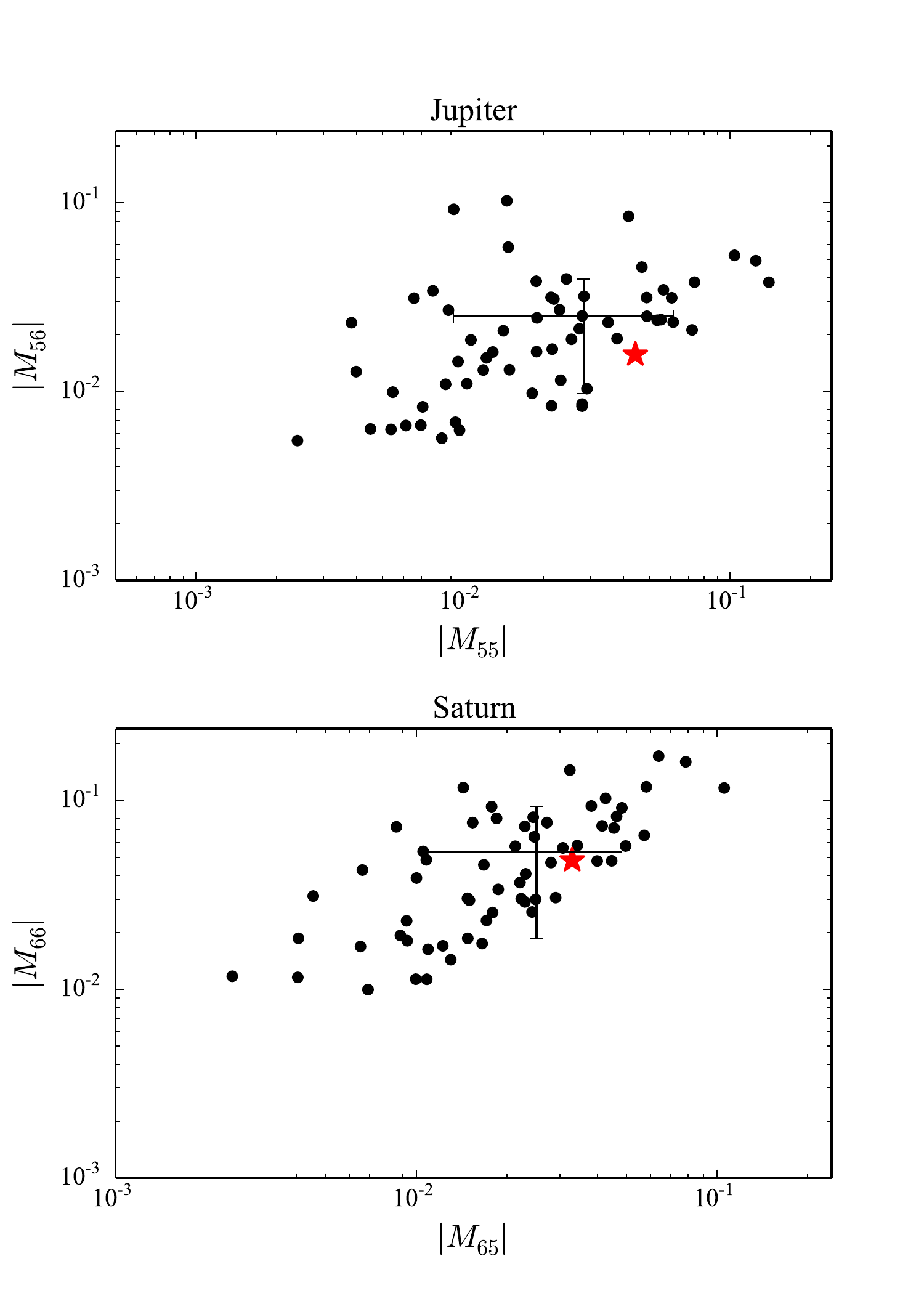}
	\caption{Eccentric magnitudes, $M_{ij}$ ($i,j=$5,6), for the Jupiter-Saturn systems that satisfied criterion \textbf{D} in our batch of simulations beginning with five planets in a 2:1,4:3,3:2,3:2 resonant chain and $e_{J,o}=e_{S,o}=$0.05.  The top panel shows the magnitudes $M_{55}$ and $M_{56}$ in Jupiter's eccentricity, the bottom panel depicts the magnitudes $M_{65}$ and $M_{66}$ in Saturn's.  The error bars indicate one standard deviation.  The red stars correspond to solar system values.}	
	\label{fig:5gp_modes_examp}
\end{figure}

\begin{figure}
	\centering
	\includegraphics[width=.49\textwidth]{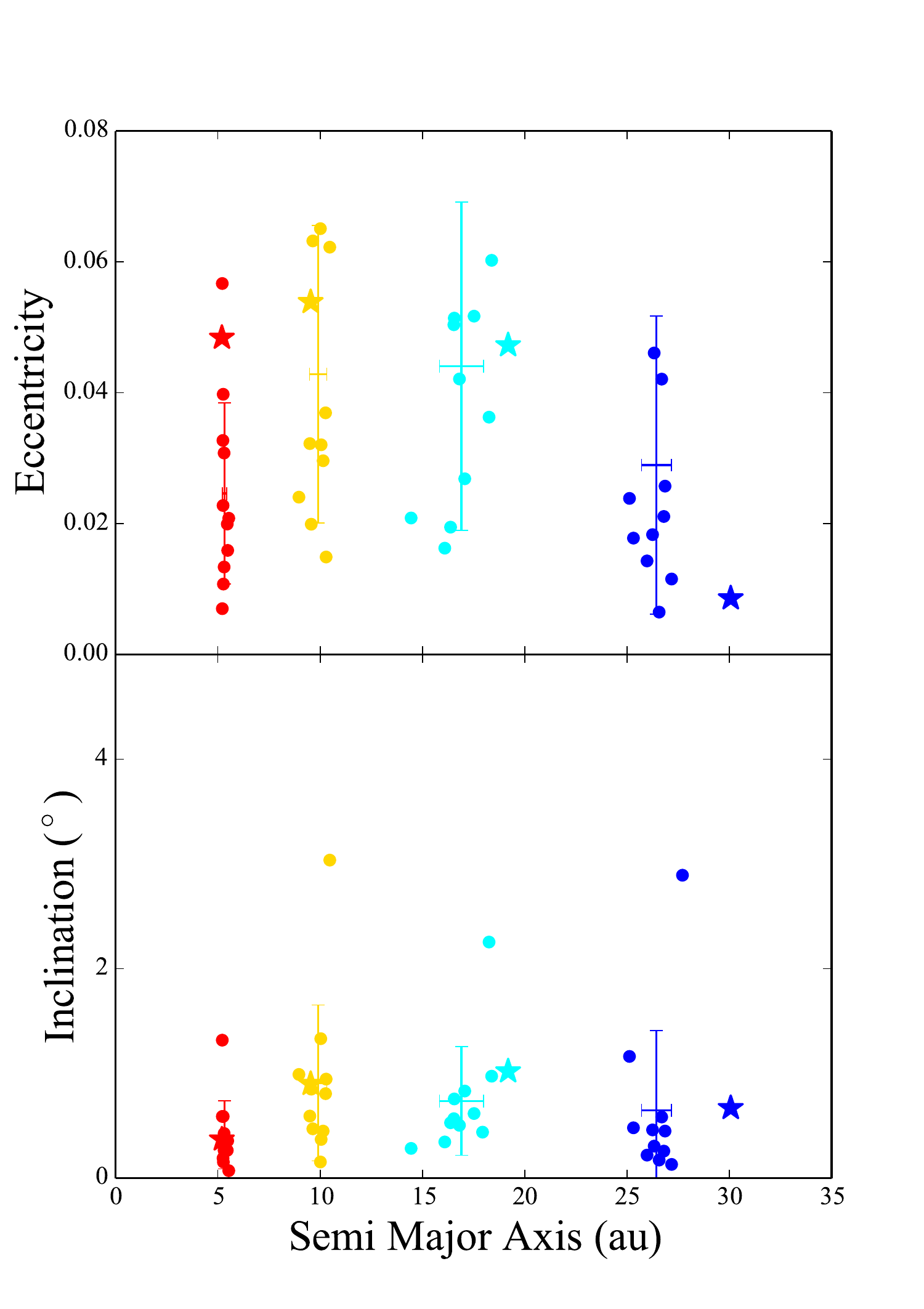}
	\caption{Final orbits for systems that satisfied criteria \textbf{A} and \textbf{B} in our batch of simulations beginning with five planets in a 2:1,4:3,3:2,3:2 resonant chain and $e_{J,o}=e_{S,o}=$0.05.  The top panel depicts $a/e$ space and the bottom panel plots $a/i$ space.  The respective planets and their simulated analogs are color coded as follows: Jupiter in red, Saturn in gold, Uranus in light blue, Neptune in dark blue.  The error bars indicate one standard deviation.  The stars correspond to solar system values.}	
	\label{fig:5gp_aei_examp}
\end{figure}

Figures \ref{fig:5gp_modes_examp} and \ref{fig:5gp_aei_examp} focus on our most successful five planet batch (2:1,4:3,3:2,3:2 with $e_{J,o}=e_{S,o}=$ 0.05).  While the solar system values of $M_{ij}$ ($i,j=$5,6) roughly fall within the 1-$\sigma$ range of outcomes depicted in figure \ref{fig:5gp_modes_examp}, it is clear that $M_{55}=$ 0.044 is somewhat closer to the extreme of the distribution for low-$M_{56}$.  Conversely, Saturn's eccentric modes are reproduced quite frequently in our simulations.  However, our numerical integrations do not fully capture the planets' residual migration phase.  Thus, we expect the distribution of Jupiter's eccentric modes in the top panel of figure \ref{fig:5gp_modes_examp} to move slightly down towards the solar system outcome as the system continues to evolve (by $\sim$0.0005 in our tests of a 100 Myr integration time; see discussion in section \ref{sect:trigger}).

Figure \ref{fig:5gp_aei_examp} demonstrates how our successful, five planet, 2:1, loose configurations systematically struggle to reproduce Jupiter's mean eccentricity and the Ice Giants' semi-major axes in systems that retain 4 planets.  While the lower values of $e_{J}$ are a consequence of the aforementioned challenges with properly exciting $M_{55}$ in weaker instabilities that eject only one planet, the ice giant's final orbital locations are  most sensitive to interactions with the remnant planetesimal disk \citep{nesvorny12}.  There are two main factors that are responsible for our simulated ice giants' inability to attain their modern semi-major axes.  First, our simulations only model 20 Myr of the residual migration phase (a compromise necessary to limit the computational cost of our work).  Second, the \textit{amount} of residual migration is a function of the total remnant mass in planetesimals; which in turn is related to the initial mass placed in the planetesimal disk.  For consistency, our simulations strictly consider $M_{KB}=$ 20 $M_{\oplus}$ disks.  However, \citet{nesvorny12} found that massive disks ($\sim$35-50 $M_{\oplus}$) are more successful in more-compact five planet configurations.  This is an obvious consequence of certain wider configurations with more planets requiring less residual migration for the planets to reach their modern orbits \citep[of note,][also prefer lighter disks in the wider 3:2,3:2,2:1,3:2]{nesvorny12}.  Furthermore, the relationship between $M_{KB}$ and simulation success is slightly more complicated because the residual migration phase also damps out the eccentric modes of the Jupiter-Saturn system.  Thus, the selection of initial $M_{KB}$ can be considered a sort of balancing act.  Too much residual migration can over-damp the gas giants and lead to failure of criterion \textbf{C}, while too little can result in the ice giants stopping short of their modern semi-major axes and lead to failure of criterion \textbf{B} \citep{gomes04}.  We explore the consequences of the particular choice of $M_{KB}$ further with additional simulations in section \ref{sect:kb_mass}.

\subsection{2:1, five planets, and tight}
\label{sect:2_1_tight}

\begin{figure*}
	\centering
	\includegraphics[width=.7\textwidth]{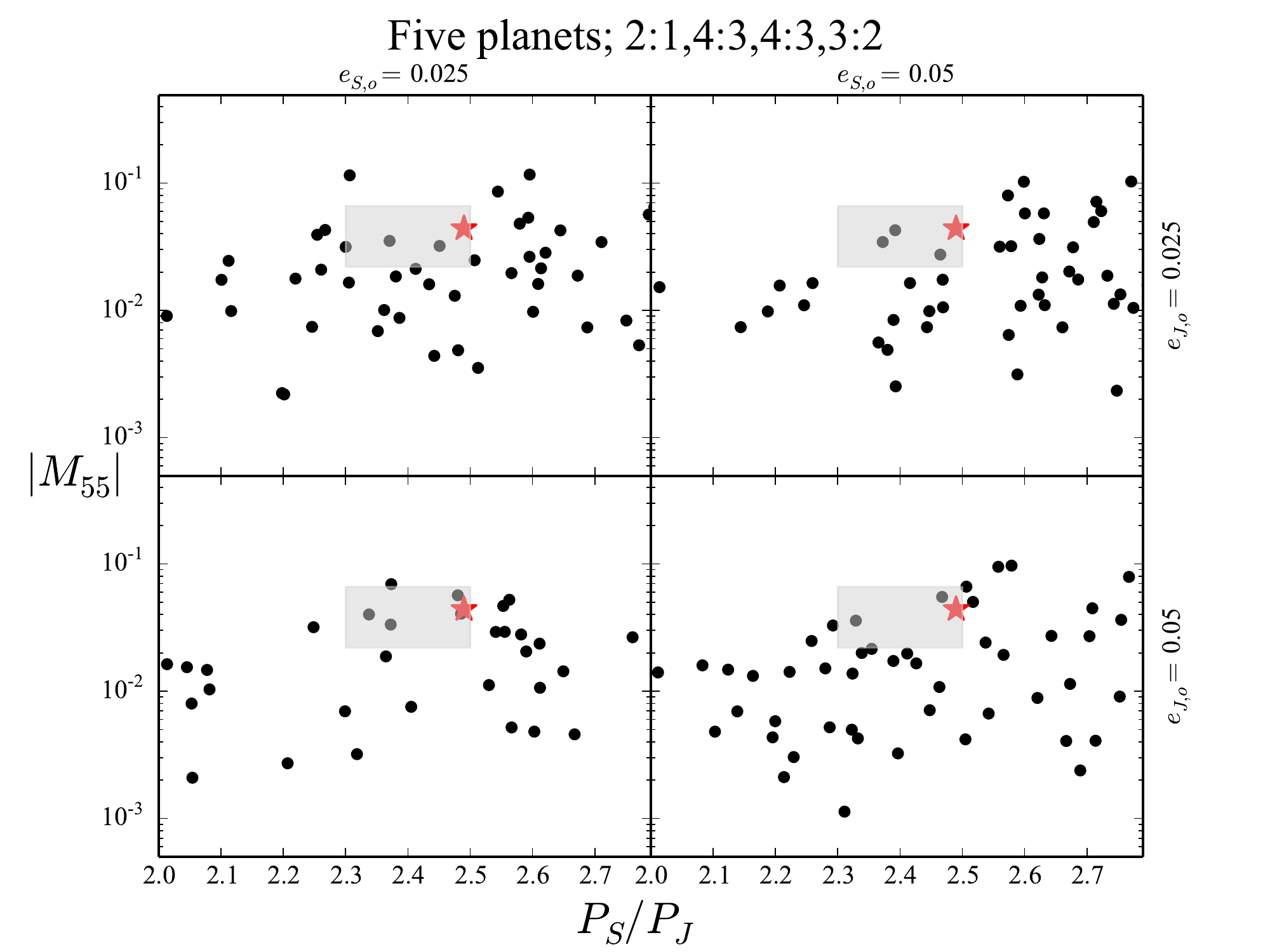}
	\caption{$M_{55}$-$P_{S}/P_{J}$ space for our simulations with five planets beginning from a 2:1,4:3,4:3,3:2 resonant chain.  The upper two panels plot simulations begin with $e_{J}=$ 0.025, and the bottom panels show systems where Jupiter's initial eccentricity is 0.05.  Similarly, the left two panels plot simulations begin with $e_{S}=$ 0.025, and the panels on the right show systems where Saturn's initial eccentricity is 0.05.  The shaded grey area delimits the region of 2.3 $<P_{S}/P_{J}<$ 2.5 (criterion \textbf{D}) and 0.022 $<M_{55}<$ 0.066 (criterion \textbf{C}).}	
	\label{fig:5_2ig_e55}
\end{figure*}

\begin{figure*}
	\centering
	\includegraphics[width=.7\textwidth]{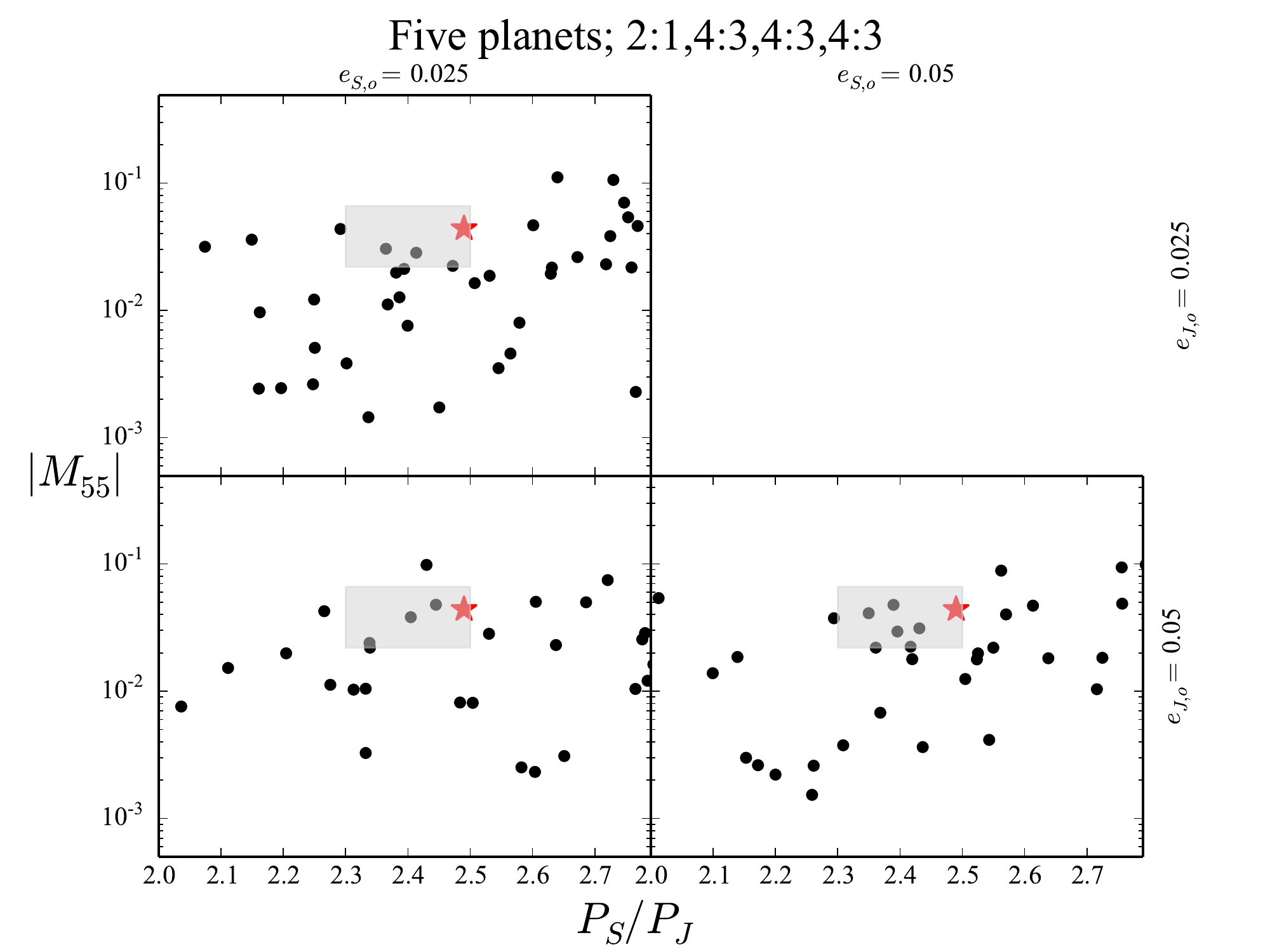}
	\caption{$M_{55}$-$P_{S}/P_{J}$ space for our simulations with five planets beginning from a 2:1,4:3,4:3,4:3 resonant chain.  The upper two panels plot simulations begin with $e_{J}=$ 0.025, and the bottom panels show systems where Jupiter's initial eccentricity is 0.05.  Similarly, the left two panels plot simulations begin with $e_{S}=$ 0.025, and the panels on the right show systems where Saturn's initial eccentricity is 0.05.  The shaded grey area delimits the region of 2.3 $<P_{S}/P_{J}<$ 2.5 (criterion \textbf{D}) and 0.022 $<M_{55}<$ 0.066 (criterion \textbf{C}).  \textit{Recall that the upper right quadrant is blank as we were unable to generate a stable chain with $e_{S}>e_{J}$ for this configuration}.}	
	\label{fig:5_3ig_e55}
\end{figure*}

\begin{figure}
	\centering
	\includegraphics[width=.49\textwidth]{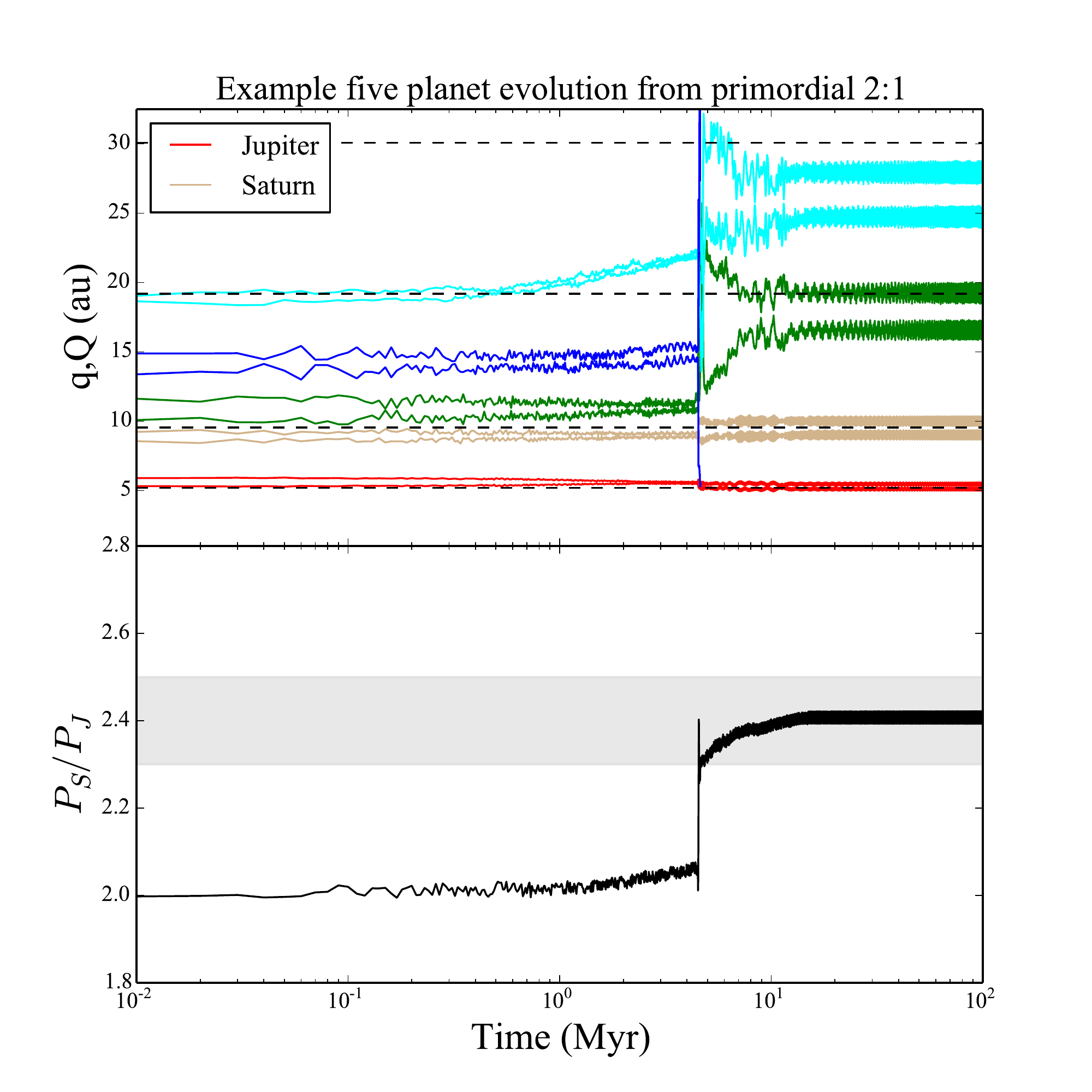}
	\caption{Example instability evolution beginning with five planets in a 2:1,4:3,3:2,3:2 resonant chain (figure \ref{fig:5_1ig_e55}).  The simulation finished with $P_{S}/P_{J}=$ 2.41, $M_{55}=$.034, $M_{56}=$.018, $M_{65}=$.027, $M_{66}=$.052 (all four success criteria are satisfied).  The top panel plots the perihelion and aphelion of each planet over the length of the simulation.  The bottom panel shows the Jupiter-Saturn period ratio.  The horizontal dashed lines in the upper panel indicate the locations of the giant planets' modern semi-major axes.  The shaded region in the lower panel delimits the range of 2.3 $<P_{S}/P_{J}<$ 2.5.}
	\label{fig:5gp_good}
\end{figure}

The results of our tighter configurations of 2:1, five planet instabilities are plotted in figures \ref{fig:5_2ig_e55} (2:1,4:3,4:3,3:2) and \ref{fig:5_3ig_e55} (2:1,4:3,4:3,4:3).  While these sets still produce successful systems (including several simulations that simultaneously satisfy all four success criteria; table \ref{table:results}), the rates of success for every success criterion except \textbf{C} are systematically lower than for our looser, 2:1,4:3,3:2,3:2 configuration.  In particular, a greater number of these more compact systems tend to devolve into extremely violent instabilities.  The net result of this is more planet ejections (lower success rates for criterion \textbf{A}), and larger scattering events (lower success rates for criteria \textbf{B} and \textbf{D}).  As with the primordial 3:2 resonance (section \ref{sect:con_results}), the ice giants in our 2:1,4:3,4:3,3:2 and 2:1,4:3,4:3,4:3 configurations attain higher eccentricities before the instability, and tend to experience stronger mutual encounters within the chaos of the instability.  To illustrate this, we analyzed the close encounter histories for Saturn in our tightest and loosest five planet chains with $e_{J,o}=e_{S,o}=$ 0.05.  On average, we find that encounters less than 3 Hill radii between Saturn and the ice giants are $11\%$ more frequent and 3$\%$ closer in the tight batch than the corresponding simulations beginning from a looser configuration.  An example evolution for a system that satisfies all four success criteria from our 2:1,4:3,3:2,3:2, $e_{J,o}=e_{S,o}=$0.05 set is plotted in figure \ref{fig:5gp_good}.  It is clear that, even in the most successful system, Uranus and Neptune are over-excited in the instability.

Another way to inspect the differences between our respective chains is by analyzing the rates at which a given batch satisfies a specific subset of our four success criteria simultaneously.  For example, our various 2:1,4:3,4:3,4:3 simulation batches have success rates of only 1-4$\%$ for criterion \textbf{B}.  As criterion \textbf{B} can only be satisfied when criterion \textbf{A} is already met, we can easily compare the ratio of criterion \textbf{B} satisfying simulations to all those that finish with $N_{GP}=$ 4 (\textbf{A}) between our various different resonant chains.  The two more compact configurations already possess lower success rates for criterion \textbf{A} (7-19$\%$ vs. $\sim$30$\%$) by virtue of the instabilities being more violent and ejecting planets more frequently.  However, only $\sim$10-20$\%$ of these four planet systems in the tighter batches also satisfy \textbf{B}, compared to $\gtrsim$60$\%$ of those that originated in the looser, 2:1,4:3,3:2,3:2 chain.  On closer inspection, we find that the majority of these criterion \textbf{A} satisfying systems experience uncharacteristically weak instabilities that leave the giant planets in a final orbital configuration that is too compact.  An additional several runs satisfy criterion \textbf{A}, but fail \textbf{B} as a result of a series of scattering events between the ice giants that drive the Neptune analog's semi-major axis into the distant Kuiper belt.  The tendency of these compact systems to only finish with 4 planets in weak instabilities is also evidenced by the various percentages of systems that meet all but one of our constraints.  For instance, only 5$\%$ of the 2:1,4:3,4:3,3:2, $e_{J,o}=e_{S,o}=$ 0.05 batch successfully meet both \textbf{A} and \textbf{B}. Of this subset of 10 simulations, nine systems experience a small jump and satisfy \textbf{D}, while only one is strong enough to properly reproduce $M_{55}$ (\textbf{C}).  In summary, while our more compact five planet configurations do yield successful evolutionary schemes, they also suffer multiple systematic issues that do not affect our wider five planet configurations as severely.

\subsection{2:1, six planets}
\label{sect:6gp_tight_res}

\begin{figure*}
	\centering
	\includegraphics[width=.7\textwidth]{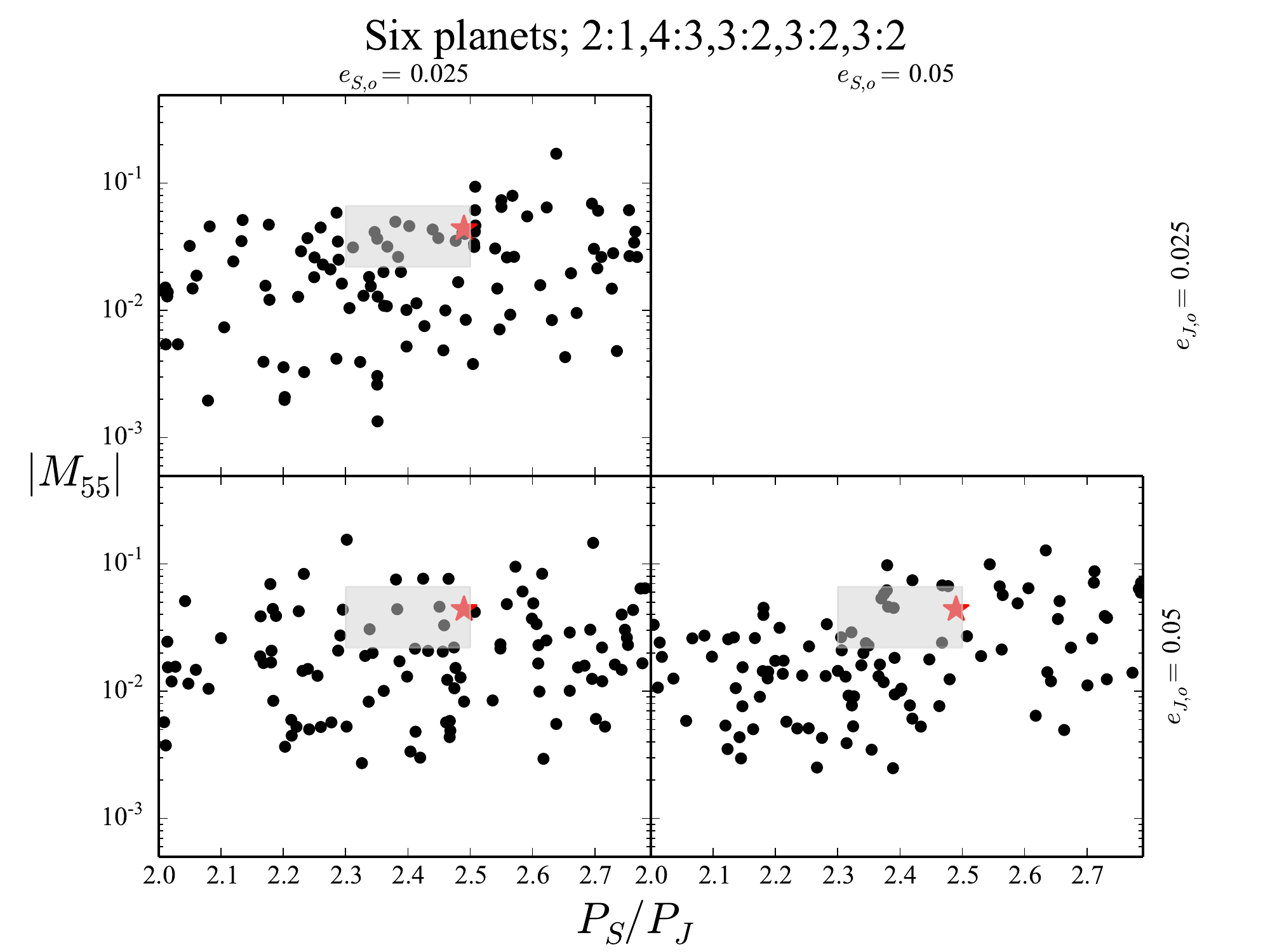}
	\caption{$M_{55}$-$P_{S}/P_{J}$ space for our simulations with six planets beginning from a 2:1,4:3,3:2,3:2,3:2 resonant chain.  The upper two panels plot simulations begin with $e_{J}=$ 0.025, and the bottom panels show systems where Jupiter's initial eccentricity is 0.05.  Similarly, the left two panels plot simulations begin with $e_{S}=$ 0.025, and the panels on the right show systems where Saturn's initial eccentricity is 0.05.  The shaded grey area delimits the region of 2.3 $<P_{S}/P_{J}<$ 2.5 (criterion \textbf{D}) and 0.022 $<M_{55}<$ 0.066 (criterion \textbf{C}).  \textit{Recall that the upper right quadrant is blank as we were unable to generate a stable chain with $e_{S}>e_{J}$ for this configuration}.}	
	\label{fig:6_1ig_e55}
\end{figure*}

Perhaps the most striking difference between our five and six planet cases is the six planet sets' higher success rates for criteria \textbf{A} and \textbf{D}.  These two results are not mutually inclusive, in that $N_{GP}=$ 3 systems satisfy criterion \textbf{D} at roughly equal rates as $N_{GP}=$ 4 systems.  Our six planet instabilities are typically weaker, and tend to behave more consistently than the counterpart five planet runs (note, however, that our six planet configurations are rather artificial in terms of our inclusion of two additional, $M_{IG}=$ 8.0 $M_{\oplus}$ planets).  While  five and six planet cases are roughly equally successful at exciting $M_{55}$ \citep[$\sim$55-65$\%$ finish with $M_{55}>$ 0.022, depending on the initial conditions, i.e.: the original criterion \textbf{C} or][]{nesvorny12}, six planet instabilities are less likely to \textit{over-excite} $M_{55}$. Thus, our six planet runs possess success rates for our updated criterion \textbf{C} that are systematically the same or better than those of our five planet simulations.  This result, and the prevalence of $N_{GP}=$ 4 systems are partially a consequence of the fact that the scattering ice giants have lower masses than in our five planet instabilities.  As with our six planet, 3:2 control runs (section \ref{sect:con_results}), the outer two, more massive, ice giants typically survive the instability and become Uranus and Neptune analogs.  In fact, this is the case in every one of our six planet simulations that satisfy all four of our constraints.  As the two outermost planets begin the simulation more dynamically isolated from the gas giants, and with lower eccentricities (e$\lesssim$0.05) in six planet configurations, they are also more resilient to loss by ejection during the instability.

\begin{figure*}
	\centering
	\includegraphics[width=.7\textwidth]{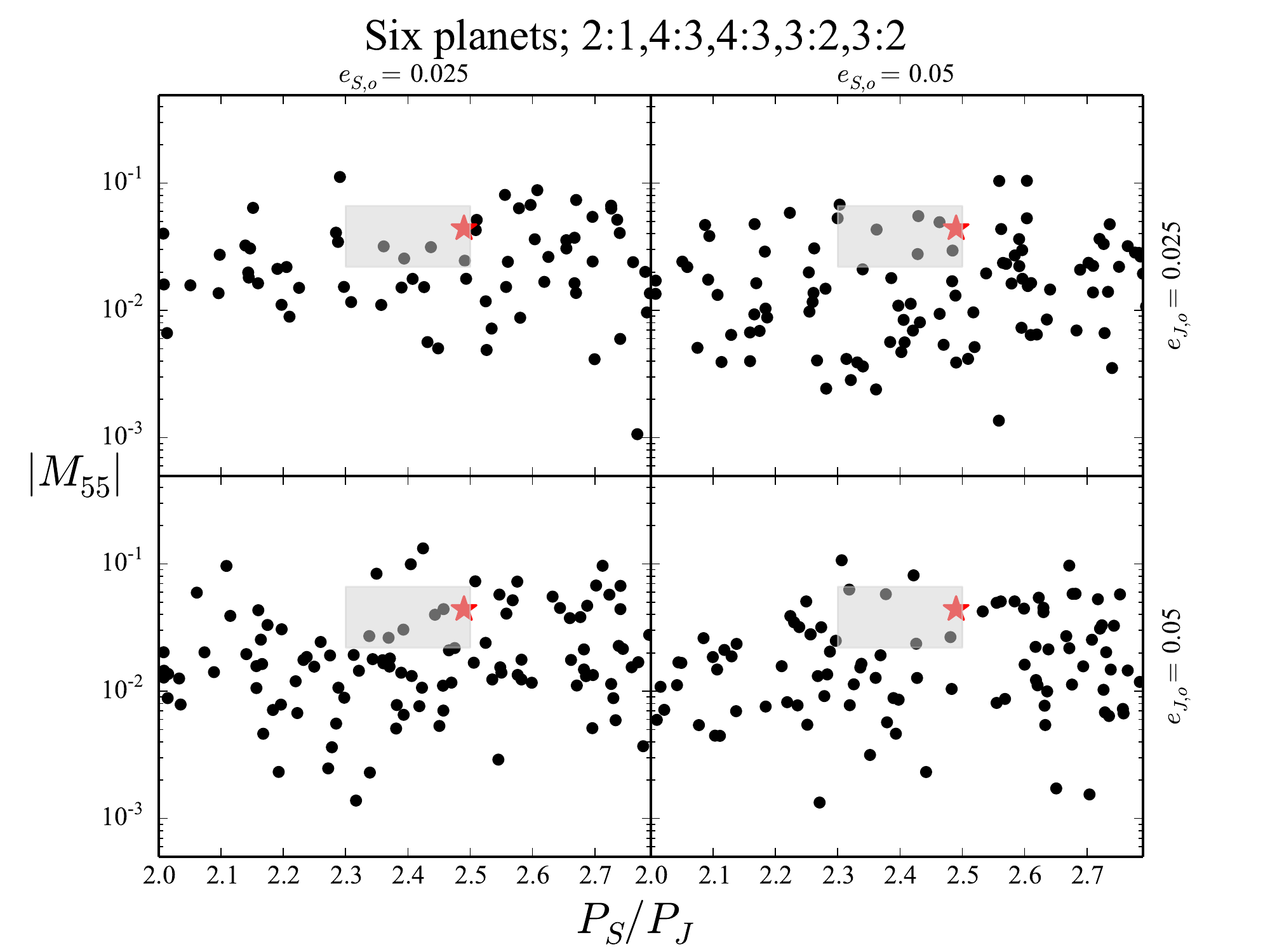}
	\caption{$M_{55}$-$P_{S}/P_{J}$ space for our simulations with six planets beginning from a 2:1,4:3,4:3,3:2,3:2 resonant chain.  The upper two panels plot simulations begin with $e_{J}=$ 0.025, and the bottom panels show systems where Jupiter's initial eccentricity is 0.05.  Similarly, the left two panels plot simulations begin with $e_{S}=$ 0.025, and the panels on the right show systems where Saturn's initial eccentricity is 0.05.  The shaded grey area delimits the region of 2.3 $<P_{S}/P_{J}<$ 2.5 (criterion \textbf{D}) and 0.022 $<M_{55}<$ 0.066 (criterion \textbf{C}).}	
	\label{fig:6_2ig_e55}
\end{figure*}

\begin{figure*}
	\centering
	\includegraphics[width=.7\textwidth]{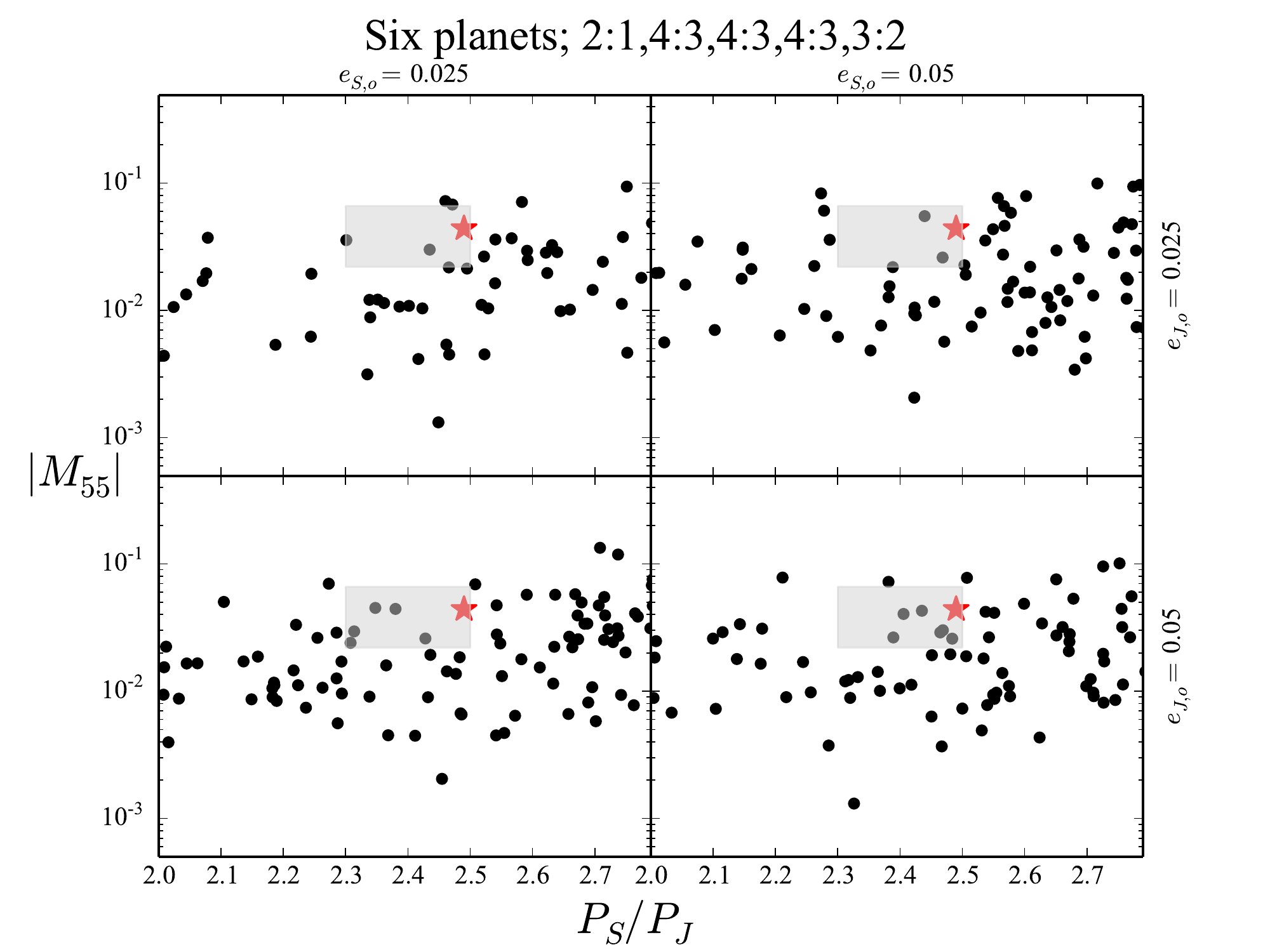}
	\caption{$M_{55}$-$P_{S}/P_{J}$ space for our simulations with six planets beginning from a 2:1,4:3,4:3,4:3,3:2 resonant chain.  The upper two panels plot simulations begin with $e_{J}=$ 0.025, and the bottom panels show systems where Jupiter's initial eccentricity is 0.05.  Similarly, the left two panels plot simulations begin with $e_{S}=$ 0.025, and the panels on the right show systems where Saturn's initial eccentricity is 0.05.  The shaded grey area delimits the region of 2.3 $<P_{S}/P_{J}<$ 2.5 (criterion \textbf{D}) and 0.022 $<M_{55}<$ 0.066 (criterion \textbf{C}).}	
	\label{fig:6_3ig_e55}
\end{figure*}

Figures \ref{fig:6_1ig_e55}, \ref{fig:6_2ig_e55} and \ref{fig:6_3ig_e55} plot the results of our various six planet instability batches in $M_{55}$-$P_{S}/P_{J}$ space.  While, the overall distributions of possible outcomes are quite similar, the differences between the respective simulation sets are best characterized by small tradeoffs in success rates for criteria \textbf{A}-\textbf{D}.  For instance, our looser configuration (2:1,4:3,3:2,3:2,3:2) is more likely to experience a small jump (36-43$\%$ of systems satisfying criterion \textbf{D}), but less successful at meeting criterion \textbf{B}.  Utilizing a more compact initial configuration boosts the probability of success in terms of the planets final semi-major axes (20-30$\%$ of systems satisfying \textbf{B}), but reduces the total sample of systems experiencing small jumps.  Thus, the choice of initial configuration is a compromise between more efficiently limiting the gas giants' jump (with a looser chain, and weaker instability), and placing the eventual Uranus and Neptune analogs at the right initial semi-major axes to produce a successful final orbital configuration (a tighter chain).  A more compact initial orientation of the planets also tends to be more successful at adequately exciting the secular eigenmodes of the Jupiter-Saturn system, while looser chains are more likely to under-excite Jupiter or Saturn's eccentricity (specifically, $\gtrsim$75$\%$ of these systems under-excite $M_{55}$ or $M_{65}$).

\begin{figure}
	\centering
	\includegraphics[width=.49\textwidth]{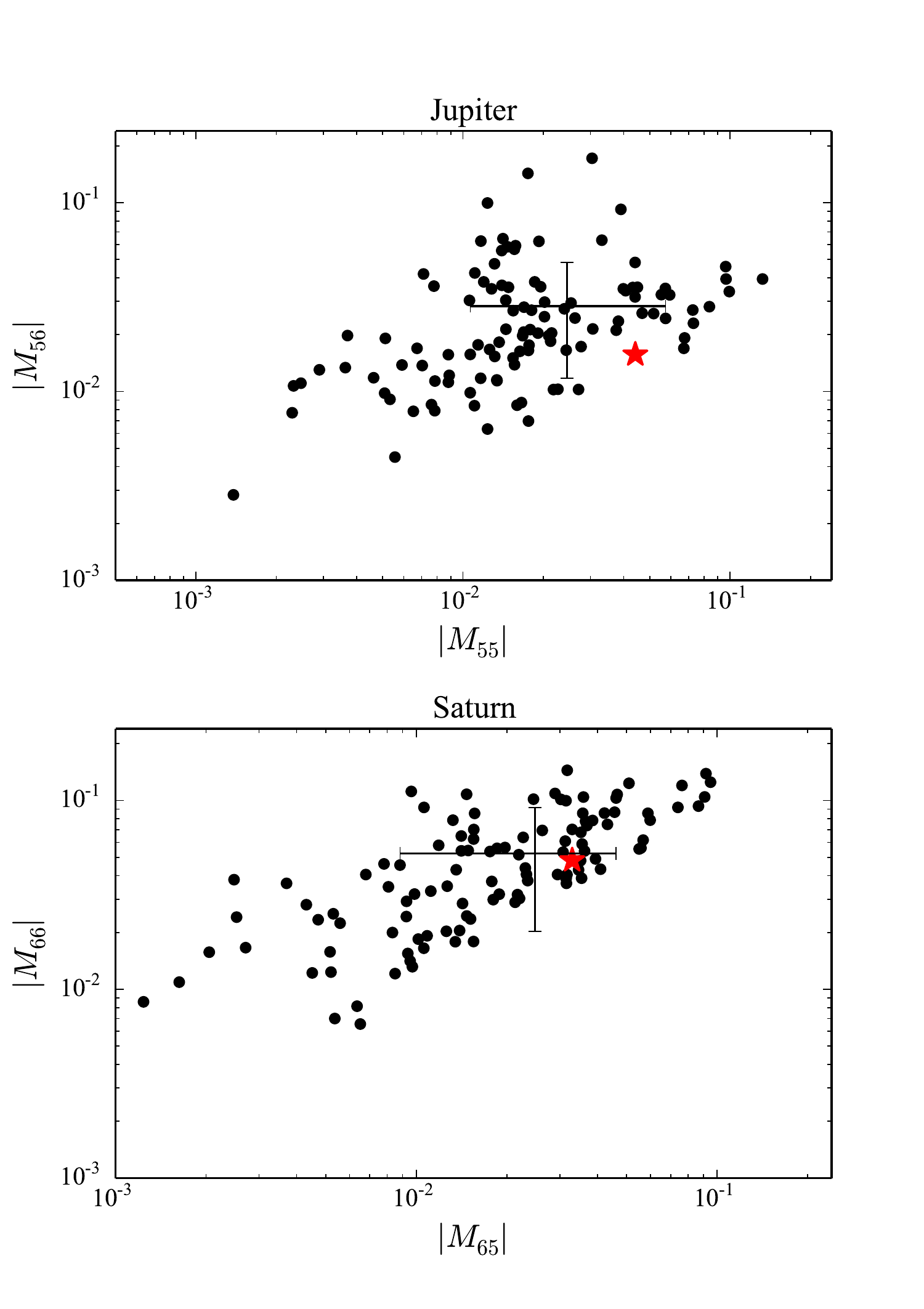}
	\caption{Eccentric magnitudes, $M_{ij}$ ($i,j=$5,6), for the Jupiter-Saturn systems that satisfied criterion \textbf{D} in our batch of simulations beginning with six planets in a 2:1,4:3,4:3,3:2,3:2 resonant chain and $e_{J,o}=e_{S,o}=$0.05.  The top panel shows the magnitudes $M_{55}$ and $M_{56}$ in Jupiter's eccentricity, the bottom panel depicts the magnitudes $M_{65}$ and $M_{66}$ in Saturn's.  The error bars indicate one standard deviation.  The red stars correspond to solar system values.}	
	\label{fig:6gp_modes_examp}
\end{figure}

\begin{figure}
	\centering
	\includegraphics[width=.49\textwidth]{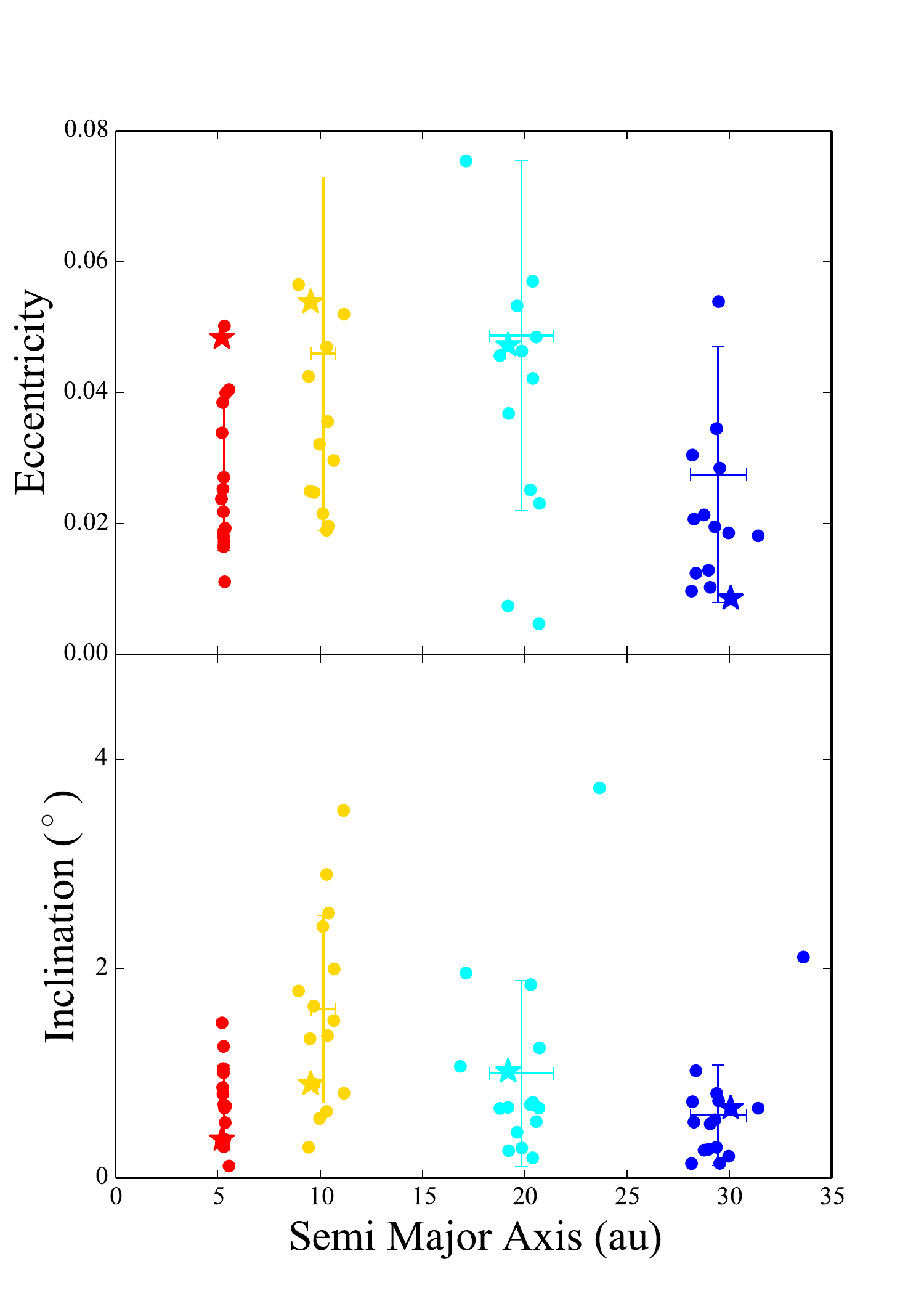}
	\caption{Final orbits for systems that satisfied criteria \textbf{A} and \textbf{B} in our batch of simulations beginning with six planets in a 2:1,4:3,4:3,3:2,3:2 resonant chain and $e_{J,o}=e_{S,o}=$0.05.  The top panel depicts $a/e$ space and the bottom panel plots $a/i$ space.  The respective planets and their simulated analogs are color coded as follows: Jupiter in red, Saturn in gold, Uranus in light blue, Neptune in dark blue.  The error bars indicate one standard deviation.  The stars correspond to solar system values.}	
	\label{fig:6gp_aei_examp}
\end{figure}

The relationship between $e_{J,o}$, $e_{S,o}$ and simulation success is also more complicated in our six planet batches than those that study five planets.  Specifically, higher initial values of $e_{J,o}$ ($\simeq$ 0.05) correlate with higher success rates for criterion \textbf{C}, as they are more likely to lead to adequately excited values of $M_{55}$.  Conversely, slightly lower initial eccentricities for Saturn ($\simeq$ 0.025) seem to improve the likelihood of the gas giants' semi-major axes staying interior to their mutual 5:2 MMR (criterion \textbf{D}).  However, it is important to note that this combination of primordial eccentricities is only produced from a specific combination of disk parameters in hydrodynamic models \citep[][discussed further in section \ref{sect:ecc_results}]{pierens14}.  Figures \ref{fig:6gp_modes_examp} and \ref{fig:6gp_aei_examp} plot the final distributions of secular eigenmodes and 4 planet system orbits for this ``ideal'' combination of $e_{J,o}=$ 0.05, $e_{S,o}=$ 0.025 for our 2:1,4:3,4:3,3:2,3:2 resonant chain.  The overall distributions of the magnitudes, $M_{ij}$ ($i,j=$ 5.6), is similar to that of our preferred five planet orientation (figure \ref{fig:5gp_modes_examp}), and a significant improvement from the 3:2 version of the Nice Model (e.g.: figure \ref{fig:control_e55}).  The final planet orbits in this simulation set also provide a fairly good match to the actual solar system (figure \ref{fig:6gp_aei_examp}), thus implying that a 20 $M_{\oplus}$ planetesimal disk provides our six planet configuration with sufficiently strong encounters to drive Uranus and Neptune towards their modern orbits in the residual migration phase.

\begin{figure}
	\centering
	\includegraphics[width=.49\textwidth]{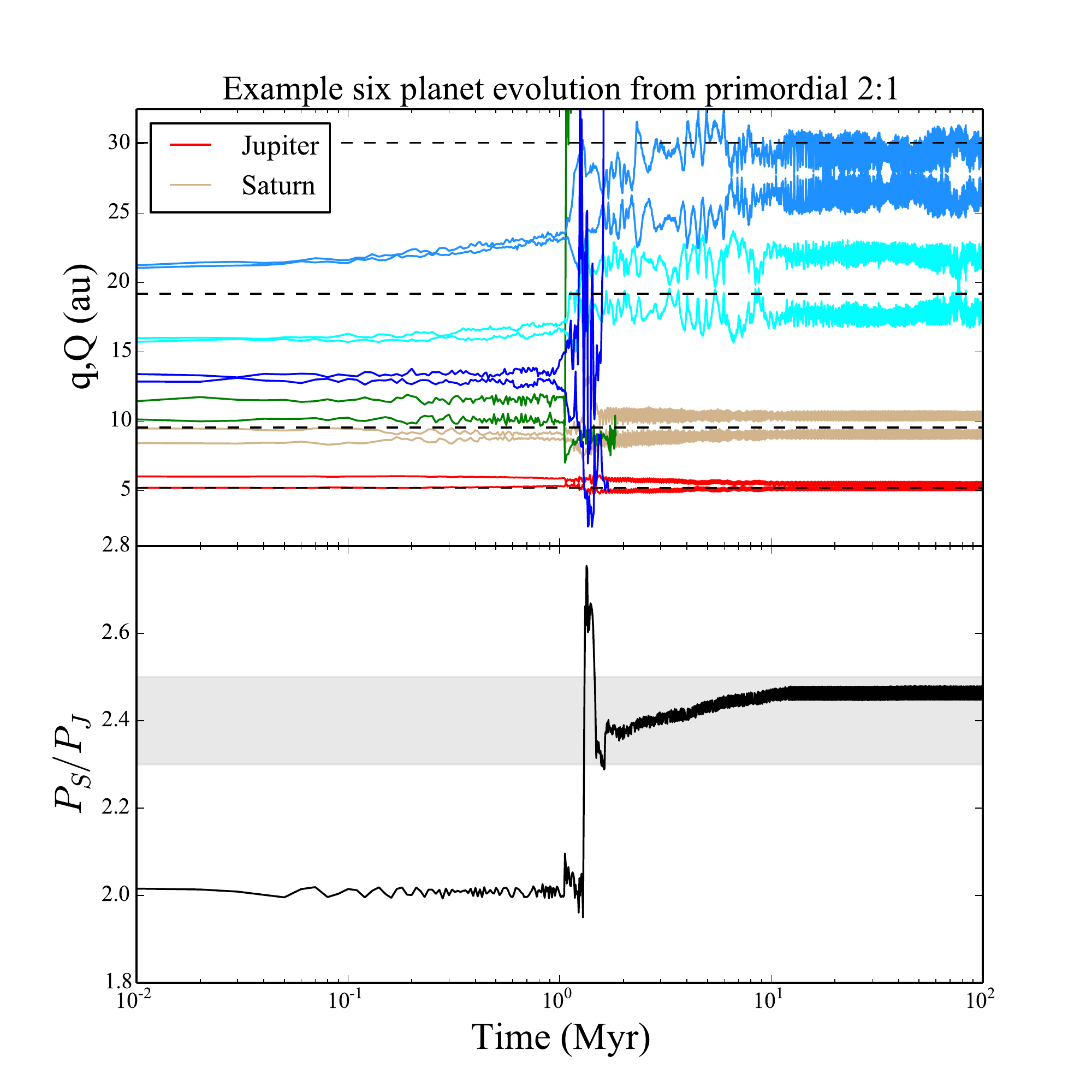}
	\caption{Example instability evolution beginning from the 2:1,4:3,4:3,4:3,3:2 resonant chain (figure \ref{fig:6_3ig_e55}).  The simulation finished with $P_{S}/P_{J}=$ 2.46, $M_{55}=$.029, $M_{56}=$.022, $M_{65}=$.022, $M_{66}=$.068 (all four success criteria are satisfied).  The top panel plots the perihelion and aphelion of each planet over the length of the simulation.  The bottom panel shows the Jupiter-Saturn period ratio.  The horizontal dashed lines in the upper panel indicate the locations of the giant planets' modern semi-major axes.  The shaded region in the lower panel delimits the range of 2.3 $<P_{S}/P_{J}<$ 2.5.}
	\label{fig:6gp_good}
\end{figure}

Overall, our six planet configurations are similar to the five planet cases in terms of their ability to reproduce important aspects of the Jupiter-Saturn (e.g. figures \ref{fig:5gp_modes_examp} and \ref{fig:6gp_modes_examp}).  However, their tendency to experience small jumps and finish with the correct number of planets makes the six planet, 2:1 instability compelling.  Indeed, five of 171 simulations in our 2:1,4:3,4:3,4:3,3:2, $e_{J,o}=e_{S,o}=$ 0.05 satisfy all four of our success criteria.  An example of such a successful simulation is plotted in figure \ref{fig:6gp_good}.

\subsection{Jupiter versus Saturn's primordial eccentricity}
\label{sect:ecc_results}

Our results \citep[in comparison to previous study of the 2:1:][]{nesvorny12,deienno17} clearly illustrate that primordial eccentricity excitation is necessary for the 2:1 Jupiter-Saturn resonance to be viable \citep{pierens14}. It is worthwhile to point out that our simulations do not indicate a significant dependency of successful outcomes on the particular choice of $e_{J,o}$ or $e_{S,o}$.  This is not extremely surprising, given the stochasticity of an event like the Nice Model, it is reasonable to expect small differences in initial conditions to be largely erased during the violent event.  However, it is readily apparent from table \ref{table:results} that higher initial values of $e_{J,o}$ tend to manifest as improved rates of success for criterion \textbf{C} (the most difficult constraint to satisfy) after systems are evolved through the Nice Model instability.  In comparison, \citet{deienno17} also studied the 2:1,4:3,3:2,3:2 configuration without including primordial excitation and reported a 0$\%$ success rate for exciting $M_{55}$ to greater than half the modern value.  Thus, it is clear that \textit{some degree of primordial excitation for Jupiter} is critical for achieving successful instability outcomes from the primordial 2:1 Jupiter-Saturn resonance.  However, we find that higher eccentricities for Saturn lead to systematically larger jumps, and correspondingly low success rates for criterion \textbf{D}.  Therefore, our simulations indicate that a moderate value of $e_{J,o}$ ($\gtrsim$0.05) and a low-moderate value of $e_{S,o}$ ($\lesssim$0.05) are most advantageous in terms of boosting success probabilities. A notable caveat of this result is that such a combination of eccentricities is only produced from a specific combination of disk parameters in hydrodynamical models.  \citet{pierens14} found that, in most cases, Saturn attains a higher eccentricity than Jupiter when the planets are captured in the mutual 2:1 MMR.  However, the authors reported an outcome with $e_{J,o} > e_{S,o}$ for their $\alpha=$ 0.01, $f=$ 0.3 disk (figure 4 in that paper).  While the cases we test that consider $e_{J,o} < e_{S,o}$ do produce successful simulations (table \ref{table:results}), the outcomes of our more successful sets of initial conditions ($e_{J,o} \gtrsim e_{S,o}$) should be taken in the appropriate context given that hydrodynamical models typically yield the opposite combination.

On the higher side of the range of possible primordial eccentricities for Jupiter, we were unable to generate stable chains with $e_{J,o}\gtrsim$0.08 by exciting the planet's orbits with artificial forces \citep[recall that][found that Jupiter and Saturn can attain eccentricities as high as $\sim$0.20 during the gas disk phase when trapped in the 2:1 MMR]{pierens14}.  Thus, our work does not consider eccentricities as high as proposed by \citet{pierens14}.  However, if the instability indeed occurred early (appendix \ref{sect:timing}), there is no strict requirement that the primordial giant planet configuration be stable in the absence of nebular gas.  In such a scenario, the instability would ensue as soon as the gas density is no longer high enough to prevent the planets from strongly perturbing one another.  This series of events might also be advantageous for stunting Mars' growth \citep{clement18}.

In summary, our results broadly indicate that the 2:1 Jupiter-Saturn resonance with primordial eccentricity pumping \citep{pierens14} is a compelling alternative to the 3:2 because it more consistently generates the modern Jupiter-Saturn system (specifically excites $M_{55}$) without exceeding $P_{S}/P_{J}=$ 2.5.  The most significant systematic issues with our simulations are (1) inaccurate final ice giant semi-major axes, (2) a preponderance of strong instabilities that tend to eject too many planets, and (3) total integration times that are inadequate to fully capture the residual migration phase.  As (1) is likely related to our initial selection of planetesimal disk mass and (2) is probably a result of the massive innermost ice giants beginning the simulation on nearly overlapping orbits, we conclude our study by varying $M_{KB}$ and $M_{IG}$ in an additional suite of simulations.  The following two sections discuss the results of this follow-on set of integrations.

\subsection{Varying the planetesimal disk's mass}
\label{sect:kb_mass}

\begin{table*}
	\centering
	\begin{tabular}{c c c c c c c c c c}
	\hline
	$N_{pln}$ & Resonant Chain & $e_{J,o}$ & $e_{S,o}$ & $M_{KB}$ ($M_{\oplus}$) & \textbf{A} & \textbf{B} & \textbf{C} & \textbf{D} & \textbf{ALL} \\
	\hline
	5 & 2:1,4:3,3:2,3:2 & 0.05 & 0.05 & 20.0 & 31 & 20 & 9 & 17 & 2 \\
	& & & & 40.0 & 58 & 31 & 10 & 23 & 0 \\
	& & & & 10.0 & 14 & 14 & 0 & 0 & 0 \\
	\hline
	6 & 2:1,4:3,4:3,3:2,3:2 & 0.05 & 0.025 & 20.0 & 60 & 28 & 6 & 37 & 1 \\
	& & & & 40.0 & 57 & 24 & 11 & 23 & 0 \\
	& & & & 10.0 & 41 & 22 & 9 & 29 & 3 \\
	\hline
	\end{tabular}
	\caption{Summary of results for our simulations that vary the initial planetesimal disk mass.  The columns are as follows: (1) the initial number of giant planets, (2) the resonant chain, beginning with the Jupiter-Saturn resonance, (3-4) the initial eccentricities of Jupiter and Saturn, (5) the mass of the external planetesimal disk, (6) the percentage of systems satisfying criterion \textbf{A} ($N_{GP}=$4), (7) criterion \textbf{B} (the planets' final semi-major axes within 20$\%$ of the real ones), (8) criterion \textbf{C} ($|\Delta M_{ij}/M_{ij,ss}|<$0.50 ($i,j=$5,6), $M_{55}>M_{56}$), (9) criterion \textbf{D} ($P_{S}/P_{J}<$2.5), and (10) the percentage of systems satisfying all four success criteria simultaneously.}
	\label{table:results_kb}
\end{table*}

To study the dependence of our results on $M_{KB}$, we perform four additional batches of 200 simulations based off our most successful five (2:1,4:3,3:2,3:2, $e_{J,o}=e_{S,o}=$ 0.05, see figures \ref{fig:5gp_modes_examp} and \ref{fig:5gp_aei_examp}) and six (2:1,4:3,4:3,3:2,3:2, $e_{J,o}=$ 0.05, $e_{S,o}=$ 0.025, see figures \ref{fig:6gp_modes_examp} and \ref{fig:6gp_aei_examp}) planet configurations.  We conduct 200 simulations for each set of initial conditions where we utilize $M_{KB}=$ 40.0 $M_{\oplus}$, and 200 runs that study 10.0 $M_{\oplus}$ disks.  These simulations are performed in the exact same manner (i.e.: simulation time, time-step, etc.) as our initial simulations described in section \ref{sect:methods}.  Table \ref{table:results_kb} summarizes the results of these additional simulations, and the percentages of systems that satisfy each of our success criteria, \textbf{A}-\textbf{D}

\begin{figure*}
	\centering
	\includegraphics[width=.45\textwidth]{5gp_2_1_05_05_1ig_aei.pdf}
	\includegraphics[width=.45\textwidth]{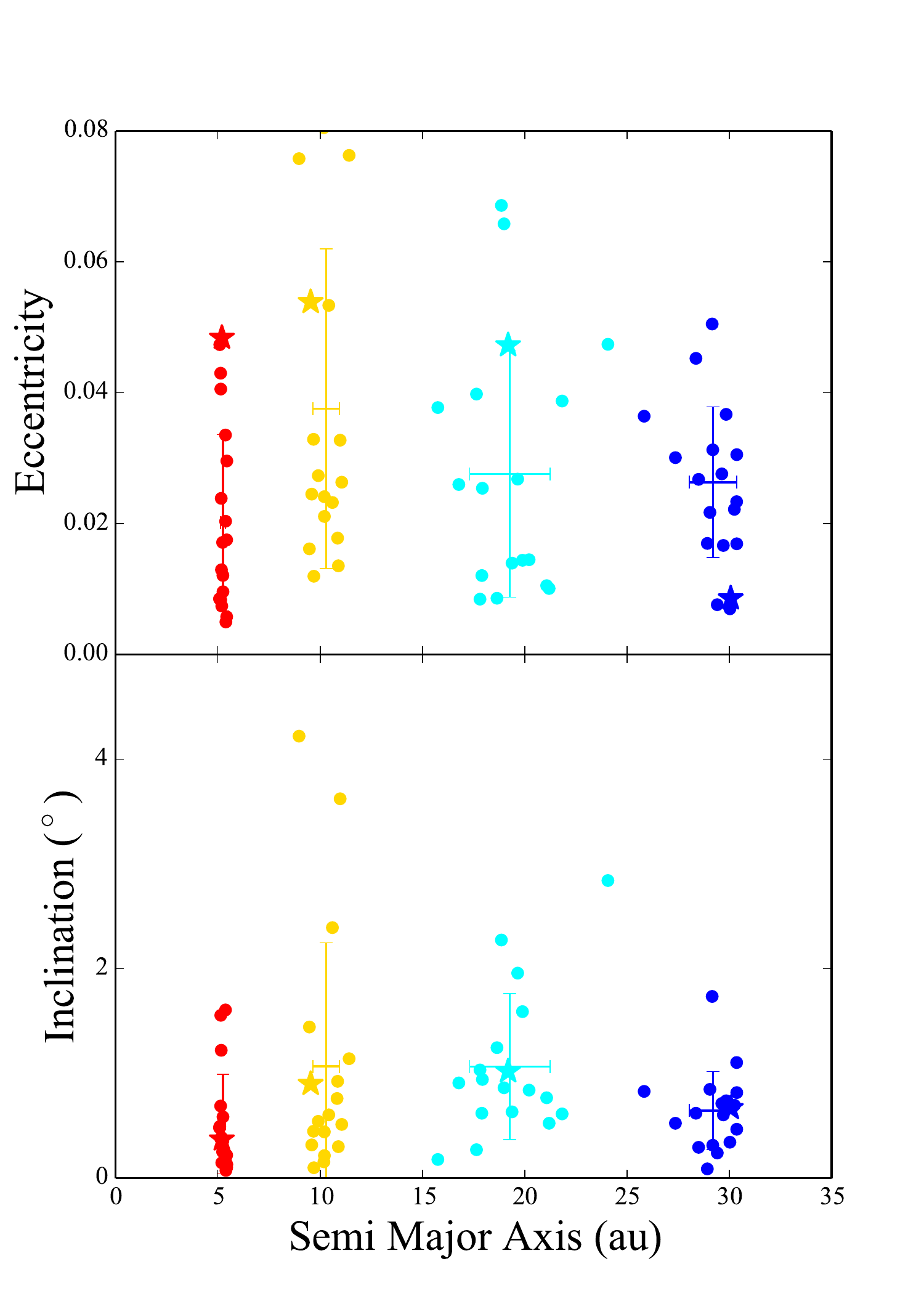}
	\caption{\textbf{Left Panel:} Reproduction of figure \ref{fig:5gp_aei_examp} for comparison.  \textbf{Right Panel:} Final orbits for systems that satisfied criteria \textbf{A} and \textbf{B} in our batch of simulations beginning with five planets in a 2:1,4:3,3:2,3:2 resonant chain, $e_{J,o}=e_{S,o}=$0.05 and a disk with $M_{KB}=$ 40.0 $M_{\oplus}$ (the same initial conditions as in the left panel except for $M_{KB}$ being doubled).  The top panel depicts $a/e$ space and the bottom panel plots $a/i$ space.  The respective planets and their simulated analogs are color coded as follows: Jupiter in red, Saturn in gold, Uranus in light blue, Neptune in dark blue.  The error bars indicate one standard deviation.  The stars correspond to solar system values.}	
	\label{fig:5gp_aei_high}
\end{figure*}

As we speculated throughout the previous sections, heavier disk masses significantly boost the probability that five planet systems retain four planets (58$\%$ versus 31$\%$) with the proper semi-major axes (meet criterion \textbf{B}; 31$\%$ versus 20$\%$).  Figure \ref{fig:5gp_aei_high} depicts how the final orbits for these criteria \textbf{A} and \textbf{B} satisfying systems provide a much better match to the actual solar system than our corresponding $M_{KB}=$ 20.0 $M_{\oplus}$ systems (figure \ref{fig:5gp_aei_examp}).  In contrast, a lighter, 10.0 $M_{\oplus}$ primordial Kuiper belt produced systematically poor solar system analogs.  Specifically, none of these systems satisfied criterion \textbf{C} or \textbf{D}.

While the success rates for each individual criterion are improved in our simulations considering a heavier disk (table \ref{table:results_kb}), no system satisfies all four criteria simultaneously.  A major reason for this is that interactions with the remnant planetesimal disk cause Saturn to migrate as well.  When an instability is mild enough to retain four giant planets, it also leaves behind enough mass in the Kuiper belt to drive Saturn past $P_{S}/P_{J}=$ 2.5.  15$\%$ of all runs in our original, $M_{KB}=$ 20.0 $M_{\oplus}$ batch satisfied criterion \textbf{A} and \textbf{D} simultaneously ($N_{GP}=$ 4 and $P_{S}/P_{J}<$ 2.5).  Conversely, with a more massive primordial Kuiper belt, this subset is only 9$\%$ of the total number of simulations.  Thus, while more systems in our new, $M_{KB}=$ 40.0$M_{\oplus}$ batch finish with Jupiter and Saturn inside of their mutual 5:2 resonance, such successful outcomes occur preferentially in systems that experience violent instabilities, lose too many planets, eject the majority of the primordial Kuiper belt objects, and leave behind a planetesimal disk with a total mass that is insufficient to drive any appreciable residual migration for Saturn.  Therefore, the vast majority of systems that satisfy both \textbf{A} and \textbf{B} also fail \textbf{D}.  Indeed, only one of our simulations considering a heavier planetesimal disk meets these three constraints simultaneously.  Interestingly, the same phenomenon is responsible for the increased success of our six planet configurations that utilize $M_{KB}=$ 10.0 $M_{\oplus}$.  While these systems achieve lower rates of success for criteria \textbf{A}-\textbf{C} than the $M_{KB}=$ 40.0 $M_{\oplus}$ set, the lighter disk tends to curtail Saturn's residual migration.  This manifests as more systems satisfying criterion \textbf{D}, and a larger percentage of simulations (3$\%$) satisfying all four success criteria simultaneously.  Though the individual rates of success are worse for most of our criteria in our $M_{KB}=$ 10.0 $M_{\oplus}$ simulations, the mutual exclusivities between the various constraints are lessened.  Specifically, only three of our nominal runs ($M_{KB}=$ 20.0 $M_{\oplus}$) satisfy the subsets of (\textbf{A},\textbf{C},\textbf{D}) or (\textbf{A},\textbf{B},\textbf{D}) simultaneously.  However, twice as many of our light disk systems are successful in this manner; thus leading to a greater number of systems satisfying all constraints simultaneously.  

In summary, within our tested parameter space, we are unable to identify an ``ideal'' value of $M_{KB}$ for our preferred five planet configuration.  Specifically, the selection of initial disk mass represents a compromise between boosting the probability of matching Uranus and Neptune's modern semi-major axes with a heavier primordial belt, and adequately curtailing Saturn's residual migration with a lower initial value of $M_{KB}$.  In contrast, our additional simulations indicate that a lighter primordial Kuiper belt ($\simeq$ 10.0 $M_{\oplus}$) might be advantageous for the six planet case.

\subsection{Varying the innermost ice giants' masses}
\label{sect:ig_mass}

\begin{table*}
	\centering
	\begin{tabular}{c c c c c c c c c c}
	\hline
	$N_{pln}$ & Resonant Chain & $e_{J,o}$ & $e_{S,o}$ & $M_{IG}$ ($M_{\oplus}$) & \textbf{A} & \textbf{B} & \textbf{C} & \textbf{D} & \textbf{ALL} \\
	\hline
	5 & 2:1,4:3,3:2,3:2 & 0.05 & 0.05 & 16.0 & 31 & 20 & 9 & 17 & 2 \\
	& & & & 24.0 & 14 & 10 & 5 & 8 & 1 \\
	& & & & 8.0 & 69 & 55 & 10 & 56 & 4 \\
	\hline
	6 & 2:1,4:3,4:3,3:2,3:2 & 0.05 & 0.025 & 8.0 & 60 & 28 & 6 & 37 & 1 \\
	& & & & 12.0 & 37 & 8 & 8 & 11 & 1 \\
	& & & & 6.0 & 60 & 37 & 10 & 57 & 2 \\
	\hline
	\end{tabular}
	\caption{Summary of results for runs that vary the inner ice giant masses.  The columns are as follows: (1) the initial number of giant planets, (2) the resonant chain, beginning with the Jupiter-Saturn resonance, (3-4) the initial eccentricities of Jupiter and Saturn, (5) the mass of the innermost ice giant(s), (6) the percentage of systems satisfying criterion \textbf{A} ($N_{GP}=$4), (7) criterion \textbf{B} (the planets' final semi-major axes within 20$\%$ of the real ones), (8) criterion \textbf{C} ($|\Delta M_{ij}/M_{ij,ss}|<$0.50 ($i,j=$5,6), $M_{55}>M_{56}$), (9) criterion \textbf{D} ($P_{S}/P_{J}<$2.5), and (10) the percentage of systems satisfying all four success criteria simultaneously.}
	\label{table:results_ig}
\end{table*}

While multiple previous authors have investigated the effects of varying $M_{KB}$ \citep[e.g.:][]{nesvorny12,batbro12}, similar attempts to fine-tune instability statistics by altering the ejected planet's mass ($M_{IG}$) are conspicuously absent from the literature.  \citet{nesvorny11} mentioned experimenting with higher masses, but eventually concluded that the resulting stronger scattering events systematically lead to excessively violent instabilities.  Subsequent work by \citet{nesvorny12} speculated that it might be possible to calibrate instability results by adjusting $M_{IG}$, and favored a lower mass for the ejected planet ($M_{IG}=$8.0 $M_{\oplus}$) in 2:1 cases because of the improved rates of success for criterion \textbf{D}.  Indeed, the selection of a Neptune-mass additional planet throughout the majority of the contemporary literature is largely based off the finding that an encounter with a $M_{IG}=$ 15.0 $M_{\oplus}$ planet is required to consistently excite Jupiter's primordially circular orbit \citep{morby09}.  As our scenario considers non-zero initial eccentricities for the gas giants, alternative values of $M_{IG}$ might prove to be advantageous.  To test this hypothesis, we perform an additional array of 800 simulations that vary the innermost ice giants' masses.  In the same manner as our study of different $M_{KB}$ values, we take our most successful five and six planet configurations as a starting point (table \ref{table:results_ig}).  For five planet architectures, where our nominal simulations evaluate $M_{IG}=$ 16.0 $M_{\oplus}$, we perform 200 simulations that test a mass of 8.0 $M_{\oplus}$, and an additional 200 that utilize $M_{IG}=$ 24.0 $M_{\oplus}$.  In the six planet case (nominal mass of 8.0 $M_{\oplus}$), we perform simulations testing $M_{IG}=$ 6.0 and 12.0 $M_{\oplus}$.

The results of this additional batch of integrations are summarized in table \ref{table:results_ig}.  Our new simulations clearly indicate that lighter-mass ejected ice giants improve instability statistics, while higher values of $M_{IG}$ lead to reduced success rates for criterion \textbf{A}, \textbf{B} and \textbf{D}.  Broadly speaking, a more massive planet generates a stronger scattering event and typical final $M_{55}$ values that are correspondingly higher.  Indeed, for the five planet case, 68$\%$ of our systems that consider a more massive additional planet finish with $M_{55}>$ 0.022.  Comparatively, this is only achieved in 63$\%$ and 39$\%$ of our simulations considering $M_{IG}=$ 16.0 and 8.0 $M_{\oplus}$, respectively.  Our six planet configurations yield similar results.  Specifically, 73$\%$, 47$\%$ and 35$\%$ of these systems attain $M_{55}>$ 0.022 for $M_{IG}=$ 12.0, 8.0 and 6.0 $M_{\oplus}$, respectively.  While the instabilities considering more massive planets are often increasingly violent, and therefore have greater potential to adequately excite the secular modes of the Jupiter-Saturn system, this advantage is outweighed by these systems' systematically lower success rates for criterion \textbf{A}, \textbf{B} and \textbf{D} (though several systems still manage to satisfy all four criteria simultaneously; table \ref{table:results_ig}).

As pointed out in the previous discussion, perhaps the most challenging set of constraints for our primordially excited resonant chains to meet simultaneously are \textbf{A} and \textbf{D} (retain four planets and not exceed $P_{S}/P_{J}=$ 2.5).  While only a few percent of the majority of our simulation sets are successful in this regard, 52$\%$ of our five planet configurations that utilize $M_{IG}=$ 8.0 $M_{\oplus}$ meet both metrics simultaneously.  Thus, by greatly increasing the likelihood of Jupiter and Saturn maintaining semi-major axes interior to their mutual 5:2 resonance, our systems testing the lowest values of $M_{IG}$ yield markedly improved rates of satisfying criteria \textbf{A}-\textbf{D} concurrently.  Therefore, future study of the 2:1 Jupiter-Saturn resonance with primordial eccentricity pumping should investigate ejected ice giants with lower masses than those considered in the bulk of our present manuscript.

\section{Discussion}
\label{sect:discuss}

The main conclusion of our study is not to disprove the viability of the primordial 3:2 Jupiter-Saturn resonance, or the work of previous studies.  Indeed, instabilities' originating from the 3:2 still yield reasonable success rates (even when scrutinized against our updated, stricter constraints: table \ref{table:success}).  However, we have shown that it is systematically challenging to limit Saturn's semi-major axis jump, and post-instability eccentricity when the two gas giants are born in closer proximity to one another.  This problem is obviously lessoned when Jupiter and Saturn emerge from the nebular disk phase in the wider, 2:1 MMR.  If the two planets also acquire elevated eccentricities during the gas disk phase by virtue of having carved larger gaps in the disk's radial profile \citep{pierens14}, systematic challenges with exciting the giant planet's secular modes in 2:1 Jupiter-Saturn instabilities \citep[e.g.:][]{nesvorny12,deienno17} might also be alleviated. 

It is worthwhile to acknowledge that, in spite of all efforts made, certain aspects of giant planets' modern architecture remain low-probability outcomes in our favored scenarios.  While our work uncovers a potential scheme for replicating the Jupiter-Saturn system's secular architecture with a moderately improved likelihood, such successful systems often fail to retain two ice giants and adequately replicate their modern orbits.  For this reason, many of the Jupiter analogs in the $N_{GP}=$ 4 systems plotted in figures \ref{fig:5gp_aei_examp}, \ref{fig:6gp_aei_examp} and \ref{fig:5gp_aei_high} tend to possess $M_{55}$ values closer to our lower limit of 0.022 than the modern value of 0.044.  While it is indeed important to avoid trading one low-probability outcome (e.g.: matching $M_{ij}$ for $i,j=$ 5,6, and $P_{S}/P_{J}$ in the same system with the primordial 3:2 Jupiter-Saturn resonance) for another, we argue that a successful scenario for producing the gas giants' orbital architecture is perhaps more compelling for a number of reasons.  Most significantly, in simulations where $M_{55}$ is matched, our 2:1 cases tend to yield significantly improved orbits for Saturn (namely lower values $e_{S}$ and $P_{S}/P_{J}$) than our best 3:2 sets.  Conversely, evolutions beginning from the primordial 3:2 Jupiter-Saturn resonance often possess over-excited Saturn analogs when Jupiter's eccentricity is adequately excited.  This is evidenced by low success rates for criterion \textbf{C} for our 3:2 simulations, and the distributions of eccentric magnitudes for our 2:1 batches plotted in figures \ref{fig:5gp_modes_examp} and \ref{fig:6gp_modes_examp}. Additionally, we have shown that successful replication of the ice giant system can be fine-tuned by varying $M_{IG}$ and $M_{KB}$.  The ice giant's final semi-major axes (a problem in some of our sets of initial conditions) are sensitive to the range of migration that occurs \textit{prior to} the instability.  As our study utilizes an artificial instability trigger to minimize computing time, we do not evaluate the full spectrum of possible primordial migration paths.  Furthermore, tighter resonances between the innermost ice giant and Saturn (e.g. 5:4 or 6:5) might provide additional fidelity in favorably modifying instability statistics.  Finally, the specific number of primordial ice giants is also a free parameter.  We find that our most successful outcomes occur in configurations that maximize the dynamical spacing between the eventual Uranus and Neptune analogs and the chaos that ensues when Jupiter and Saturn interact with the additional ice giants. Therefore, it is possible that a configuration that includes a larger number of less massive additional planets in a compact resonant chain located between the gas giants and modern ice giants might also be successful.

As discussed in section \ref{sect:ecc_results}, our work is limited to the study of chains with initial gas giant eccentricities $\lesssim$0.05 because we require our configurations exhibit stability in the absence of an external planetesimal disk.  However, there is no reason to enforce such a constraint on the actual solar system.  If the high-eccentricity resonant chain of primordial giant planets simply was not self-stable, the instability would ensue rapidly; once the gas density was low enough such that the planets began to strongly perturb one another.  From a philosophical standpoint, such a scenario might be preferable as it does not invoke additional generations of stability.  While we leave the complete investigation of higher-eccentricity configurations to future work, there are other reasons to prefer an extremely early instability.  A major potential pitfall of the primordial 2:1 version of the Nice Model presented in this manuscript is its effects on the asteroid belt and terrestrial-forming regions.  Specifically, a number of our systems spend a significant amount of time inhabiting the $P_{S}/P_{J}\simeq$ 2.1-2.3 regime where the $g_{1}=g_{5}$ and $g_{2}=g_{5}$ secular resonances would be encountered.  This may not be particularly consequential if the instability occurred early \citep[appendix \ref{sect:timing}:][]{izidoro16,morb18,nesvorny18,clement18,deienno18}.  However, follow-on studies must throughly investigate the consequences of our proposed scenario on the inner solar system.

Finally, it is clear that our proposed scenario represents somewhat of a paradigm shift; particularly given that the original motivation for the Nice Model's development was to provide a mechanism for exciting the giant planets' eccentricities.  If the giant planets were indeed ``born eccentric,'' it would be logical to question whether an epoch of instability is necessary to explain the solar system's dynamical state.  However, we strongly assert that our work should not be interpreted in such a manner.  While we have shown that planetary encounters within the instability might not be the sole source of the giant planets' modern eccentricities, the prevalence of irregular moons in the outer solar system and the asteroid belt's dynamical structure \citep[among other aspects:][]{nesvorny18_rev} strongly conflict with alternative giant planet migration scenarios \citep[i.e. smooth migration:][]{morb09,walshmorb11}.

\section{Conclusions}
\label{sect:conclude}

We presented a statistical analysis of more than 6,000 simulations of the Nice Model instability.  Our work investigated the possibility that Jupiter and Saturn inhabited a 2:1 MMR with inflated eccentricities \citep{pierens14} around the time of nebular gas dispersal.  Our investigation of the primordial 2:1 Jupiter-Saturn resonance was partially motivated by a detailed analysis of issues with the 3:2 version of the Nice Model typically invoked in the literature \citep[e.g.:][]{bat10,nesvorny11,batbro12,nesvorny12,deienno17,clement18} that we argued might be systematic problems.  Specifically, there is a strong anti-correlation between the adequate excitation of Jupiter's $M_{55}$ eccentric mode and maintaining the gas giant's period ratio less than 2.5 when the planets' begin in the more compact, 3:2 MMR.  We used an updated series of success metrics (table \ref{table:success}) that fully evaluate all the important eigenmodes of the Jupiter-Saturn secular system to show that the 2:1 version of the Nice Model improves the probability of proper replication of these qualities of the solar system.  However, a caveat of our findings is that our favored combinations of eccentricities for Jupiter and Saturn ($e_{J,o}>e_{S,o}$) are slightly at odds with the results of most hydrodynamical studies \citep[][although we find reasonable success rates for the opposite combination]{pierens14}.  While our simulations also indicated a minor anti-correlation between acceptable outcomes for the gas and ice giants, we used an additional suite of simulations to show that certain instability statistics related to Uranus and Neptune can be fine-tuned by varying the masses of the primordial Kuiper belt and innermost ice giants.  Though our work shows that the primordial 2:1 Jupiter-Saturn resonance is a viable evolutionary path for the solar system, future work is still required to fully validate our presumed initial conditions, and robustly analyze the consequences of such a scenario on the solar system's fragile populations of small bodies.  In particular, follow-on investigation of the giant planets' instability evolution with Jupiter and Saturn in a primordial 2:1 MMR with enhanced eccentricities must consider longer integration times ($\gtrsim$100 Myr), higher resolution disks ($\gtrsim$10,000 particles), and account for the dissipating gaseous nebula.

\section*{Acknowledgments}

The authors are grateful for informative discussions with Jaehan Bae, and insightful reviews of the manuscript provided by Ramon Brasser and an anonymous referee.  N.A.K. thanks the National Science Foundation for support under award AST-1615975 and NSF CAREER award 1846388.  S.N.R. acknowledges support from the CNRS’s PNP program and NASA Astrobiology Institute's Virtual Planetary Laboratory Lead Team, funded via the NASA Astrobiology Institute under solicitation NNH12ZDA002C and cooperative agreement no. NNA13AA93A.  R.D. acknowledges support from the NASA SSW program.  A.I. acknowledges NASA grant 80NSSC18K0828 to Rajdeep Dasgupta, during preparation and submission of the work.  The majority of computing for this project was performed at the OU Supercomputing Center for Education and Research (OSCER) at the University of Oklahoma (OU).  Some of the computing for this project was performed on the Memex cluster. We would like to thank Carnegie Institution for Science and the Carnegie Sci-Comp Committee for providing computational resources and support that contributed to these research results.  The authors acknowledge the Texas Advanced Computing Center (TACC) at The University of Texas at Austin for providing {HPC, visualization, database, or grid} resources that have contributed to the research results reported within this paper. URL: http://www.tacc.utexas.edu.
\bibliographystyle{apj}
\newcommand{\sci}{$Science$ }
\bibliography{e55.bib}

\appendix
\section{The timing of the instability}
\label{sect:timing}

As we neglect pre-instability migration, our work investigates an instability scenario that occurs rather quickly after nebular gas dissipation.  Therefore, it is worthwhile to comment on the various proposed timings for the instability. In its original formulation \citep{gomes05}, the Nice Model instability was proposed to have occurred coincident with the Late Heavy Bombardment \citep[LHB;][]{tera74}.  Thus, a \textit{late} instability implies a rather specific timing for the event; $\simeq$ 650 Myr after gas disk dispersal in order to provide a trigger for the cratering spike in the inner solar system as part of a LHB scenario.  However, it is worth noting that the timescale of giant planet migration is significantly shorter than 650 Myr for many sets of initial disk conditions.  Specifically, maintaining a system of giant planets stable in resonance for hundreds of Myr requires larger radial spacings between Neptune and the primordial Kuiper belt \citep{levison11,nesvorny12,deienno17}.  Furthermore, average instability delays in simulations that account for Kuiper belt self-gravity \citep{quarles19} are typically less than $\sim$100 Myr regardless of initial offset from Neptune.  In a similar manner, \citet{ribeiro20} found late instabilities to be almost non-existent in giant planet formation models that successfully generate the large obliquities of Uranus and Neptune via embryo-embryo impacts \citep{izidoro15_ig}.

While late instabilities do occur in numerical simulations, albeit less frequently than early destabilizations, authors in recent years have begun invoking earlier versions of the Nice Model for a variety of reasons.  Perhaps most significantly, the very existence of a cratering spike has been called into question as a result of updated isotopic dating methods used in the analysis of Lunar samples.  The refined basin ages \citep[many utilizing $^{40}Ar/^{39}Ar$ dataing; e.g.:][]{boehnke16} reported in contemporary studies seem to imply a smoother decline in bombardment, rather than a terminal cataclysm \citep{zellner17}.  These include, for example, $\sim$4.2 Gyr melt breccia sampled at North Ray crater on Apollo 16 \citep{norman06}, $\sim$3.92 Gyr melts in Apollo 12 soil samples \citep{liu12}, evidence of $\sim$1.4 and 1.9 Gyr resetting events on an Apollo 15 melt breccia \citep{grange13}, a more precise 3.938 Gyr age of the Imbrium impact from Ca-phosphate dating of Apollo 14 samples \citep{merle14}, and two Apollo 17 melt breccias indicating impact ages ranging from $\sim$3.8-3.3 Gyr \citep{mercer15}.   Moreover, it is now clear from high-resolution LRO (Lunar Reconnoissance Orbiter) imagery that the Serenitatis basin is both highly contaminated with Imbrium ejecta, and more densely cratered than the Imbrium basin \citep{spudis11}.  Thus, the Serenitatis event probably occurred significantly earlier than the impact that formed the Imbrium basin, and Apollo 17 samples that were thought to represent Serenitatis ejecta in the 1970s are likely representative of the 3.9 Gyr Imbrium event.  

From a dynamical standpoint, it is important to understand that these geochemical results do not imply that the Nice Model-envisioned evolution of the giant planets must have occurred early, rather that it need not be constrained by any specific timing.  However, in the past few years, several additional aspects of the solar system have been shown to be at odds with a late instability.  \citet{morb18} posited that the discrepancy between the highly siderophile element (HSEs: elements that should partition into iron during core formation and thus represent material delivered after core closure) inventories of the Earth and Moon are better explained by the slow crystallization of the Lunar magma ocean \citep{elkins11} than by a delayed cratering spike.  Additionally, \citet{nesvorny18} argued that the Patroclus-Menoetius binary of Jupiter trojan asteroids would not have survived a delayed version of the Nice Model.  In a similar manner, two known collisional families in the asteroid belt \citep{delbo17,delbo19} with determined ages $\gtrsim$4.0 Gyr seem to imply that the instability occurred early.  In addition, Mercury or Mars are often ejected or over-excited in simulations of  a late instability \citep{bras09,agnorlin12,kaibcham16}.  This problem disappears if the instability took place in conjunction with terrestrial planet formation: instead of destroying the already-formed inner planets, the instability would therefore have played a role in sculpting their masses and orbits \citep{clement18_frag}.  Indeed, the Early Instability model is broadly consistent with the small mass and rapid formation of Mars \citep{clement18}, the compositional evolution of the respective planets \citep{mojzsis19,brasser20}, as well as the total mass \citep{clement18_ab} and dynamical structure \citep{deienno18} of the asteroid belt.

\section{Secular evolution of the solar system}
\label{sect:secular}

The secular dynamics of a system can be studied when the planets are sufficiently far from MMR by expanding the mutual interaction components to order one in mass \citep{michtchenko04}:
\begin{equation}
	\mathcal{H} = \mathcal{H}_{0} + \epsilon \mathcal{H}_{1}
\end{equation}
Here, $\mathcal{H}_{0}$ is the integrable approximation, $\mathcal{H}_{1}$ is the secular normal form, and $\epsilon$ represents the \textit{order} of the perturbation.  The standard procedure involves writing the secular Hamiltonian in terms of the modified Delaunay variables, assuming the planets' eccentricities and inclinations are small, and expanding in Taylor series.  For complete derivations, we direct the reader to chapter 7 of \citet{dermott99} or chapter 7 of \citet{morby02}.  The so-called Lagrange-Laplace solution is as follows:
\begin{equation}
    \begin{array}{l}
    e_{i}\cos{\varpi_{i}} = \sum_{j}^{8}M_{ij}\cos{(g_{j}t + \beta_{j})} \\
    e_{i}\sin{\varpi_{i}} = \sum_{j}^{8}M_{ij}\sin{(g_{j}t + \beta_{j})} \\
    \sin{\frac{i_{i}}{2}}\cos{\Omega_{i}} = \sum_{j}^{8}M_{ij}\cos{(s_{j}t + \beta_{j})} \\
    \sin{\frac{i_{i}}{2}}\sin{\Omega_{i}} = \sum_{j}^{8}M_{ij}\sin{(s_{j}t + \beta_{j})}
	\label{eqn:sec}
	\end{array}
\end{equation}
The fundamental frequencies $g_{j}$ and $s_{j}$ ($i=$ 1-8, with $s_{5}$ necessarily set to zero by convention) represent the dominant eigenfrequencies for the precession of perihelia and longitudes of nodes, respectively.  The solar system's frequencies, as well as their respective amplitudes in each planets' orbit, are typically calculated via Fourier analysis of the outputs of high-accuracy numerical simulations \citep[e.g.:][]{nobili89,laskar90,nesvorny96,laskar11,rein19}.  Tables \ref{table:gi} and \ref{table:Mij} summarize the current values of the dominant eigenfrequencies, $g_{j}$, and amplitudes, $M_{ij}$, in the outer solar system.

\begin{table}
	\centering
	\begin{tabular}{c c c c}
	\hline
	$g_{5}$ & $g_{6}$ & $g_{7}$ & $g_{8}$ \\
	\hline 
	4.257519 & 28.2449 & 3.087946 & 0.673019 \\
	\hline
	\end{tabular}
	\caption{Secular precession frequencies ($''/yr$) as calculated in \citet{laskar11}.}
	\label{table:gi}
\end{table}

\begin{table}
\centering
\begin{tabular}{c c c c c}
	\hline
	$_{j}\backslash^{i}$ & 5 & 6 & 7 & 8 \\
	\hline
	5 & 0.0441872 & 0.0329583 & -0.0375866 & 0.00188143 \\
	6 & -0.0157002 & 0.0482093 & -0.0015471 & -0.00010309 \\
	7 & 0.0018139 & 0.0015113 & 0.0290330 & -0.00369711 \\
	8 & 0.0000580 & 0.0000575 & 0.0016665 & 0.00911787 \\
	\hline
\end{tabular}
\caption{Amplitudes, $M_{ij}$ (equation \ref{eqn:sec}), for the dominant eigenmodes $g_{5}$-$g_{8}$ (rows) in the orbit in each of the giant planets (columns).  All values are reproduced from the work of \citep{nobili89}.}
\label{table:Mij}
\end{table}

\begin{figure}
 	\centering
 	\includegraphics[width=0.48\textwidth]{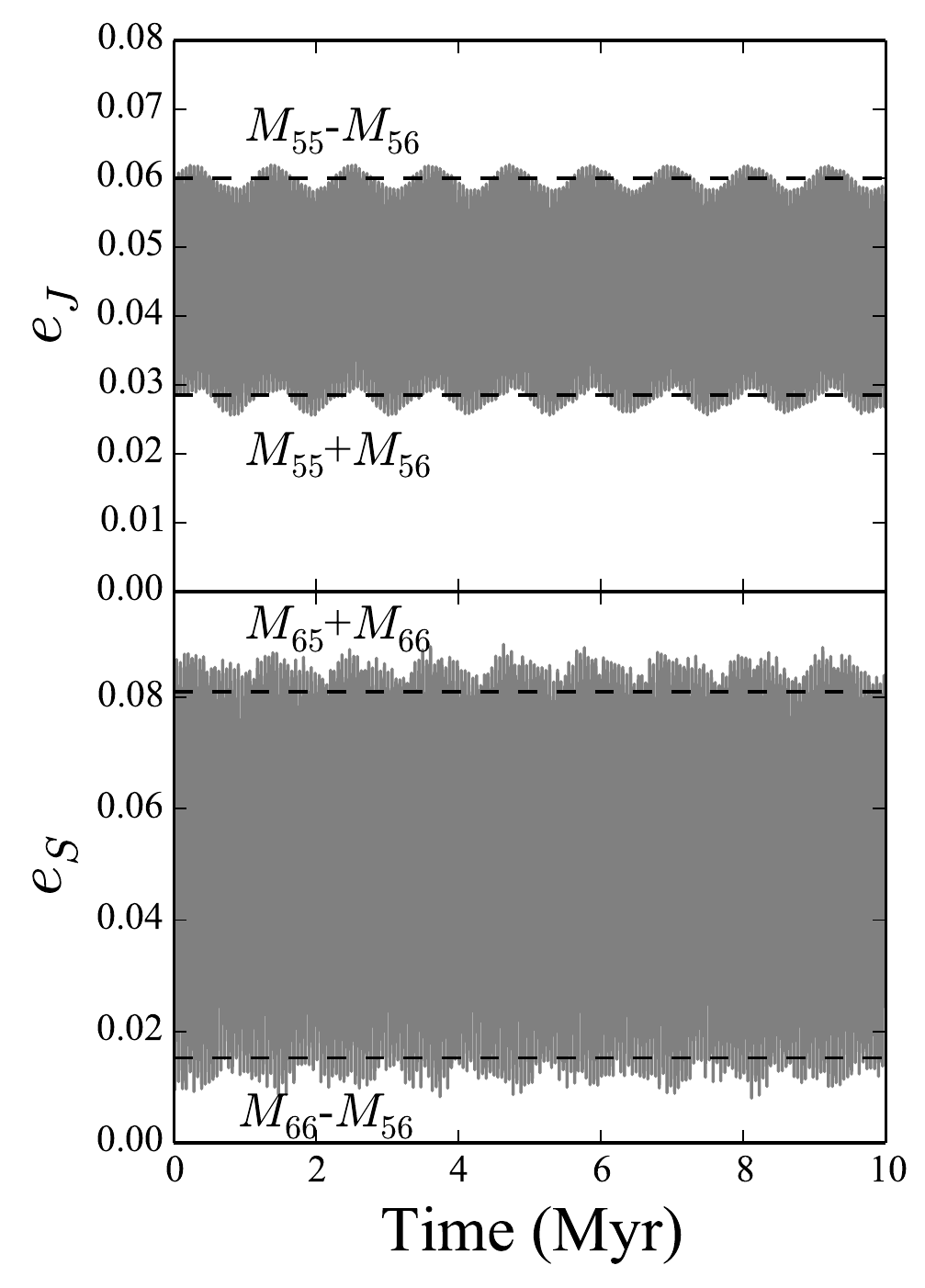}
 	\caption{Eccentricity variations in the orbits of Jupiter and Saturn from a 10 Myr integration of the Sun and 8 planets with the $Mercury6$ hybrid integrator \citep{chambers99}.  The dashed lines denote the magnitudes of the dominant $g_{5}$ and $g_{6}$ eigenfrequency in each planets orbit: $M_{55}$, $M_{56}$, $M_{65}$, and $M_{66}$.}
 	\label{fig:jupsat}
 \end{figure}

\begin{figure}
	\centering
	\includegraphics[width=0.48\textwidth]{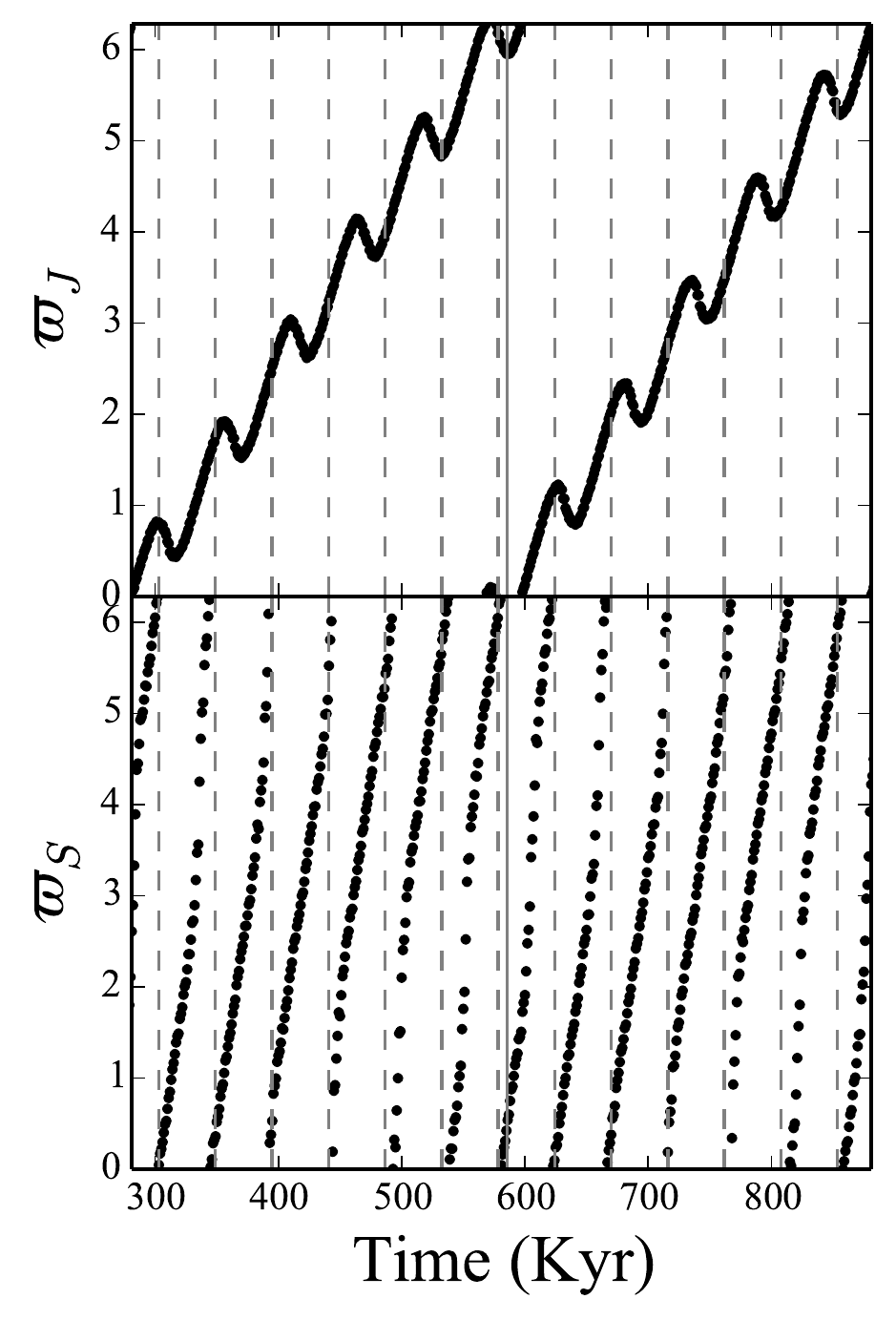}
 	\caption{Precession of Jupiter and Saturn's orbital longitude of perihelia ($\varpi$) in a numerical integration of all 8 planets.  The vertical lines are spaced in equal increments of $g_{5}$ (solid grey) and $g_{6}$'s (dashed grey) periods.}
 	\label{fig:jupsat2}
\end{figure}

The precise numerical values of the secular precession frequencies in the outer solar system ($g_{5}$-$g_{8}$) are largely a function of the orbital spacing between the respective planets \citep{dermott99,nesvorny12}.  However, the various planets' eccentric magnitudes, $M_{ij}$, were acquired via dynamical processes \citep{bras09,morby09}.  Therefore, the dominant secular eigenfrequencies, along with their respective amplitudes within the relevant planets' eccentricities and inclinations \textit{must be broadly reproduced in any successful evolutionary model}.  Given the hierarchical nature of the solar system's mass distribution, one can simplify the equations in \ref{eqn:sec} by studying only the Jupiter-Saturn secular problem for a simple approximation of the solar system's global secular dynamics \citep{morby09}.  Thus, the eccentricity evolution of the two gas giants can roughly be described by:
\begin{equation}
    \begin{array}{l}
    e_{J}\cos{\varpi_{J}} = M_{55}\cos{(g_{5}t + \beta_{5})} - M_{56}\cos{(g_{6}t + \beta_{6})} \\
    e_{J}\sin{\varpi_{S}} = M_{55}\sin{(g_{5}t + \beta_{5})} - M_{56}\sin{(g_{6}t + \beta_{6})} \\
    e_{S}\cos{\varpi_{J}} = M_{65}\cos{(g_{5}t + \beta_{5})} + M_{66}\cos{(g_{6}t + \beta_{6})} \\
    e_{S}\sin{\varpi_{S}} = M_{65}\sin{(g_{5}t + \beta_{5})} + M_{66}\sin{(g_{6}t + \beta_{6})} \\
    \end{array}
\end{equation}
In this Jupiter-Saturn approximation, the time evolution of each planet's eccentricity is given by:
\begin{equation}
	\begin{array}{l}
		e_{J}(t) = \sqrt{M_{55}^2 + M_{56}^2 - 2M_{55}M_{56}\cos{\big((g_{5}-g_{6})t - \beta_{56}}\big)} \\
		e_{S}(t) = \sqrt{M_{65}^2 + M_{66}^2 + 2M_{65}M_{66}\cos{\big((g_{5}-g_{6})t - \beta_{56}}\big)}
	\end{array}
	\label{eqn:js}
\end{equation}
where $\beta_{56} = \beta_{5} - \beta_{6}$.  Thus, noting that $M_{56}$ is negative, Jupiter's eccentricity oscillates between an approximate minimum of  $M_{55}$+$M_{56}\simeq$ 0.029 and a maximum of $M_{55}$-$M_{56}\simeq$ 0.060.  Similarly, the eccentricity of Saturn is bounded by $M_{65}$-$M_{66}=$ 0.015 and $M_{65}$+$M_{66}=$ 0.081, and the characteristic period of these oscillations is $\propto
 |g_{5}-g_{6}|^{-1} \simeq$ 54,000 years.  Figure \ref{fig:jupsat} depicts the eccentricity evolution of Jupiter and Saturn in a 10 Myr integration of the modern solar system.  While there are obvious lower-magnitude perturbations from the other planets (namely $g_{7}$; see table \ref{table:Mij}), this figure illustrates how the secular structure of the solar system is reasonably approximated by the 2 frequencies and 4 amplitudes of the Jupiter-Saturn system.  This is important because Jupiter is the most significant eccentric perturber for all the planets except Mars and Neptune (as quantified by the magnitude of $g_{5}$ in each planets' orbit).  In addition to driving variations in the planets' eccentricities, the dominant $g_{5}$ and $g_{6}$ frequencies provide a good first order approximation of Jupiter and Saturn's eccentricity vector precessions (figure \ref{fig:jupsat2}).
 
\section{Alternative avenues for adequately exciting $M_{55}$}
\label{sect:alternatives}

\subsection{The effect of approaching the 5:2 MMR}

\begin{figure}
	\centering
	\includegraphics[width=.49\textwidth]{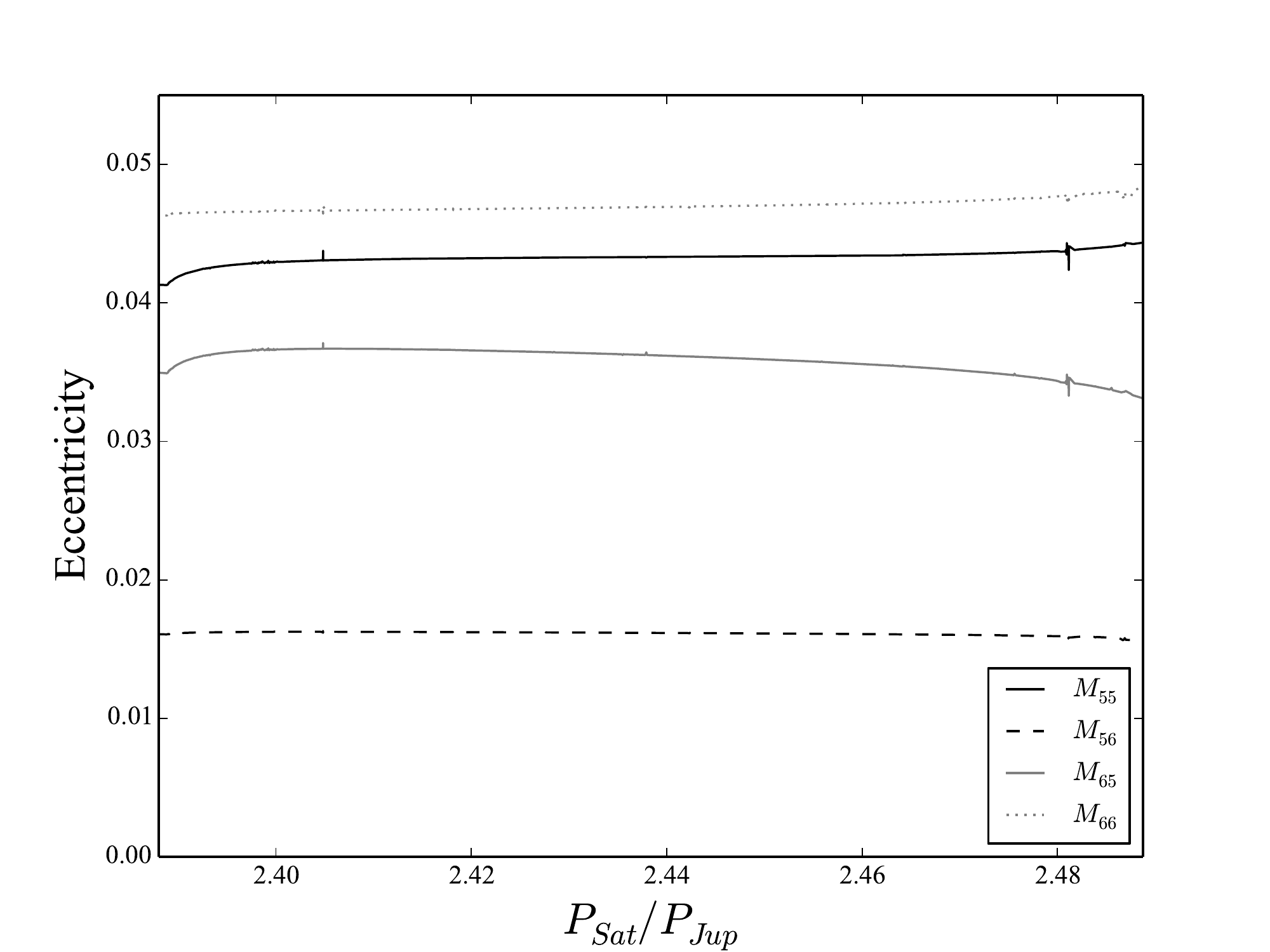}
	\caption{Evolution of the $M_{ij}$ (i,j$=$5,6) amplitudes as Jupiter and Saturn period ratio is adjusted.  This figure is generated by performing 3,200 integrations \citep[see][]{clement20_mnras} of the modern solar system with the $Mercury6$ hybrid integrator \citep{chambers99}.  In each run, Saturn's semi-major axis is decreased by 0.005 au such that $P_{S}/P_{J}$ varies from 2.39-2.49, and all other orbital elements are left unchanged.  Each system is integrated for 10 Myr, and the secular amplitudes and frequencies are calculated via Fourier analysis of the simulation time outputs \citep{nesvorny96}.  The figure is limited to the range of 2.39$<P_{S}/P_{J}<$2.49 that is just outside of the 3:1 Saturn-Uranus MMR where the eigenmodes' amplitudes evolve chaotically.}
	\label{fig:modes}
\end{figure}

We also explored the possibility that Jupiter's eccentricity can be further excited \textit{after} the instability occurs.  Indeed, the planets orbits evolve chaotically over Gyr timescales \citep[though this departure from quasi-periodic behavior is mostly confined to the orbits of the terrestrial worlds, e.g.:][]{laskar90}.  It is known that the giant planets continue to migrate smoothly over some radial range after the instability \citep{deienno17} as they continue to interact with and clear out debris in the vicinity of their orbits.  Through this process, the eccentricities and inclinations of Uranus and Neptune can damp rather substantially via secular friction \citep{nesvorny12}.  While this mechanism tends to damp the eccentric modes of the Jupiter/Saturn system as well, the planets' mutual secular interactions are known to enhance in close proximity to the 5:2 MMR \citep{clement20_mnras}.  Figure \ref{fig:modes} quantifies the degree to which $M_{55}$ can be further excited within this phase of approach.  In general, as Jupiter and Saturn's mutual interactions weaken with increased radial separation, their eccentric forcing on one another lessens correspondingly.  Thus, the magnitude of $g_{6}$ in Jupiter's eccentricity lowers, and so does $g_{5}$ in Saturn's.  This results in $M_{55}$ rising slightly relative to $M_{56}$, and $M_{66}$ increasing minimally with respect to $M_{65}$.  However, the net total change in $M_{55}$ is only a few percent over the range of mutual separations depicted in figure \ref{fig:modes}.  We conclude that $M_{55}$ likely cannot evolve appreciably after the instability via residual migration towards the 5:2 MMR \citep[the subsequent section addresses the possibility of a resonant crossing, see also][]{morb09}.

\subsection{A late dynamical event}

To search for potential avenues for the post-instability excitation of $M_{55}$, we integrate 371 system formed in \citet{clement18} for an additional 1 Gyr.  These systems, born out of the 3:2 mutual Jupiter-Saturn resonance (table \ref{table:gp}), finished with a range of final values for $P_{S}/P_{J}$ and $M_{55}$ (e.g.: figure \ref{fig:c18_19}).  \textit{It is important to note that many of these systems differ from perfect solar system analogs for a variety of reasons}. However, we intentionally study a wide range of systems that are similar to the solar system in order to conduct a broad exploration of possible additional excitation mechanisms for $M_{55}$.  After the complete inner and outer solar systems are integrated for 1 Gyr with the $Mercury6$ Hybrid integrator \citep[using a 6 day timestep,][]{chambers99}, we calculate the secular frequencies and amplitudes via frequency modulated Fourier Transform \citep{nesvorny96}.  Of these 371 systems, 22 undergo a late, secondary instability (at a median time of $\sim$165 Myr) that results in the ejection of one of the ice giants (i.e.: a six giant planet system that only lost one planet during the first instability ejects an another ice giant; thus becoming a four giant planet system).  When this is the case, Jupiter and Saturn experience a corresponding ``jump'' in semi-major axis, leading to a large absolute change in $M_{55}$.  We also observe systems that lose an additional terrestrial planet during this phase of evolution.  Typically, this occurs when the eccentricity an additional $\sim$Mars-mass planet that formed between Mars and the asteroid belt \citep[e.g.:][]{chambers07} is excited to the point where it collides with one of the other planets or is scattered into the Sun.  We find that, when a terrestrial world is lost by either ejection or collision (often with an Earth or Venus analog), $M_{55}$ does not change appreciably.  In contrast, the average change in $M_{55}$ is 81$\%$ when an additional ice giant is ejected late.  

Large absolute decreases (by at least 50$\%$) in $M_{55}$ are about twice as likely as increases (20 versus 12 total systems, respectively) in our systems that do not experience a collision or planetary ejection.  Thus, our simulations indicate that it is substantially more difficult to further excite $M_{55}$ after the instability than it is to damp Jupiter's eccentricity.  Moreover, the majority of the larger relative changes in $M_{55}$ (up or down in magnitude) are in systems with lower initial $M_{55}$ magnitudes in the $M_{56}>M_{55}$ regime.  Indeed, the average change in $M_{55}$ for any system beginning with $M_{55}>$ 0.01 is just 8$\%$.  Thus, we are left with just two systems with an initial value of $M_{55}>$ 0.01 that exhibit at least a 50$\%$ increase in $M_{55}$ over 1 Gyr, without ejecting a giant planet.  On closer inspection, both of these systems begin with Jupiter and Saturn just outside of a MMR (5:2 and 3:1).  At some point in the simulation, the gas giants fall into resonance with each other, leading to chaotic and irregular behavior of the secular eigenmodes.  

We find similar results for the other amplitudes of the Jupiter-Saturn secular system (equation \ref{eqn:js}).  The magnitudes $M_{56}$ and $M_{66}$ only change by an average of 18$\%$ and 9$\%$, respectively, in systems without a late ice giant ejection.  Moreover, large magnitude decreases are around three times more likely than increases for these two modes.  However, after carefully analyzing each simulation that experiences a large change in one of the eccentric magnitudes, $M_{ij}$, we find that either a late dynamical instability or a giant planet MMR crossing occurred.  Thus, we cannot discount the possibility that Jupiter and Saturn's eccentricities were under-excited during the instability, and subsequently further excited up to their modern values by a late dynamical event.  However, such a scenario would imply that the fragile dynamical structures in the solar system \citep[e.g.: the terrestrial system, asteroid belt and binary trojans:][]{kaibcham16,delbo17,nesvorny18} survive an additional violent episode ($>$200 Myr after gas disk dissipation in our simulations).  For instance, the consequences of Jupiter and Saturn having temporarily entered the 5:2 MMR are unclear, as is whether or not they could have been perturbed back out of the resonance.  Moreover \citet{mojzsis19} found that a dynamical instability occurring after $\sim$80 Myr is not consistent with the asteroid belt's inferred geochronology.  We therefore proceed with our study under the assumption that the Jupiter-Saturn system must be effectively established by the end of the Nice Model instability.

\end{document}